\documentclass[journal = jacsat,manuscript = article]{achemso}
\usepackage{graphicx}
\usepackage{lineno}
\usepackage{chemformula} 
\usepackage[T1]{fontenc} 
\usepackage{esdiff}
\usepackage[colorlinks]{hyperref}
\usepackage[nameinlink,noabbrev]{cleveref}
\usepackage[utf8]{inputenc} 
\usepackage[T1]{fontenc}
\usepackage{soul}
\usepackage{geometry}
\author{Deepak. K. Agrawal}
\affiliation{Department of Bioengineering, University of Colorado Denver, Anschutz Medical Campus, Aurora, CO 80045, USA}
\author{Bradford J. Smith}
\affiliation{Department of Bioengineering, University of Colorado Denver, Anschutz Medical Campus, Aurora, CO 80045, USA}
\alsoaffiliation{Department of Pediatrics, University of Colorado Denver, Anschutz Medical Campus, Aurora, CO 80045, USA}
\author{Peter D. Sottile}
\affiliation{Division of Pulmonary Sciences and Critical Care Medicine, Department of Medicine,University of Colorado School of Medicine, Aurora, CO 80045, USA}
\author{David J. Albers}
\email{david.albers@cuanschutz.edu}
\affiliation{Section of Informatics and Data Science, Department of Pediatrics, School of Medicine, University of Colorado Denver, Anschutz Medical Campus, Aurora, CO 80045, USA}
\alsoaffiliation{Department of Bioengineering, University of Colorado Denver/Anschutz Medical Campus, Aurora, CO 80045, USA}
\alsoaffiliation{Department of Biomedical Engineering, Columbia University, New York, NY 10032}

\title {A damaged-informed lung model for ventilator waveforms}

\abbreviations{IR,NMR,UV}

\begin{document}
\begin{abstract}
\singlespacing{
The acute respiratory distress syndrome (ARDS) is characterized by the acute development of diffuse alveolar damage (DAD) resulting in increased vascular permeability and decreased alveolar gas exchange. Mechanical ventilation is a potentially lifesaving intervention to improve oxygen exchange but has the potential to cause ventilator-induced lung injury (VILI). A general strategy to reduce VILI is to use low tidal volume and low-pressure ventilation, but optimal ventilator settings for an individual patient are difficult for the bedside physician to determine and mortality from ARDS remains unacceptably high. Motivated by the need to minimize VILI, scientists have developed models of varying complexity to understand diseased pulmonary physiology. However, simple models often fail to capture real-world injury while complex models tend to not be estimable with clinical data, limiting the clinical utility of existing models. To address this gap, we present a physiologically anchored data-driven model to better model lung injury. Our approach relies on using clinically relevant features in the ventilator waveform data that contain information about pulmonary physiology, patients-ventilator interaction and ventilator settings. Our lung model can reproduce essential physiology and pathophysiology dynamics of differently damaged lungs for both controlled mouse model data and uncontrolled human ICU data. The estimated parameters values that are correlated with a known measure of lung physiology agree with the observed lung damage. In future endeavors, this model could be used to phenotype ventilator waveforms and serve as a basis for predicting the course of ARDS and improving patient care.
}
\end{abstract}

\section{Introduction}
\small{
\singlespacing{The acute respiratory distress syndrome (ARDS) is characterized by diffuse alveolar damage resulting in increased vascular permeability and decreased alveolar gas exchange.~\cite{RN1,RN2,RN3,RN4} Mechanical ventilation is an essential lifesaving therapy for ARDS that has the potential to worsen lung injury through barotrauma, volutrauma, and atelectrauma that are referred to collectively as ventilator induced lung injury (VILI).~\cite{RN6,RN7,RN8,RN12,RN5,RN71,RN72}Identifying lung-protective ventilation to avoid VILI can be challenging because of the complex interplay between ventilator mechanics, patient-ventilator interactions, and the underlying pulmonary physiology.~\cite{RN54,RN55,RN61,RN62} The current standard of care dictates a formulaic application of low tidal volumes to reduce overdistension and positive end expiratory pressure to maintain patency. This approach reduces VILI but does not prevent it in all cases.~\cite{RN82, RN59,RN60} One example is the ARDS Network protocol which can be used to guide ventilator settings to minimize VILI. While such protocols are very helpful, but because they are not personalized, such protocols can always be improved. This is due partially to the heterogeneity of ARDS, both between patients and in different regions of the same lung. In addition, management of patients with ARDS is further complicated by variable patient respiratory effort that may lead to patient self-inflicted lung injury.~\cite{RN81}

Modern mechanical ventilators produce time-dependent pressure, volume, and flow waveforms that contain a wealth of information about respiratory mechanics, patient-ventilator interactions, and ventilator settings. These data can be used to trouble-shoot and optimize mechanical ventilation.~\cite{RN25,RN26} However, ventilator waveforms are typically analyzed heuristically by visual inspection and, therefore, the outcome of such an analysis is limited by individual expertise.~\cite{RN25,RN26} Therefore, our goal is to develop a model-inference system to quantify the characteristics of the pressure and volume waveforms of healthy and injured lungs. This type of analysis decomposes the complex characteristics of the pressure-volume waveforms into numerical values to allow tracking changes over time. One example of this approach that is currently used in clinical care is the driving pressure, which serves as a readout of both patient condition and ventilator settings.~\cite{ RN84} We seek to expand on that methodology to provide a more comprehensive description of lung injury severity and ongoing VILI. 

Waveform-based analysis is a departure from traditional methods that utilize mathematical models to link the measured pressure and flow, such as the well-recognized single compartment model that lumps the spatially heterogeneous lung mechanical properties into single values of resistance and compliance.~\cite{RN15,RN34, RN11,RN17,RN73} Due to this straightforward formulation, the single compartment model is computationally efficient but may not be able to reproduce all of the features in measured data. On the other hand, complex multi-compartment models use many states and parameters that cannot be directly measured, such as recruitment pressure distributions, causing identifiability problems where there is no unique solution. As such, those model require more expansive data to estimate with any success, and require substantially more computational resources. Even then, complex multi-compartment models may not produce all the relevant features present in the pressure and volume data.~\cite{RN13, RN14,RN16,RN18,RN19,RN20,RN51,RN75,RN76}. 

Our novel waveform-based approach offers the potential to overcome these limitations because all of the data necessary for high-fidelity analysis is contained in the pressure and volume waveforms. We bridge the gap between identifiability and fidelity by developing a systematic framework to quantify physiological and pathophysiological lung dynamics using mathematical models that have interpretable parameters. We anticipate that this approach will find applications in real-time clinical readouts of ventilation safety, long-term monitoring to detect changes in patient condition, and as a quantitative outcome measure for clinical trials. In addition, the relationship between components of the pressure and volume waveforms may be used to identify specific physiologic features, just as the quasi-static compliance is defined as the ratio of tidal volume and driving pressure. 

In the current study, we first identify clinically important features in typical pressure and volume waveform data. We then separately define the pressure and volume waveforms as the sum of a set of essential features. This approach allows independent modeling of the components of damage so that clinical and physiologic knowledge can be used to constrain the model. The pressure and volume models are validated in a simulation study by demonstrating that the model has sufficient flexibility to produce relevant pressure and volume features. Model evaluation~\cite{RN50} is conducted with both mouse model and human ICU ventilator data~\cite{RN35} by comparing measurements and model predictions for pressure and volume waveforms. We also relate changes in the model parameters to assessments of injury severity as well as qualitative features of the pressure and volume waveforms.

\section{Methods}
\subsection{ Identifying relevant and realistic variables for the model} 

Our goal is to develop a lung model that can reproduce all the physiologically relevant features present in the waveforms data such that the model could be used to understand lung pathophysiology in clinical settings. Therefore, it is critical to identify the appropriate complexity of the model that is necessary to achieve the desired outcome.
\\
Mechanical ventilation is characterized using three state variables, volume, pressure and flow, and dozens of parameters that could be used to characterize a diversity of features including physiology and ventilator settings. In a clinical setting, ventilators are initially setting pressure or a flow pattern, and as such, pressure, flow and volume are conceptualized according to this ordering. Here, for the purposes of constructing the model, it is advantageous to begin with the less complex volume model, followed by the more complex pressure model. The flow can be derived from volume and typically these two variables contains much of the same information about the underlying lung mechanics in certain ventilation modes.~\cite{RN29,RN51} Therefore, in this study, we focused on two state variables, pressure and volume. Moreover, depending on the ventilator mode there can be a controlled variable, volume or pressure, depending on whether volume-controlled or pressure-controlled ventilation is set. There are also hybrid ventilation modes where there is not one single controlled state variable. Generally, only the independent variables contain direct information about the respiratory mechanics of the patient.~\cite{RN29,RN51} Here, we construct models of pressure and volume such that the models can represent observed pathophysiology present in all of these aforementioned situations.

\subsection{Identifying and modeling important features in the volume and pressure waveform}

The volume waveform has a characteristic shape that is typically independent on the ventilation mode and can be divided into two subprocesses (Fig.~\ref{fig1}a). The first subprocess is the inspiration, denoted as A in Fig.~\ref{fig1}a, which continues until the desired -- either by the patient or according to a ventilator setting -- tidal volume (the amount of gas delivered in that breath) is reached. The second subprocess is expiration, denoted as B in Fig.~\ref{fig1}a. 

The features in the volume waveforms that we use to delineate lung damage are directly related to variability of these two subprocesses. Depending on the ventilator settings and lung mechanics, the gradient of the rising and falling signals can vary widely not only among patients but in the same patient over time. Therefore, the model must be able to control each of these features independently. Accordingly, the gradients of inspiration and expiration in volume waveforms are features that must be variable and estimable within the volume model.

The characteristic shape of the pressure waveform can vary more dramatically than the volume waveform depending on lung mechanics and ventilation mode. In the case of pure pressure control ventilation (PCV), typically, a rectangular or trapezoidal waveform is observed.~\cite{RN25,RN26} When pressure is an independent variable, such as in volume-controlled ventilation, the pressure waveform has several important features that convey information about lung mechanics and ventilator-patient interaction (Fig.~\ref{fig1}b). 

Based on the knowledge of physiology, clinical experience, and observation of the data, we identified five features in the pressure waveform that must be captured by the model. Features one and two determine the gradient of the inspiration, which are denoted as A1 and A2 in Fig.~\ref{fig1}b. The time-varying graph of inspiration can have two distinct modes where the gradient of the signal may increase (breath 1) or decrease (breath 2) during inspiration. This might correspond to nonlinear volume-dependent lung compliance (breath 1) or an increase in compliance, indicating recruitment (breath 2).~\cite{RN11,RN17} Features three and four are related to shape of the waveform at the start and end of the plateau pressure, which is a period of constant pressure, are denoted as B1 and B2 in Fig.~\ref{fig1}b. There may be peaks at the beginning (B1) and/or at the end (B2) of the plateau pressure which may correspond to inspiratory flow resistance and patient effort, respectively.~\cite{RN25,RN51} The fifth feature is related to the onset of the expiration process, and in particular, corresponds to the gradient of expiration, denoted as C in Fig.~\ref{fig1}b. It is worth noting that we do model the constant baseline pressure is the Positive End-Expiratory Pressure (PEEP) because it is a key independent variable in ARDS management,~\cite{RN33} but this was not an additional feature we had to add to the model.

\begin{figure}[pt!]
\centerline{\includegraphics[width = 1\textwidth]{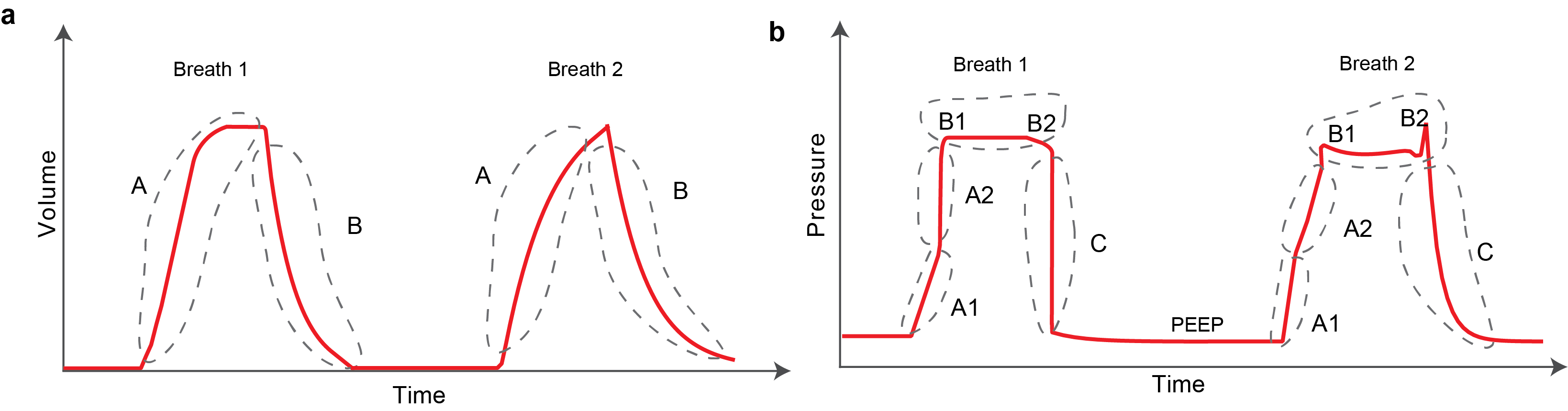}}
\caption{\small{ {\bf Graphical representation of typical volume and pressure waveforms.} {\bf (a)} Characteristic shape of the volume waveform is generally independent of the ventilation mode and has two distinct subprocesses. The rising and falling of the volume signal during inspiration and expiration, respectively, are denoted as A and B. {\bf (b)} When pressure is an independent variable, there can be multiple features in the waveform that contain useful information. The gradient of the rising signal in which pressure continues to increase during inspiration can have two distinct features, denoted as A1 and A2. These two features define the gradient of the rising signal before and after the inflection point such that there may be abrupt increases (breath 1) or decreases (breath 2) in the signal gradient. The shape of the plateau pressure is captured using features B1 and B2 such that there may be a peak at the beginning (B1-breath 2) and/or at the end (B2-breath 2) of plateau. Finally, the gradient of the falling signal is captured using feature C that represents the expiration process. The baseline pressure is known as positive end-expiratory pressure (PEEP), and often used in ARDS patient to maintain an open lung.~\cite{RN57}}}
\label{fig1}
\end{figure}

\subsection{Model validation and evaluation}
In order to establish the effectiveness of our approach, we validate and evaluate the volume and pressure models in three steps. Model validation is necessary to show that model output have enough fidelity to capture the desired variability, which is often seen in the data. Model evaluation allows the model output to represent the data via optimum parameter estimation and test whether it can be used to extract the desired outcome. Following this, first, we validate that the our volume and pressure models have the flexibility such that they can produce all the claimed variability in the waveform data. We do this by showing how model parameters allow to alter the important features in the waveform data and then how many of model parameters correspond to interpretable pathophysiology. We then evaluate that the models are indeed able to estimate data well, or in other words the model output can represent the wide variety of waveform data accurately by estimating volume and pressure ventilation data from individuals -- mouse models and humans -- with injured and healthy, or relatively healthy in the case of the human ICU data, lungs. Finally, we demonstrate the model parameters capture and represent the desired physiology and are interpretable corresponding to different lungs condition. The validation is done with model simulations without estimating data. The evaluations are done by estimating mouse and human model data.~\cite{RN50}

\subsection{Constructing the damage-informed lung model}

\emph{Construction of the volume model}: Irrespective of the state variable, the models have periodic dynamics with a frequency defined by the respiratory rate (breaths/min). In addition to this constraint, the volume model must have two additional features -- the rate of inspiration and expiration -- that must be changeable. We begin the volume model development by modeling the respiratory rate with a sinusoidal function ($f_{s_{1}}$): 
\begin{align}
f_{s_{1}} = \text{sin}( 2 \pi \theta t- \phi_1) - b_1.
\label{eq:v1}
\end{align}
Here, the respiratory rate (breaths/s) is set by $\theta$ and $t$ represents time in seconds while parameter $\phi_1$ allows to control the starting point in the respiratory cycle. Typical inspiration and expiration -- as either volume or pressure -- are not well represented by a sinusoid due to the abrupt rise from a baseline volume or pressure as shown in Fig.~\ref{fig1}. To control the rate of inspiration or expiration while maintaining the periodicity, we create a periodic rectangular waveform function $f_{b_{1}}$ by combining the sinusoidal function with hyperbolic tangent function: 
\begin{equation}
f_{b_{1}} = \frac{1}{2}\{\text{tanh}(a_1 f_{s_{1}})+1\}.
\label{eq:v2}
\end{equation}
To control the smoothness of the rectangular waveform, we added a smoothing parameter $a_1$. The other terms (1/2, +1) are added to generate a rectangular waveform that has a zero-base value and unit amplitude. To control the duty cycle of the rectangular waveform that sets inspiratory:expiratory ratio, we used parameter $b_1$ such that zero value of $b_1$ corresponds to 1:1 I:E ratio. 

In Fig.~\ref{fig1}a shows additional model features: the rate of inspiration and expiration. To represent these two rates independently we created two separate submodels that define the volume ($V$) using the rectangular waveform as a base waveform: 
\begin{align}
V = A_{v} ( f_{v_{1}} +f_{v_{2}}),
\label{eq:v3}
\end{align}
where
\begin{align}
f_{v_{1}} = [\sum_{i = 1}^{n} \{ \frac{1}{\beta_1}f_{b_{1}}(i) + (1- \frac{1}{\beta_1}) f_{v_{1}}(i-1)\} ] \frac{f_{b_{1}}}{\text{max}(f_{v_{1}})},
\label{eq:v4}
\end{align}
and
\begin{align}
f_{v_{2}} = [\sum_{i = 1}^{n} \{ \frac{1}{\beta_2} f_{b_{1}}(i) + (1- \frac{1}{\beta_2}) f_{v_{2}}(i-1)\}] \frac{(1-f_{b_{1}})}{\text{max}(f_{v_{2}})}.
\label{eq:v5}
\end{align}
Here, $\beta_1$ and $\beta_2$ control the gradient of the inspiration and expiration, respectively, while $A_{v}$ controls the amplitude of the volume waveform.

\emph{Construction of the pressure model}: We begin building the pressure model as we did the volume model, by modeling the respiratory rate and the I:E ratio. The pressure model has five features that must be changeable, the gradient of the rising signal during inspiration at low (1) and high (2) volume, the shape of the peaks at the beginning (3) and end (4) of the plateau pressure, and the rate of change of the pressure during expiration (5). While volume and pressure are coupled in several ways, the most foundational coupling is via their period. We enforce this constraint by requiring that both models have the same respiratory frequency ($\theta$) in their base periodic sinusoid: 
\begin{align}
f_{s_{2}} = \text{sin}( 2 \pi \theta t- \phi_2) - b_2.
\label{eq:p1}
\end{align}
Because the pressure may lag or lead the volume, we include a phase shift term, $\phi_2$ in the sinusoid. Additionally, to account for variations in the I:E ratio we added the parameter $b_2$. We then create a rectangular waveform submodel $f_{b_{2}}$ as we did for the volume model using the hyperbolic tangent, or: 
\begin{align}
f_{b_{2}} = \frac{1}{2}\{\text{tanh}(a_2 f_{s_{2}})+1\}.
\label{eq:p2}
\end{align}
The five key features in pressure are represented with three submodels: (i) $f_{p_{13}}$ defines the rates of pressure change during inspiration and expiration, (ii) $f_{p_{24}}$ determines the peaks at the beginning and end of the pressure plateau, and (iii) $f_{p_{33}}$ specifies the gradient of the initial rising signal during inspiration, leaving us with the full the pressure model ($P$):
\begin{equation}
P = f_{p_{13}} + f_{p_{24}} + f_{p_{33}} + A_{p_{4}}.
\label{eq:p3}
\end{equation}
The constant parameter $A_{p_{4}}$ corresponds to the baseline pressure value (PEEP). The rates of pressure change during inspiration and expiration (see A2, and C in Fig.~\ref{fig1}b, respectively) are:
\begin{align}
f_{p_{13}} = A_{p_{1}} (f_{p_{11}} + f_{p_{12}}),
\label{eq:p4}
\end{align}
where
\begin{align}
f_{p_{11}} = \sum_{i = 1}^{n} \{ \frac{1}{\beta_3}f_{b_{2}}(i) + (1- \frac{1}{\beta_3}) f_{p_{11}}(i-1)\} \frac{f_{b_{2}}}{\text{max}(f_{p_{11}})},
\label{eq:p5}
\end{align}
and
\begin{align}
f_{p_{12}} = \sum_{i = 1}^{n} \{ \frac{1}{\beta_4} f_{b_{2}}(i) + (1- \frac{1}{\beta_4}) f_{p_{12}}(i-1)\} \frac{(1-f_{b_{2}})}{\text{max}(f_{p_{12}})}.
\label{eq:p6}
\end{align}
Here, $\beta_3$ and $\beta_4$ control the gradient during inspiration and expiration, respectively. The next set of features, the peaks at the beginning and end of plateau pressure (see B1, and B2 in Fig.~\ref{fig1}b), are modeled by: 
\begin{equation}
f_{p_{24}} = A_{p_{2}} \frac{|(f_{p_{23}})|}{\text{max}(|(f_{p_{23}})|)},
\label{eq:p7}
\end{equation}
where
\begin{equation}
f_{p_{21}} = \frac{1}{\beta_5} \sum_{i = 1}^{n} [ f_{p_{21}}(i-1) +\{f_{b_{2}} (i)-f_{b_{2}}(i-1) \},
\label{eq:p8}
\end{equation}
\begin{equation}
f_{p_{22}} = f_{p_{21}}f_{b_{2}}, 
\label{eq:p9}
\end{equation}
and
\begin{equation}
f_{p_{23}} = \frac{1}{\beta_6} \sum_{i = 1}^{n} [ f_{p_{23}}(i-1) + \{f_{p_{22}} (i)-f_{p_{22}}(i-1) \}].
\label{eq:p10}
\end{equation}
The parameters $\beta_5$ and $\beta_6$ control the shape of both the peaks. Finally, the gradient of the initial rate of inspiration, (A1 in Fig.~\ref{fig1}b), is modeled by: 
\begin{equation}
f_{p_{33}} = A_{p_{3}} \frac{f_{p_{32}} \{1- (f_{p_{11}}+ f_{p_{12}})\}}{\text{max}[f_{p_{32}} \{1- (f_{p_{11}}+ f_{p_{12}})\}]},
\label{eq:p11}
\end{equation}
where
\begin{equation}
f_{p_{31}}(t) = \text{sin}( 2\pi \theta t - \phi_3) - b_3,
\label{eq:p12}
\end{equation}
and
\begin{equation}
f_{p_{32}} = \frac{1}{2}\{\text{tanh}(a_3 f_{p_{31}})+1\}.
\label{eq:p13}
\end{equation}
The position, shape and gradient of the rising signal, produced by $f_{p_{33}}$ submodel are controlled using the parameters $\phi_3$, $b_3$ and $a_3$, respectively. 

\subsection{Mouse Mechanical Ventilation Experiments}

A nine week old female BALB/c mouse (Jackson Laboratories, Bar Harbor, ME, USA) was studied under a University of Colorado Anschutz Medica Campus Institutional Animal Care and used Committee (IACUC)-approved protocol (\#00230). Anesthesia was induced with an intraperitoneal (IP) injection of 100~mg/kg Ketamine and 16~mg/kg Xylazine, a tracheostomy was performed with a 18~ga metal cannula, and ventilation was started on the flexiVent small animal ventilator (SCIREQ, Montreal, QC, Canada). Anesthesia was maintained with 50~mg/kg Ketamine or 50~mg/kg Ketamine with 8~mg/kg Xylazine at 30~min intervals along with 50~$\mu$L IP 5\% dextrose lactated Ringer’s solution. Respiratory efforts were suppressed with 0.8~mg/kg pancuronium bromide administered at 90 min intervals. Heart rate was monitored via electrocardiogram.
\\
Baseline ventilation, consisting of a tidal volume (Vt) = 6~ml/kg, PEEP = 3~cmH$_2$O, and respiratory rate (RR) = 250~BPM, was applied for 10~mins with recruitment maneuvers at 3~min intervals. Pressure and volume were recorded with a custom flowmeter based on our previously published design (REF SAMER PAPER). Three types of ventilation were recorded for analysis: LowVT-PEEP0, consisting the baseline ventilation with PEEP = 0~cmH$_2$O, LowVT-PEEP12 that was the baseline ventilation with PEEP = 12~cmH$_2$O, and HighPressure that consisted of (Pplat) = 35~cmH$_2$O at PEEP = 0~cmH$_2$O with RR = 60~BPM. Lung injury was induced with a 0.15~ml lavage with warm saline. This fluid was pushed into the lung with an additional 0.3~ml air, and suction was applied to the tracheal cannula with an approximate return of 0.05~ml. The mouse was ventilated for 10 mins with a plateau pressure (Pplat) = 35~cmH$_2$O, PEEP = 0~cmH$_2$O, and respiratory rate (RR) = 60~BPM and the LowVT-PEEP0, LowVT-PEEP12, and HighPressure ventilation was recorded again. 

\subsection{Human Data Collection}

Between June 2014 and January 2017, adult patients admitted to the University of Colorado Hospital medical intensive care unit (MICU) at risk for or with ARDS and requiring mechanical ventilation were enrolled within 12 hours of intubation.~\cite{RN39} At risk patients were defined as intubated patients with hypoxemia and a mechanism of lung injury known to cause ARDS, who had not yet met chest x-ray or oxygenation criteria for ARDS. To facilitate the capture of continuous ventilator data, only patients ventilated with a Hamilton G5 ventilator were included. Patients requiring mechanical ventilation only for asthma, COPD, heart failure, or airway protection were excluded. Additionally, patients less than 18 years of age, pregnant, or imprisoned were excluded. The University of Colorado Hospital utilizes a ventilator protocol that incorporated the ARDS network low tidal volume protocol with the low PEEP titration table. The Colorado Multiple Institutional Review Board approved this study and waived the need for informed consent.

Baseline patient information including age, gender, height, and initial P/F ratio were collected. Continuous ventilator data were collected using a laptop connected to the ventilator and using Hamilton DataLogger software (Hamilton, v5.0, 2011) to obtain pressure, flow, and volume measurements. Additionally, the DataLogger software allowed collection of ventilator mode and ventilator settings based on mode (i.e.: set tidal, respiratory rate, positive end-expiratory pressure (PEEP), and fraction inspired oxygen (FiO$_2$)). Data were collected until extubation or for up to seven days per patient.

\subsection{Parameter estimation methodology}

Estimating model parameters is relatively straightforward when the model is \emph{identifiable} given data, or, the model is constructed such that every state and parameter is uniquely estimable and there are enough data to uniquely estimate every state and parameter uniquely.~ \cite{RN46,RN67,RN68} In practice, most models are not identifiable even with ideal data. Moreover, in clinical settings -- where we eventually want to use this model -- the data are often noisy and difficult to use \cite{RN42,RN43,RN44}. Given this reality, we must use care to set up the inference task such that we can ensure robust results with quantifiable uncertainty \cite{RN45}. This forces three issues, how to choose and limit model features estimated, how to choose an inference methodology, and how to manage uncertainty quantification.

First issue of limiting model features estimated is important to minimize identifiability failure where there is no unique solution in terms of best parameters values for a given data. We employ two approaches for managing identifiability failure.~\cite{RN46,RN47} In the first approach, we estimate all parameters but constrain their ranges to lie within physically possible values while in the second approach, we fix many low-impact, low-sensitivity parameters, and estimate a limited number of parameters that are chosen based on features present in the waveform data.~\cite{RN40} For example, in the mouse-model data, shown in Fig.~\ref{fig6} and Supplementary Fig.~S6, the peaks at the plateau pressure did not appear, and because of this, we did not estimate parameters that control those peaks ($\beta_5$, $\beta_6$ and $A_{p_{2}}$). Similarly, for the the human data, shown in Fig.~\ref{fig7}, the characteristic shape of the volume and pressure waveforms remain the same at different time points except for significant variations in the peak amplitudes. Therefore, for the first breath we estimated all the parameters but kept certain parameters ($\beta_1$, $\beta_2$, $a_3$, $b_3$, $\beta_3$) constant in the second breath to maintain the characteristic shape of the volume and pressure waveforms between the two breaths. 

 Second and third issues are choosing an inference methodology that would allow to estimate states and parameters of the model effectively, and the respective uncertainties in the estimated parameters.~\cite{RN48,RN49}. While stochastic methods, e.g., Markov Chain Monte Carlo (MCMC)~\cite{RN70}, might guarantee to find global minima and quantifying uncertainty in the estimated parameters values, they are generally quite slow. On the other hand, deterministic methods, e.g., Nelder-Mead optimization~\cite{RN69}, are substantially faster and by choosing many initial conditions, a robust solution may be obtained. Therefore, here we focused on a smoothing or optimization task that employ deterministic inference scheme. 

In this study, we used MATLAB FMINCON function, which is a gradient-based minimization algorithm for nonlinear functions. To ensure a robust solution and to quantify uncertainty we additionally used MATLAB MULTISTART function that performs optimization starting from multiple start points. MULTISTART effectively boostraps the optimization, uniformly sampling optimization initial conditions across a provided interval. We determine realistic lower and upper bound values (constraints) for each case using an iterative method and these bounds define the constraints employed by the parameter estimation problem in the optimization scheme. A full description for the computational and mathematical aspects and implementation of parameter estimation methodology can be found in ref. 56.~\cite{RN40} This approach not only allows to determine the best fit parameter values but the respective uncertainties as well while trying to find global or multiple minima depending on the solution surface for each parameter.

\section{Results}

We validate and evaluate the lung models using numerical simulations and measured data, respectively.~\cite{RN50} In the validation step, we demonstrate that the models have the flexibility to the desired variability through simulations and identify the parameters that correspond to interpretable pathophysiology by analyzing simulated pressure-volume waveforms. In the evaluation step, we demonstrate that the model parameters capture and represent the desired physiology and are interpretable by estimating volume and pressure ventilation data. 

\subsection{Validation of volume and pressure models} 
\begin{figure}[h!]
\centerline{\includegraphics[width = 0.4\textwidth]{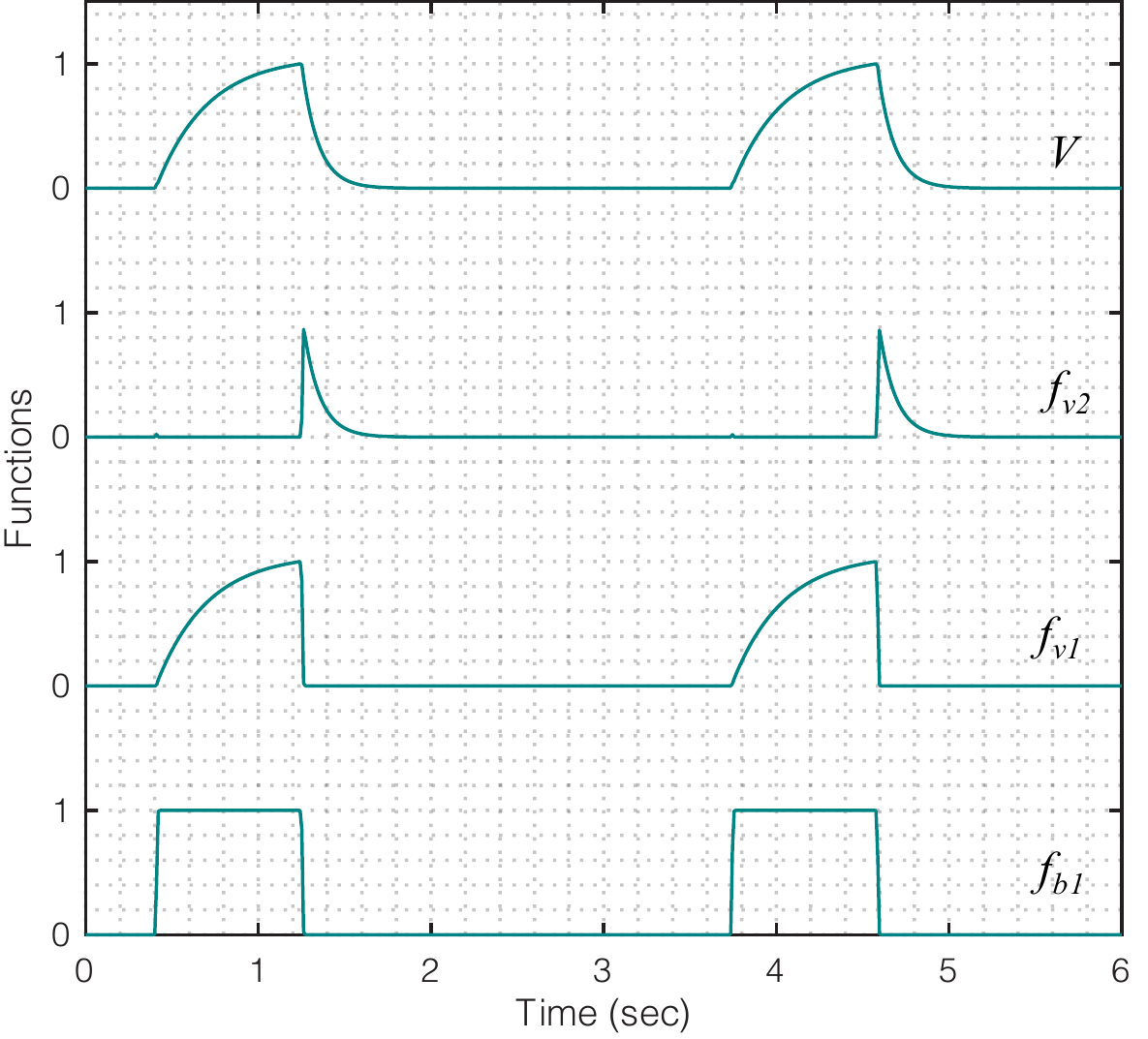}}
\caption{\small{ {\bf Simulated response of various submodels that make up the injury-inclusive volume model ($V$).} A periodic rectangular waveform submodel $f_{b_{1}}$ is used to create two more submodels ($f_{v_{1}}$ and $f_{v_{2}}$) through which the gradient of the rising and falling signals in the volume waveform are controlled, respectively. Equations~\ref{eq:v1}-\ref{eq:v5} were used to simulate the response of each submodel with parameter values $\theta$ = 0.3, $a_1$ = 200, $b_1$ = 0.7, $\phi_1$ = 0, $\beta_1$ = 30, $\beta_2$ = 10, $A_{v}$ = 1.}}
\label{fig2}
\end{figure}
\begin{figure}[h!]
\centerline{\includegraphics[width = 0.75\textwidth]{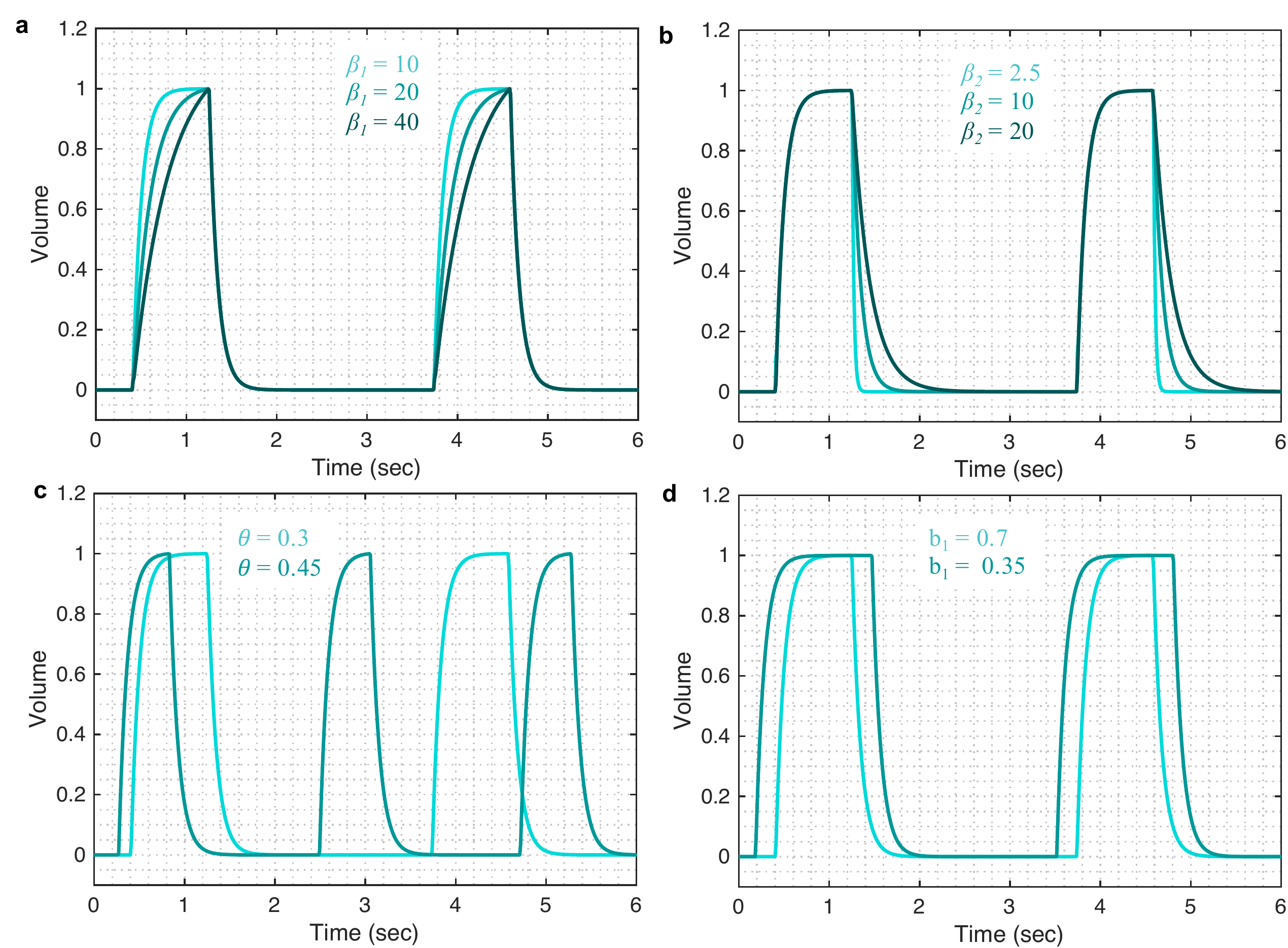}}
\caption{\small{ {\bf Demonstrating the volume model flexibility by varying specific parameters that allow altering the gradient of the rising and falling signals, respiratory rate and I:E ratio.} The gradient of the rising and falling signals can be altered using the {\bf (a)} $\beta_1$ and {\bf (b)} $\beta_2$ parameters, respectively. Increased values of these parameters increase the transient time for the signal to reach the same volume level. {\bf (c)} Changes in the respiratory frequency ($\theta$) change the period of the breath while {\bf (d)} the I:E ratio (inspiratory to expiratory time ratio) can be modified using the $b_1$ parameter. The output of the model ($V$) was calculated using Eqns.~(1)-(5) while considering $\theta$ = 0.3, $a_1$ = 200, $b_1$ = 0.7, $\phi_1$ = 0, $\beta_1$ = 10, $\beta_2$ = 10, $A_{v}$ = 1. The respective variation in the submodels that make the volume model is shown in Fig.~S1 for each case. Additional control on these features is shown in the Supplementary Fig.~S2}}
\label{fig3}
\end{figure}
\emph{Validation of volume model}: Figure \ref{fig2} shows the volume model and the three submodels it is constructed from, detailed in Eqns.~\ref{eq:v1}-\ref{eq:v5}. The volume model is a sum of the inspiration and expiration submodels, and is shown as the top plot of Fig.~\ref{fig2}. 
The effective variability in rates of inspiration and expiration, specified by $\beta_1$ and $\beta_2$ respectively, is shown in Fig.~\ref{fig3}a and~\ref{fig3}b. The respective variation in the submodels is shown in Supplementary Fig.~S1a and b, respectively. Additionally, the peak amplitude value of the volume waveform can be changed by altering $A_v$; this variability is shown in the Supplementary Fig.~S2a. Variations in respiratory rate are controlled by the respiratory frequency ($\theta$) and is shown in Fig.~\ref{fig3}c and in the Supplementary Fig.~S1c. The I:E ratio is represented through the parameter $b_1$ that changes the duty cycle of the rectangular base waveform, and is shown in Fig.~\ref{fig3}d and in the Supplementary Fig.~S1d. Finally, the starting point of the breath in the breathing cycle and the smoothness of the volume waveform are set by $\phi_1$ and $a_1$ respectively, and are shown in the Supplementary Figs.~S2b and c, respectively. 

\emph{Validation of pressure model}: Figure \ref{fig1}b demonstrates the features of the pressure waveform that we deem important for understanding lung function and lung damage. Each of these features in Fig.~\ref{fig1}b is controlled by a specific submodel with associated parameters that dictate the shape of that feature while its contribution is controlled via the respective amplitude term. Figure \ref{fig4} shows the pressure model and the ten submodels it is constructed from, detailed in Eqns.~\ref{eq:p1}-\ref{eq:p13}. 

\begin{figure}[h!]
\centerline{\includegraphics[width = 0.5\textwidth]{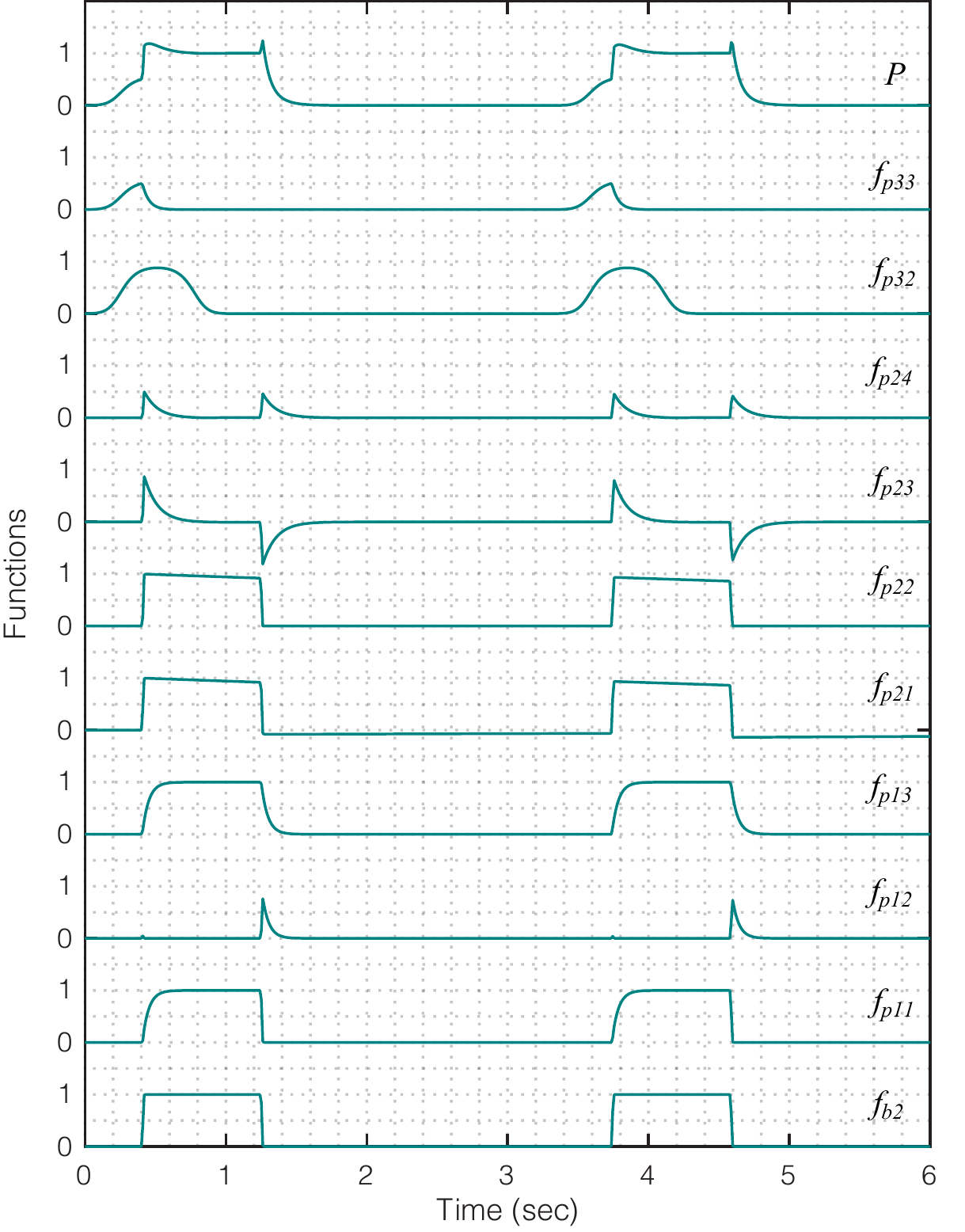}}
\caption{\small{ {\bf Simulated timing response of various submodels that make up the injury-inclusive pressure model ($P$).} A periodic rectangular waveform ($f_{b_{2}}$) serves as a basis to create other submodels that contribute to the pressure model. The overall shape of the pressure waveform, which defines gradient of the inspiration and expiration signals are formed using $f_{p_{13}}$ submodel comprised of the rising signal of $f_{p_{11}}$ (A2) and falling signal of $f_{p_{12}}$ (C). The shape of the plateau pressure is defined by $f_{p_{24}}$, where the output of $f_{b_{2}}$ is processed via $f_{p_{21}}$, $f_{p_{22}}$ and $f_{p_{23}}$ to produce peaks at the beginning (B1) and end (B2) of the plateau pressure. The shape of the rising signal at low volume (A1) is defined by $f_{p_{33}}$, where a short pulse is produced via $f_{p_{31}}$ and reshaped via $f_{p_{32}}$. Note that the amplitude terms $A_{p_{1}}$, $A_{p_{2}}$ and $A_{p_{3}}$ control the amplitude of $f_{p_{13}}$, $f_{p_{24}}$ and $f_{p_{33}}$ submodels, respectively. Equations~\ref{eq:p1}-\ref{eq:p13} were used to simulate the response of each submodel with parameter values $\theta$ = 0.3, $a_2$ = 200, $b_2$ = 0.7, $\phi_2$ = 0, $a_3$ = 10, $b_3$ = 0.9, $\phi_3$ = -0.6, $\beta_3$ = $\beta_4$ = 5, $\beta_5$ = 1.001, $\beta_6$ = 1.1111, $A_{p_{1}}$ = 1, $A_{p_{2}}$ = 0.5, $A_{p_{3}}$ = 0.5, $A_{p_{4}}$ = 0.}}
\label{fig4}
\end{figure}

The validation of the model flexibility is shown in Fig.~\ref{fig5} and Supplementary Fig.~S2d-i. We carry out this validation by varying five features of the pressure waveform; variation in the rate of change of the pressure before (A1 in Fig.~\ref{fig1}b) and after (A2 in Fig.~\ref{fig1}b) the inflection point during inspiration; the shape of the peaks at the beginning (B1 in Fig.~\ref{fig1}b) and end (B2 in Fig.~\ref{fig1}b) of the plateau pressure; and variation in the rate of change of the pressure during expiration (C in Fig.~\ref{fig1}b). In brief, these features are controlled by the following parameters. 

The initial gradient of the pressure during inspiration (A1) is controlled by the $a_3$ parameter such that higher values of $a_3$ result in a slower rising signal as seen in Fig.~\ref{fig5}a and in the Supplementary Fig.~S3a. The shape of the initial gradient signal before inflection point can be altered using the $b_3$ parameter as shown in the Supplementary Fig.~S2d. And the amplitude of the initial gradient alteration is controlled by the $A_{p_{3}}$ parameter as shown in the Supplementary Fig.~S2e. The gradient of pressure dynamics at inspiration after the inflection point (A2) is specified by $\beta_3$ such that higher values of $\beta_3$ result in a slower rising signal as seen in Fig.~\ref{fig5}b and in the Supplementary Fig.~S3b. The shapes of the peaks at the beginning (B1) and end (B2) of the plateau pressure are controlled by several parameters. The overall shape of the peaks is controlled by the $\beta_5$ parameter for a given $\beta_6$ as can be observed in Fig.~\ref{fig5}c and in the Supplementary Fig.~S3c. The sharpness of these peaks can be altered further by the $\beta_6$ parameter for a given shape of the peaks as shown in the Supplementary Fig.~S2f. The amplitude of the peaks is controlled by the $A_{p_{2}}$ parameter whose effect can be seen in the Supplementary Fig.~S2g. Additionally, we can control individual peaks by the parameter $\beta_3$ as shown in the Supplementary Fig.~S2h and i. And finally, variation in the gradient of pressure dynamics at expiration (C) is specified by $\beta_4$ such that higher values of $\beta_4$ result in a slower falling signal as seen in Fig.~\ref{fig5}d and in the Supplementary Fig.~S3d. Along with these, the I:E ratio is characterized by the $b_2$ parameter in the same way that parameter $b_1$ controls the I:E ratio in the volume model, cf Fig.~\ref{fig3}d.
\begin{figure}[t!]
\centerline{\includegraphics[width = 0.8\textwidth]{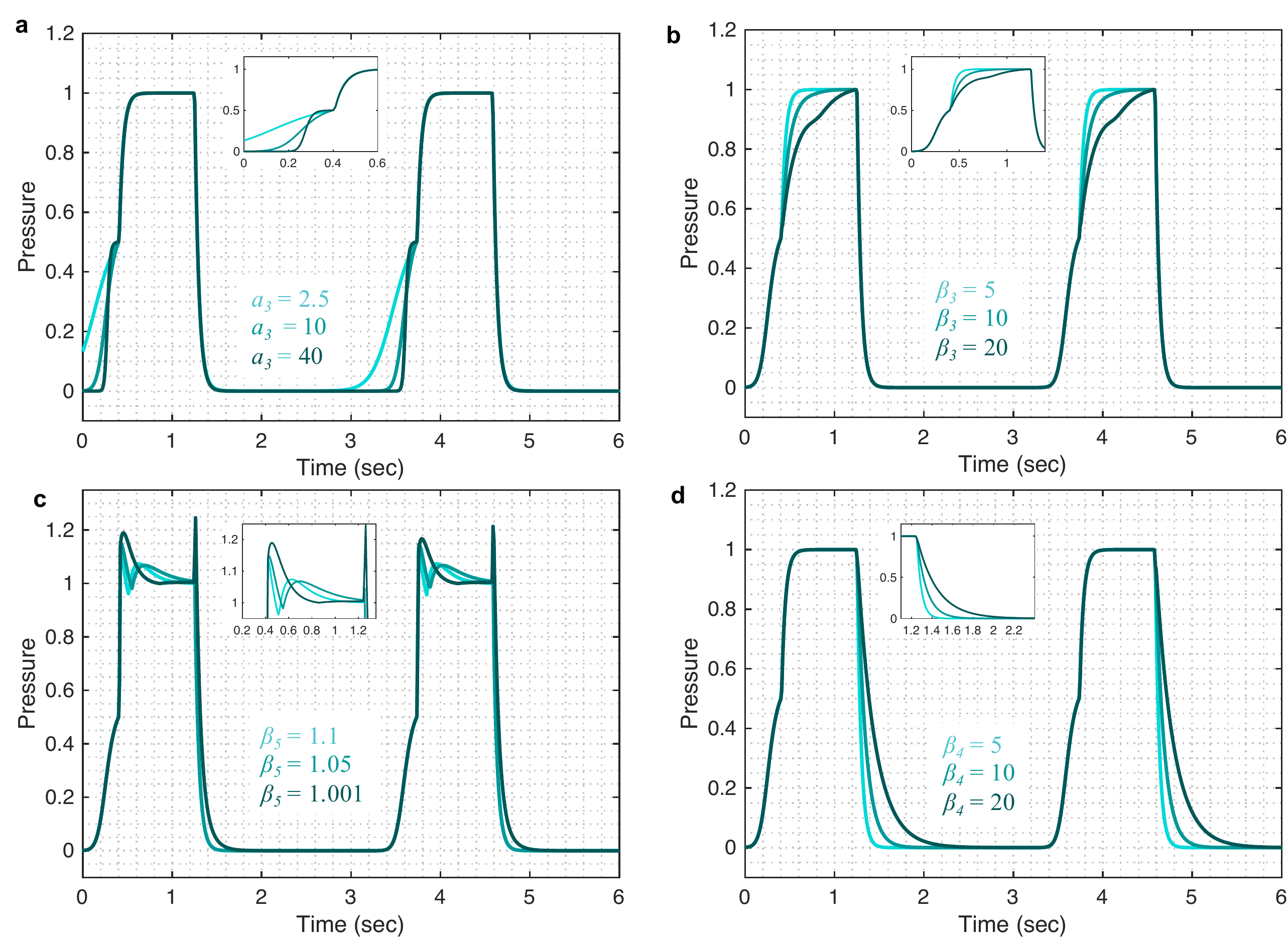}}
\caption{\small{ {\bf Demonstrating the pressure model flexibility by altering physiologically relevant features.} {\bf (a)} The initial gradient of the pressure signal during inspiration at low volume (A1) is controlled by the $a_3$ parameter {\bf (b)} The gradient of the rising signal after the inflection point (A2), is controlled by the $\beta_3$ parameter. {\bf (c)} The shapes of the peaks at the beginning (B1) and at the end (B2) of the plateau are regulated by the $\beta_5$ parameter when $A_{p_{4}}$ = 0.5. {\bf (d)} The gradient of the falling signal (C) during expiration can be modified by the $\beta_4$ parameter. Equations~\ref{eq:p1}-\ref{eq:p13} were used to simulate the response of the pressure model while considering $\theta$ = 0.3, $a_2$ = 200, $b_2$ = 0.7, $\phi_2$ = 0, $a_3$ = 10, $b_3$ = 0.9, $\phi_3$ = -0.6, $\beta_3$ = $\beta_4$ = 5, $\beta_5$ = 1.001, $\beta_6$ = 1.1111, $A_{p_{1}}$ = 1, $A_{p_{2}}$ = 0, $A_{p_{3}}$ = 0.5, $A_{p_{4}}$ = 0. A zoomed-in view of each plot is shown inside the respective plot to highlight the changes in the waveform. The respective variations in the submodels that make the pressure model is shown in Fig.~S3 for each case. Additional control on these features is shown in the Supplementary Fig.~S2}}
\label{fig5}
\end{figure}

\subsection{Linking model parameters to lung function}
The next step of our model validation is to demonstrate how the model parameters can be related to physiology. In the volume model, we focus on three parameters that have physiological meaning: $\beta_1$, $\beta_2$ and $A_v$. The rate of inspiration is controlled by the $\beta_1$ parameter, which is shown as feature A in Fig.~\ref{fig1}a. Higher values of $\beta_1$ result in a lower inspiratory flow rate (Supplementary Fig.~S4). During pressure control ventilation (PCV), inspiratory flow rate can change due to reduction in lung compliance and/or increase in lung resistance. Alternatively, during volume-controlled ventilation (VCV), this feature corresponds to the set inspiratory flow rate. The gradient of expiration is controlled by the $\beta_2$ parameter and is captured as feature B in Fig.~\ref{fig1}a. Higher values of $\beta_2$ result in a longer expiration (Supplementary Fig.~S4). This parameter is directly proportional to the expiratory time constant which is the product of resistance and compliance. Finally, the tidal volume in VCV is represented by the amplitude parameter, $A_v$. In PCV, higher values of $A_v$ for the same pressure waveform would suggest an increase in the overall compliance (Supplementary Fig.~S4). There are several other parameters in the volume model that represent settings controlled by the ventilator such as respiratory frequency, I:E ratio etc. A short description of how these, and other, model parameters contribute to the model is provided in the Supplementary Table S1.

In the pressure model, we identified five parameters that are associated with aspects of lung compliance during VCV: $a_3$, $b_3$, $\beta_3$, $A_{p_{1}}$ and $A_{p_{3}}$. During PCV, these (and other) parameters may be directly controlled via ventilator. The gradient of the initial rising pressure signal (A1) is controlled by the $a_3$ parameter and higher values of $a_3$ result in slower pressure rise at low volume while maintaining the shape of the gradient as shown in the Supplementary Fig.~S5. We can therefore directly relate this parameter to the low volume compliance during VCV such that higher values of $a_3$ would suggest an increase in the low volume compliance and vice versa. 

The shape of the initial rising pressure signal at the onset of inspiration (A1) is also controlled by the $b_3$ parameter such that higher values of $b_3$ result in slower pressure rise at low volume while changing the shape of the gradient as shown in the Supplementary Fig.~S5. Note that parameters $a_3$ and $b_3$ control the same feature in the pressure waveform (A1) but different aspects of it which might be relevant to distinguish the cases where alveoli recruitment varies substantially at low volume.

The gradient of the rising signal above the inspiratory inflection point (A2) is controlled by the $\beta_3$ parameter, and higher values of $\beta_3$ result in slower pressure rising signal as shown in the Supplementary Fig.~S5. We relate this parameter directly to the high volume compliance during VCV such that higher values of $\beta_3$ would suggest an increase in the high volume compliance and vice versa. 

The pressure value at the plateau is defined using the $A_{p_{1}}$ parameter, and higher values of $A_{p_{1}}$ result in higher values of the plateau pressure as shown in the Supplementary Fig.~S5. This parameter is inversely related to the overall lung compliance such that increasing values of $A_{p_{1}}$ would suggest a reduction in the compliance and vice versa, given the tidal volume does not change. 

Finally, change in the upper inflection point (UIP) can be directly related to the $A_{p_{3}}$ parameter such that higher values of $A_{p_{3}}$ increase the value of UIP in the waveform while maintaining the shape of the pressure waveform as shown in the Supplementary Fig.~S5. 
 
 It is important to note that these interpretations are valid only when a change is observed in one of the variables (volume or pressure) while having the other features of the waveforms fixed. There may be cases where both volume and pressure waveforms change simultaneously and, in those cases, additional interpretation is needed to establish the relationships between pressure and volume parameters. For example, when there is a change in the amplitude of volume and pressure simultaneously, $A_v$/$A_{p_{1}}$ ratio should be considered to determine the over change in the lung compliance. 

\subsection{Model evaluation with animal and human data}

In the previous sections, we validated that the model can simulate the diversity of observable volume and pressure features we had previously identified as important. The validation is carried out without data, and therefore without an inference task. Here, we begin the data-driven model evaluation by showing that the model is indeed flexible enough to estimate the pathophysiology we designed it to estimate.

\emph{Parameter selection and estimation}: As mentioned in the Methods section, we do not always estimate every parameter. In particular, we did not infer parameters that control features that were not observed in the data to reduce confounding problems. In more detail, for the mouse model experiments shown in Fig.~\ref{fig6}, we estimated $a_1$, $b_1$, $\phi_1$, $\beta_1$, $\beta_2$ $A_v$, $a_2$, $b_2$, $\phi_2$, $a_3$, $b_3$, $\phi_3$, $\beta_3$, $\beta_4$, $A_{p_1}$, $A_{p_3}$, $A_{p_4}$, and held the parameters that control the pressure plateau peaks, $\beta_5$, $\beta_6$, $A_{p_{2}}$, constant. For the retrospective human data-based evaluation, shown in Fig.~\ref{fig7}, we estimated $a_1$, $b_1$, $\phi_1$, $A_v$ $a_2$, $b_2$, $\phi_2$, $\phi_3$, $\beta_4$, $\beta_5$, $\beta_6$, $A_{p_1}$, $A_{p_2}$, $A_{p_3}$, $A_{p_4}$. We then held the variables that control the shape and the gradients at inspiration and expiration -- constant, including $\beta_1$, $\beta_2$, $a_3$, $b_3$ $\beta_3$. In order to maintain the coupling between volume and pressure models, respiratory rate, $\theta$, was kept constant for each dataset. 
\begin{figure}[t!]
\centerline{\includegraphics[width = 1\textwidth]{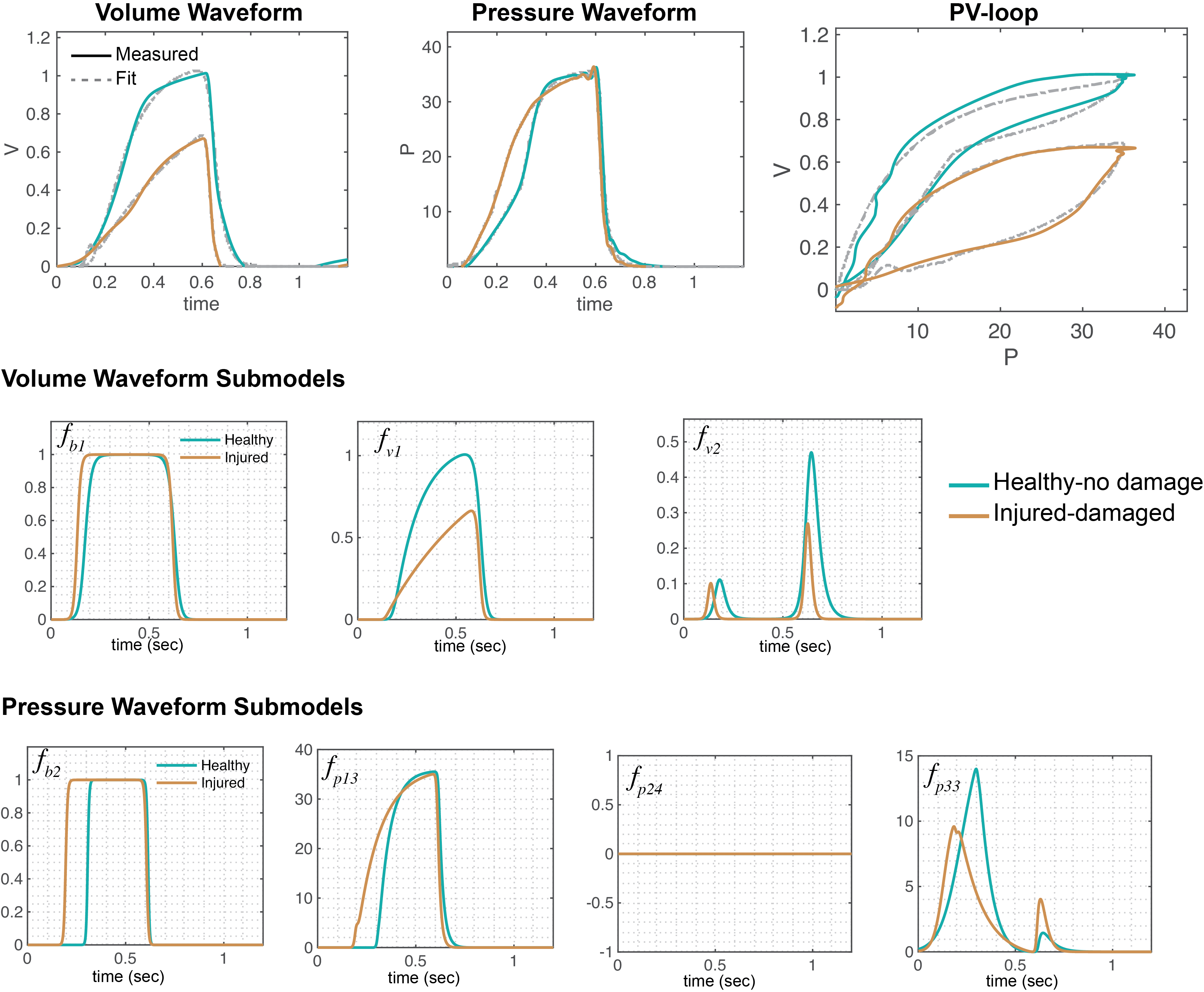}}
\caption{\small{ {\bf Volume and pressure models responses closely agree with the experimental data from a representative mouse in healthy and injured condition.} In the first row, the measured response is shown in solid lines while the model inferred response is shown in dashed lines. Changes in the volume and pressure submodels are shown in the second and third rows, respectively (in solid lines). The volume and pressure models shown in Eqns.~\ref{eq:v1}-\ref{eq:v5} and \ref{eq:p1}-\ref{eq:p13} were used to generate the best-fit model response using estimated mean parameter values shown in Table~\ref{table1}, respectively. The respective uncertainties in the parameter values are shown in Table~\ref{table1} estimations for each breath. }}
\label{fig6}
\end{figure}

\emph{Data selection}: Each data set contained thousands of breaths. In an effort to perform a more controlled evaluation, we isolated a single breath in each case that is representative of the breaths in that data set and performed the parameter estimation and evaluation on those data. The best-fit parameter values for Fig.~\ref{fig6} and~\ref{fig7} are shown in Table~\ref{table1} and for Supplementary Fig.~S6 and S7 are shown in Supplementary Table S2 with 95\% confidence intervals with respect to mean. 

\emph{Broad model evaluation}: Figure 6 shows two breaths measured in the same mouse when healthy (green) and after lung injury (orange) during ventilation with Pplat = 35~cmH$_2$O and PEEP = 0~cmH$_2$O. The model estimates are shown in dashed lines and the submodels of the volume and pressure waves are shown in the 2nd and 3rd rows, respectively. Low tidal volume ventilation measurements for these two time points are shown in the supplement for PEEP = 0 cmH$_2$O (see Supplementary Fig.~S6a) and PEEP = 12 cmH$_2$O (see Supplementary Fig.~S6b). The model states and parameters were also estimated using data from two patients with ARDS, shown in Fig.~\ref{fig7} and in the Supplementary Fig.~S7. As can be seen in the figures (Fig.~\ref{fig6},~\ref{fig7}, and Supplementary Fig.~S6, S7), the models are able to accurately estimate all data and their observed pathophysiology.
\begin{figure}[t!]
\centerline{\includegraphics[width = 1\textwidth]{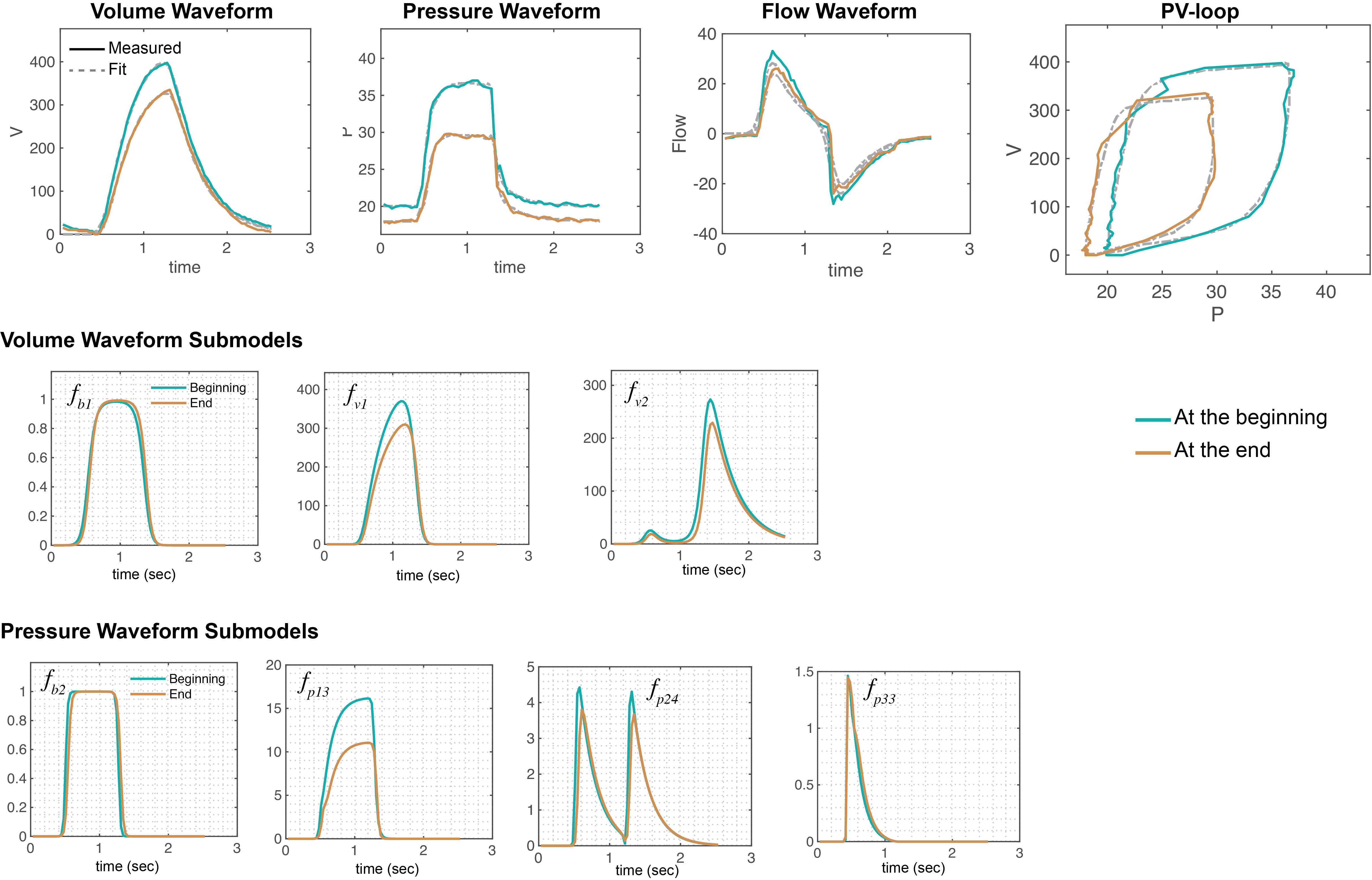}}
\caption{\small{ {\bf Damaged informed-lung model can accurately follow two different breaths of an ICU patient with ARDS. } In the first row, the measured response is shown in solid lines while the model inferred response is shown in dashed lines. Changes in the volume and pressure submodels are shown in the second and third rows, respectively (in solid lines). The volume and pressure models shown in Eqns.~\ref{eq:v1}-\ref{eq:v5} and \ref{eq:p1}-\ref{eq:p13} were used to generate the best-fit model response using estimated mean parameter values shown in Table~\ref{table1}, respectively. The respective uncertainties in the parameter values are shown in Table~\ref{table1} estimations for each breath.}}
\label{fig7}
\end{figure}
 \begin{table}[ht!]
 \begin{center}
 \begin{tabular}{ c }
 \includegraphics[width = 0.8\textwidth]{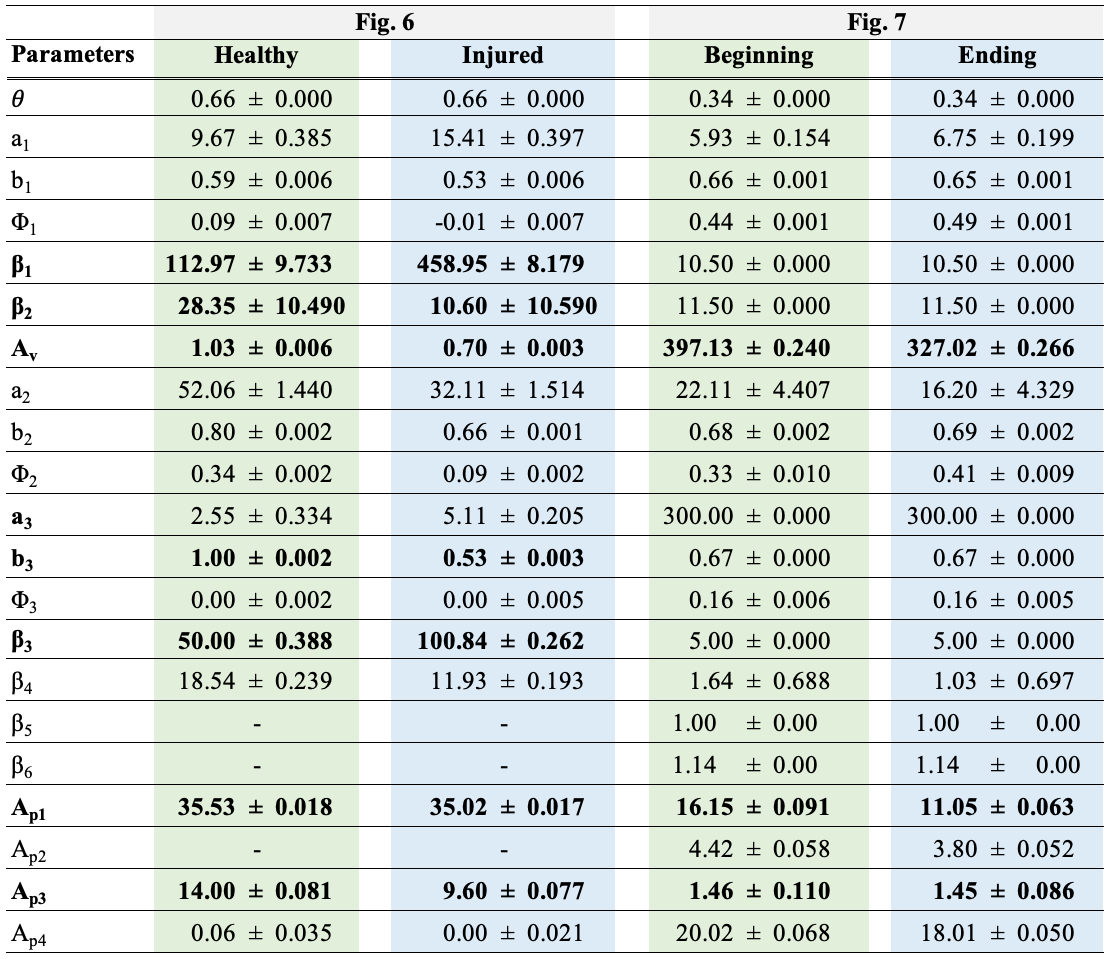}\\
 \end{tabular}
 \caption{Estimated model parameters obtained from the optimization scheme for the results shown in Fig.~\ref{fig6} and Fig.~\ref{fig7} that correspond to the mouse and human data, respectively. The error values were determined using the standard error of the mean. N = 1000. The parameters that are correlated with a known measures of lung physiology are in bold. }
 \label{table1}
 \end{center}
 \end{table}

\subsection{Estimated model parameters correspond to interpretable pathophysiology}

Our final evaluation step is to show that the values of the estimated parameters for data sets corresponding to different phenotypes -- injured/damaged versus healthy -- have physiological meaning. In other words, that differences in the estimated parameter values reflect different phenotypic states of the subject in a manner that is consistent with the pathophysiology.

\emph{Mouse model, PCV}: Figure~\ref{fig6} shows two different breaths of mouse model data, one healthy breath at the beginning of the experiment, and one injured breath at the end of the experiment with $\approx$0 PEEP during PCV. The pressure-volume loops indicate a reduction in lung compliance and increase in hysteresis that are characteristic of lung injury. Our model estimation results suggest the same interpretation. The full model estimation results are shown in Table~\ref{table1} where the bold symbols correspond to particular parameters we are focusing on for this evaluation. Model parameter interpretations are detailed in the Supplementary Table S1. 

In the volume model, we observed the injured lung showed slower estimated inspiration, quantified by an increase in $\beta_1$, and a faster expiration, quantified by a reduction in both $\beta_2$ and $A_v$ compared to the healthy lung model estimates. This leads directly to an interpretation of a reduction in lung compliance.

In the pressure model, we observed a decrease in $b_3$, which could be inferred as a reduction in the compliance in the injured versus healthy lung data. However, we also observed two parameters related to lung compliance, $a_3$ and $\beta_3$, indicate increased overall lung compliance as the lungs become more damaged. These results seem to be contradictory with each other and with the volume model. The data shown in Fig.~\ref{fig6} corresponds to PCV, where volume was an independent variable while the pressure signal was the ventilator controller variable. Therefore, any changes in the pressure waveform correspond to the ventilator settings and not the respiratory mechancs. These results, make an important point: it is essential to see the relative change in the parameters that control these features and to synthesize the model-based inference in a holistic fashion, instead of focusing on any one parameter or feature in isolation given lung mechanics depends on both pressure and volume signal mutually.

Ideally, we would expect pressure signal to be the same over time in PCV, but our mouse model ventilator is not a perfect controller since it uses a piston pump. However, larger changes in the volume signal would be expected, considering a significant change in respiratory mechanics over time. In this specific case, we observed a much greater change in $\beta_1$ compared to $a_3$ and $\beta_3$, and hence changes in the volume waveform are dominating over changes in the pressure waveform. Moreover, $A_v$/$A_{p_{1}}$ ratio is reduced in the injured case (Supplementary Table S3). By considering the model-based parameter estimates and ventilator mode in total, the conclusion is that the injured lung is estimated to have substantially lower compliance than the healthy lung.

\emph{Mouse model, VCV}: The second mouse model evaluation, which includes variations in PEEP during VCV, has PV loops indicating a reduced compliance in the injured case compared to the healthy case for both PEEPs. The full PEEP-varied results are shown in the Supplementary Fig.~S6. 
 
In the volume model, the healthy lung with $0$ PEEP has a slower rate of inspiration leading to an interpretation of mildly worse compliance in comparison to the injured lung, as quantified by the $\beta_1$ parameter value (Supplementary Fig.~S6, Table S1, S2). In contrast, the pressure model indicates a reduction in compliance in the injured lung as quantified by lower values of parameters $a_3$, $b_3$ and $\beta_3$, and elevated estimates of in $A_{p_{1}}$, cf Supplementary Table S1, S2. In contrast to the results shown in Fig.~\ref{fig6}, here, changes in parameter estimates in the pressure model were much larger in comparison to the observed differences in the volume model. 
This is expected since the tidal volumes were approximately equal during VCV, and the reduction in compliance is reflected in increased pressure. This effect can be inferred by analyzing $A_v$/$A_{p_{1}}$ ratio where we observed reduction this ratio in the injured cases at both the PEEPs (Supplementary Table S3). 

\emph{Human ICU data-driven evaluation}: Fig.~\ref{fig7} shows two different breaths of an ICU patient with ARDS that were taken near extubation when ARDS has nearly resolved (see Methods section, Table~\ref{table1}). The ventilator mode was human-triggered, a mode that is not possible in our mouse ventilators and is a commonly used ventilator mode in the ICU. Of the thousands of breaths available, we selected breaths without dyssynchrony. PV loops for these cases suggest that lung compliance is increased at the later time point. We found that the model-estimated parameters suggest the same interpretation. Between the early and later breath data respectively, we observed an increase in $A_v$/$A_{p_{1}}$ ratio indicating increase in compliance, cf Table~\ref{table1}, Supplementary Table S1-S3. The set reduction in PEEP was reflected in a reduction in $A_{p_{4}}$. 

The cases where some patient effort is present, additional model parameters might be used to understand the interaction between the ventilator and the respiratory mechanics. Such a case is shown in the Supplementary Fig.~S7, where PV loops for these cases suggest that lung compliance is increased at the later time point. The model estimated parameters show the ratio of tidal volume to the plateau was increasing, quantified by the $A_v$/$A_{p_{1}}$ ratio increasing (Supplementary Table S3), suggesting an increase in the compliance from when the patient had acute ARDS to the point of extubation. In the pressure waveform, inspiration is happening at a slower rate in the later breath as suggested by an increase in the $b_3$ and $\beta_3$ parameters, also indicating higher compliance. But, while tidal volume appeared to be the same in the two breaths, a significant increase in the $\beta_1$ parameter indicating a reduction in the compliance. This effect is likely to be a result of patient effort to overcome the ventilator~\cite{RN42}. This can be inferred from an increased value of $A_{P_{2}}/A_{P_{1}}$ in the later breath, suggesting an increase in the inspiratory flow resistance and patient effort.

To further validate our finding, we used single-compartment model, which was developed by our research group earlier~\cite{RN11,RN17}, to extract the relevant respiratory parameters and found a qualitative agreement between the outcomes of the two models (see Supplementary Table S3) and thereby, further validating our parameter estimation and interpretation scheme.

Overall, these results suggest that our model can not only reproduce a wide variety of waveform data but also capable of extracting clinically relevant information from the waveforms that might allow to understand injured lung dynamics systematically.

\section{Discussion}
We developed a damage-informed lung model that represents pressure and volume time-series data by reconstructing the waveforms from a modular set of subcomponents. We demonstrate the efficacy of the model using a combination of simulations for validation along with mouse and human data for evaluation. The model was able to simulate desired physiology and pathophysiology, accurately estimate volume and pressure waveforms, and distinguish healthy from injured lungs based on parameter estimation. The model is novel because of the flexibility afforded by the waveform-based approach. Furthermore, we directly incorporate clinical and physiologic knowledge and hypotheses regarding important and observable lung pathophysiology into the model. The model is also limited using prior knowledge so as to not have the capability to estimate every possible variation in PV waveforms, but rather is constrained to estimate the features of the ventilator data that are the most clinically impactful. 

Our approach of developing a model that incorporates clinical insights and limits the model to represent meaningful physiology and pathophysiology appears capable of reproducing a wide range of ventilator waveform including pressure- and volume-controlled ventilation in healthy and lung-injured mice and humans (Fig.~\ref{fig6}, \ref{fig7} and Supplementary Fig. S6, S7). This approach lives between a machine learning approach, were the model is flexible enough to estimate every feature and must then discern which features are important through regularization to prevent overfitting, and the fully mechanistic lung modeling approach where the observed physiology must emerge from the proposed lung mechanics. It is possible that taking this middle path will help advance all approaches. 

The most direct application of our modeling approach is to quantify the qualitative physiological interpretation of pressure and volume data. An experienced clinician or physiology can infer the status of a patient, the safety of ongoing ventilation, the presence of ventilator dyssynchrony, and other important details from visual inspection. However, we currently do not yet have methods to quantitatively identify all of these characteristics in ventilator data. The entire waveform may be utilized and this provides a rich repository of data that is challenging and time consuming to use for diagnosis and treatment. Alternative, these data are, for example, by summarizing in scalar values for resistance and compliance and this may cast aside important details. Our approach offers a methodology for condensing the pressure-volume data to assess ongoing VILI, track changes in injury severity over time, and estimate injury phenotypes (Fig.~\ref{fig6}, \ref{fig7}, Table~\ref{table1} and Supplementary Fig. S6, S7, Table S2). These phenotypes could be used for to categorize and understand lung injury, serve as outcome measures for interventions, and may describe the impacts of VILI and dyssynchrony,~\cite{RN35} and VILI.~\cite{RN6,RN7,RN8,RN12,RN58}. This is reminiscent of current interest in the driving pressure, which is derived from the pressure waveform and has been linked to ventilation safety and ARDS outcomes.~\cite{RN86, RN87,RN88} 

Lung injury diagnosis and decision-making are based in part on interpretation of the pressure, volume, and flow waveforms, such as the aforementioned driving pressure. However, different pathophysiologic mechanisms can lead to the same observed waveform features. For example, increased driving pressure could be a result of derecruitment (alveolar collapse) or alveolar flooding.~\cite{RN89,RN90} In other words, the human-based inference using single waveform data can be ill-posed. Our modeling approach suggests that the ill-posed nature of the inference problem can be addressed in two ways. First, we can quantify the potential observed impact of different pathophysiologic-driven features in the waveforms using experimental data. Second, by estimating over many similar but varied breaths, it may be possible to better triangulate the most probable pathophysiologic drivers because the primary driver of damage will likely be present and significant despite breadth variations while more extraneous details will not be consistently expressed in every breadth.

Then in future studies we can look at the relationship between parameters. The model we present does not fully couple pressure and volume. We have taken this approach in the current study to preserve flexibility so that we can accurately recapitulate a wide variety of clinically and experimentally observed features in the pressure and volume signals, including the effects of ventilator dyssynchrony. This fidelity and flexibility is not always possible with rigid coupling between pressure and volume data like, for example, in a single compartment model where pressure is defined as the sum of linear resistive and elastic contributions. This is not to say that pressure and volume are totally independent in our model because we utilize the same respiratory rate for both. In future studies we will link specific components of the pressure and volume waveforms through physiologically-relevant parameters such as nonlinear lung elastance or inspiratory and expiratory flow resistance. 

As secondary application of our modeling approach, and a method to incorporate the physiologic coupling between pressure and volume data, is to utilize the outputs from the model presented here as inputs for compartment models. Currently, most compartment models are fit to measured data using regression. In a model with few parameters (e.g. only resistance and compliance) this is feasible for real-time analysis. However, as model complexity increases to include representations of nonlinear tissue elastance, recruitment dynamics, and other factors it is no longer possible to perform the regressions in a clinically-applicable timescale. If our waveform-based model is used to process the data prior to analysis using a compartment model then it is possible to formulate the problem entirely of ordinary differential equations and this opens up a range of more efficient inference machinery.~\cite{RN47,RN48}

Finally, our work here has several notable limitations. \emph{First}, our evaluations were performed with single, but typical, breaths of mouse and human ventilator data. We took this approach because each breath is, in some sense, a single controlled experiment and our goal was to demonstrate the functionality of the model under varied conditions. \emph{Second}, our evaluation was conducted using healthy and severely lung-injured mice as well as a single human data set. This is sufficient for proof in principle that the model can capture physiologic differences. However, establishing that the model can accurately differentiate more specifically defined phenotypes will require evaluation on much larger populations. \emph{Third}, we relied on the expert knowledge of a single critical care physician to determine the clinically important characteristics of the pressure and volume waveforms and it is likely that differing opinions will exist among intensivists. Collecting and synthesizing such information will require a different qualitative study. Moreover, that there may be differing opinions regarding what should and should not be included in the model. This does not negate our methodology or our model. Instead, it suggests future work is necessary to better understand and verify clinically important features. Alternatively, we may instead seek to link model features to patient outcomes, thus establishing the important characteristics of the model by linking those parameters to outcomes. 

In summary, we developed a physiologically anchored and data-driven lung model that can reproduce the important features pressures and volumes during mechanical ventilation. The performance of the model was verified with experimental and clinical data in healthy and injured lungs to demonstrate model efficacy in robustly estimating interpretable parameters. This methodology represents a departure from many lung modeling efforts, and suggests future directions of work that can provide another pathway for better understanding lung function during mechanical ventilation and can potentially form a bridge between experimental physiology and clinical practice. 

\section{Grants}
This work was supported by National Institutes of Health R01 “Mechanistic machine learning,” LM012734 and LM006910 “Discovering and applying knowledge in clinical databases,” along with R00 HL128944, and K24 HL069223.

\section{Author Contribution}
D.K.A., and D.J.A. conception and design of research; D.K.A., B.J.S., and P.D.S. performed experiments; D.K.A., B.J.S., and D.J.A. analyzed data; D.K.A., B.J.S., P.D.S., and D.J.A. interpreted results of experiments; D.K.A. prepared figures; D.K.A. drafted manuscript; D.K.A., B.J.S., P.D.S., and D.J.A. edited and revised manuscript; D.K.A., B.J.S., P.D.S., and D.J.A. approved final version of manuscript.
\bibliography{references_V1}

\providecommand{\latin}[1]{#1}
\makeatletter
\providecommand{\doi}
  {\begingroup\let\do\@makeother\dospecials
  \catcode`\{=1 \catcode`\}=2 \doi@aux}
\providecommand{\doi@aux}[1]{\endgroup\texttt{#1}}
\makeatother
\providecommand*\mcitethebibliography{\thebibliography}
\csname @ifundefined\endcsname{endmcitethebibliography}
  {\let\endmcitethebibliography\endthebibliography}{}
\begin{mcitethebibliography}{62}
\providecommand*\natexlab[1]{#1}
\providecommand*\mciteSetBstSublistMode[1]{}
\providecommand*\mciteSetBstMaxWidthForm[2]{}
\providecommand*\mciteBstWouldAddEndPuncttrue
  {\def\EndOfBibitem{\unskip.}}
\providecommand*\mciteBstWouldAddEndPunctfalse
  {\let\EndOfBibitem\relax}
\providecommand*\mciteSetBstMidEndSepPunct[3]{}
\providecommand*\mciteSetBstSublistLabelBeginEnd[3]{}
\providecommand*\EndOfBibitem{}
\mciteSetBstSublistMode{f}
\mciteSetBstMaxWidthForm{subitem}{(\alph{mcitesubitemcount})}
\mciteSetBstSublistLabelBeginEnd
  {\mcitemaxwidthsubitemform\space}
  {\relax}
  {\relax}

\bibitem[Ware and Matthay(2000)Ware, and Matthay]{RN1}
Ware,~L.~B.; Matthay,~M.~A. The acute respiratory distress syndrome. \emph{New
  England Journal of Medicine} \textbf{2000}, \emph{342}, 1334--1349\relax
\mciteBstWouldAddEndPuncttrue
\mciteSetBstMidEndSepPunct{\mcitedefaultmidpunct}
{\mcitedefaultendpunct}{\mcitedefaultseppunct}\relax
\EndOfBibitem
\bibitem[Force \latin{et~al.}(2012)Force, Ranieri, Rubenfeld, Thompson,
  Ferguson, and Caldwell]{RN2}
Force,~A. D.~T.; Ranieri,~V.; Rubenfeld,~G.; Thompson,~B.; Ferguson,~N.;
  Caldwell,~E. Acute respiratory distress syndrome. \emph{Jama} \textbf{2012},
  \emph{307}, 2526--2533\relax
\mciteBstWouldAddEndPuncttrue
\mciteSetBstMidEndSepPunct{\mcitedefaultmidpunct}
{\mcitedefaultendpunct}{\mcitedefaultseppunct}\relax
\EndOfBibitem
\bibitem[Network(2000)]{RN3}
Network,~A. R. D.~S. Ventilation with lower tidal volumes as compared with
  traditional tidal volumes for acute lung injury and the acute respiratory
  distress syndrome. \emph{New England Journal of Medicine} \textbf{2000},
  \emph{342}, 1301--1308\relax
\mciteBstWouldAddEndPuncttrue
\mciteSetBstMidEndSepPunct{\mcitedefaultmidpunct}
{\mcitedefaultendpunct}{\mcitedefaultseppunct}\relax
\EndOfBibitem
\bibitem[Amato \latin{et~al.}(2015)Amato, Meade, Slutsky, Brochard, Costa,
  Schoenfeld, Stewart, Briel, Talmor, and Mercat]{RN4}
Amato,~M.~B.; Meade,~M.~O.; Slutsky,~A.~S.; Brochard,~L.; Costa,~E.~L.;
  Schoenfeld,~D.~A.; Stewart,~T.~E.; Briel,~M.; Talmor,~D.; Mercat,~A. Driving
  pressure and survival in the acute respiratory distress syndrome. \emph{New
  England Journal of Medicine} \textbf{2015}, \emph{372}, 747--755\relax
\mciteBstWouldAddEndPuncttrue
\mciteSetBstMidEndSepPunct{\mcitedefaultmidpunct}
{\mcitedefaultendpunct}{\mcitedefaultseppunct}\relax
\EndOfBibitem
\bibitem[Slutsky and Ranieri(2013)Slutsky, and Ranieri]{RN6}
Slutsky,~A.~S.; Ranieri,~V.~M. Ventilator-induced lung injury. \emph{New
  England Journal of Medicine} \textbf{2013}, \emph{369}, 2126--2136\relax
\mciteBstWouldAddEndPuncttrue
\mciteSetBstMidEndSepPunct{\mcitedefaultmidpunct}
{\mcitedefaultendpunct}{\mcitedefaultseppunct}\relax
\EndOfBibitem
\bibitem[Gattinoni \latin{et~al.}(2003)Gattinoni, Carlesso, Cadringher,
  Valenza, Vagginelli, and Chiumello]{RN7}
Gattinoni,~L.; Carlesso,~E.; Cadringher,~P.; Valenza,~F.; Vagginelli,~F.;
  Chiumello,~D. Physical and biological triggers of ventilator-induced lung
  injury and its prevention. \emph{European Respiratory Journal} \textbf{2003},
  \emph{22}, 15s--25s\relax
\mciteBstWouldAddEndPuncttrue
\mciteSetBstMidEndSepPunct{\mcitedefaultmidpunct}
{\mcitedefaultendpunct}{\mcitedefaultseppunct}\relax
\EndOfBibitem
\bibitem[Dos~Santos and Slutsky(2000)Dos~Santos, and Slutsky]{RN8}
Dos~Santos,~C.; Slutsky,~A. Invited review: mechanisms of ventilator-induced
  lung injury: a perspective. \emph{Journal of applied physiology}
  \textbf{2000}, \emph{89}, 1645--1655\relax
\mciteBstWouldAddEndPuncttrue
\mciteSetBstMidEndSepPunct{\mcitedefaultmidpunct}
{\mcitedefaultendpunct}{\mcitedefaultseppunct}\relax
\EndOfBibitem
\bibitem[Bates and Smith(2018)Bates, and Smith]{RN12}
Bates,~J.~H.; Smith,~B.~J. Ventilator-induced lung injury and lung mechanics.
  \emph{Annals of translational medicine} \textbf{2018}, \emph{6}\relax
\mciteBstWouldAddEndPuncttrue
\mciteSetBstMidEndSepPunct{\mcitedefaultmidpunct}
{\mcitedefaultendpunct}{\mcitedefaultseppunct}\relax
\EndOfBibitem
\bibitem[Phua \latin{et~al.}(2009)Phua, Badia, Adhikari, Friedrich, Fowler,
  Singh, Scales, Stather, Li, and Jones]{RN5}
Phua,~J.; Badia,~J.~R.; Adhikari,~N.~K.; Friedrich,~J.~O.; Fowler,~R.~A.;
  Singh,~J.~M.; Scales,~D.~C.; Stather,~D.~R.; Li,~A.; Jones,~A. Has mortality
  from acute respiratory distress syndrome decreased over time? A systematic
  review. \emph{American journal of respiratory and critical care medicine}
  \textbf{2009}, \emph{179}, 220--227\relax
\mciteBstWouldAddEndPuncttrue
\mciteSetBstMidEndSepPunct{\mcitedefaultmidpunct}
{\mcitedefaultendpunct}{\mcitedefaultseppunct}\relax
\EndOfBibitem
\bibitem[Tobin(2001)]{RN71}
Tobin,~M.~J. Advances in mechanical ventilation. \emph{New England Journal of
  Medicine} \textbf{2001}, \emph{344}, 1986--1996\relax
\mciteBstWouldAddEndPuncttrue
\mciteSetBstMidEndSepPunct{\mcitedefaultmidpunct}
{\mcitedefaultendpunct}{\mcitedefaultseppunct}\relax
\EndOfBibitem
\bibitem[Dellaca and Veneroni(2017)Dellaca, and Veneroni]{RN72}
Dellaca,~R.~L.; Veneroni,~C. Trends in mechanical ventilation: are we
  ventilating our patients in the best possible way? \emph{Breathe}
  \textbf{2017}, \emph{13}, 84--98\relax
\mciteBstWouldAddEndPuncttrue
\mciteSetBstMidEndSepPunct{\mcitedefaultmidpunct}
{\mcitedefaultendpunct}{\mcitedefaultseppunct}\relax
\EndOfBibitem
\bibitem[Gilstrap and MacIntyre(2013)Gilstrap, and MacIntyre]{RN54}
Gilstrap,~D.; MacIntyre,~N. Patient–ventilator interactions. Implications for
  clinical management. \emph{American journal of respiratory and critical care
  medicine} \textbf{2013}, \emph{188}, 1058--1068\relax
\mciteBstWouldAddEndPuncttrue
\mciteSetBstMidEndSepPunct{\mcitedefaultmidpunct}
{\mcitedefaultendpunct}{\mcitedefaultseppunct}\relax
\EndOfBibitem
\bibitem[Blanch \latin{et~al.}(2015)Blanch, Villagra, Sales, Montanya,
  Lucangelo, Luján, García-Esquirol, Chacón, Estruga, and Oliva]{RN55}
Blanch,~L.; Villagra,~A.; Sales,~B.; Montanya,~J.; Lucangelo,~U.; Luján,~M.;
  García-Esquirol,~O.; Chacón,~E.; Estruga,~A.; Oliva,~J.~C. Asynchronies
  during mechanical ventilation are associated with mortality. \emph{Intensive
  care medicine} \textbf{2015}, \emph{41}, 633--641\relax
\mciteBstWouldAddEndPuncttrue
\mciteSetBstMidEndSepPunct{\mcitedefaultmidpunct}
{\mcitedefaultendpunct}{\mcitedefaultseppunct}\relax
\EndOfBibitem
\bibitem[Yoshida \latin{et~al.}(2017)Yoshida, Fujino, Amato, and
  Kavanagh]{RN61}
Yoshida,~T.; Fujino,~Y.; Amato,~M.~B.; Kavanagh,~B.~P. Fifty years of research
  in ARDS. Spontaneous breathing during mechanical ventilation. Risks,
  mechanisms, and management. \emph{American journal of respiratory and
  critical care medicine} \textbf{2017}, \emph{195}, 985--992\relax
\mciteBstWouldAddEndPuncttrue
\mciteSetBstMidEndSepPunct{\mcitedefaultmidpunct}
{\mcitedefaultendpunct}{\mcitedefaultseppunct}\relax
\EndOfBibitem
\bibitem[Chiumello \latin{et~al.}(2008)Chiumello, Carlesso, Cadringher,
  Caironi, Valenza, Polli, Tallarini, Cozzi, Cressoni, and Colombo]{RN62}
Chiumello,~D.; Carlesso,~E.; Cadringher,~P.; Caironi,~P.; Valenza,~F.;
  Polli,~F.; Tallarini,~F.; Cozzi,~P.; Cressoni,~M.; Colombo,~A. Lung stress
  and strain during mechanical ventilation for acute respiratory distress
  syndrome. \emph{American journal of respiratory and critical care medicine}
  \textbf{2008}, \emph{178}, 346--355\relax
\mciteBstWouldAddEndPuncttrue
\mciteSetBstMidEndSepPunct{\mcitedefaultmidpunct}
{\mcitedefaultendpunct}{\mcitedefaultseppunct}\relax
\EndOfBibitem
\bibitem[Network(2000)]{RN82}
Network,~A. R. D.~S. Ventilation with lower tidal volumes as compared with
  traditional tidal volumes for acute lung injury and the acute respiratory
  distress syndrome. \emph{New England Journal of Medicine} \textbf{2000},
  \emph{342}, 1301--1308\relax
\mciteBstWouldAddEndPuncttrue
\mciteSetBstMidEndSepPunct{\mcitedefaultmidpunct}
{\mcitedefaultendpunct}{\mcitedefaultseppunct}\relax
\EndOfBibitem
\bibitem[Grasso \latin{et~al.}(2007)Grasso, Stripoli, De~Michele, Bruno,
  Moschetta, Angelelli, Munno, Ruggiero, Anaclerio, and Cafarelli]{RN59}
Grasso,~S.; Stripoli,~T.; De~Michele,~M.; Bruno,~F.; Moschetta,~M.;
  Angelelli,~G.; Munno,~I.; Ruggiero,~V.; Anaclerio,~R.; Cafarelli,~A. ARDSnet
  ventilatory protocol and alveolar hyperinflation: role of positive
  end-expiratory pressure. \emph{American journal of respiratory and critical
  care medicine} \textbf{2007}, \emph{176}, 761--767\relax
\mciteBstWouldAddEndPuncttrue
\mciteSetBstMidEndSepPunct{\mcitedefaultmidpunct}
{\mcitedefaultendpunct}{\mcitedefaultseppunct}\relax
\EndOfBibitem
\bibitem[Khemani \latin{et~al.}(2018)Khemani, Parvathaneni, Yehya, Bhalla,
  Thomas, and Newth]{RN60}
Khemani,~R.~G.; Parvathaneni,~K.; Yehya,~N.; Bhalla,~A.~K.; Thomas,~N.~J.;
  Newth,~C.~J. Positive end-expiratory pressure lower than the ARDS network
  protocol is associated with higher pediatric acute respiratory distress
  syndrome mortality. \emph{American journal of respiratory and critical care
  medicine} \textbf{2018}, \emph{198}, 77--89\relax
\mciteBstWouldAddEndPuncttrue
\mciteSetBstMidEndSepPunct{\mcitedefaultmidpunct}
{\mcitedefaultendpunct}{\mcitedefaultseppunct}\relax
\EndOfBibitem
\bibitem[Bein \latin{et~al.}(2013)Bein, Weber-Carstens, Goldmann, M\"{u}ller,
  Staudinger, Brederlau, Muellenbach, Dembinski, Graf, and Wewalka]{RN81}
Bein,~T.; Weber-Carstens,~S.; Goldmann,~A.; M\"{u}ller,~T.; Staudinger,~T.;
  Brederlau,~J.; Muellenbach,~R.; Dembinski,~R.; Graf,~B.~M.; Wewalka,~M. Lower
  tidal volume strategy ($\approx$ 3 ml/kg) combined with extracorporeal CO 2
  removal versus ‘conventional’protective ventilation (6 ml/kg) in severe
  ARDS. \emph{Intensive care medicine} \textbf{2013}, \emph{39}, 847--856\relax
\mciteBstWouldAddEndPuncttrue
\mciteSetBstMidEndSepPunct{\mcitedefaultmidpunct}
{\mcitedefaultendpunct}{\mcitedefaultseppunct}\relax
\EndOfBibitem
\bibitem[Mellema(2013)]{RN25}
Mellema,~M.~S. Ventilator waveforms. \emph{Topics in companion animal medicine}
  \textbf{2013}, \emph{28}, 112--123\relax
\mciteBstWouldAddEndPuncttrue
\mciteSetBstMidEndSepPunct{\mcitedefaultmidpunct}
{\mcitedefaultendpunct}{\mcitedefaultseppunct}\relax
\EndOfBibitem
\bibitem[Corona and Aumann(2011)Corona, and Aumann]{RN26}
Corona,~T.~M.; Aumann,~M. Ventilator waveform interpretation in mechanically
  ventilated small animals. \emph{Journal of Veterinary Emergency and Critical
  Care} \textbf{2011}, \emph{21}, 496--514\relax
\mciteBstWouldAddEndPuncttrue
\mciteSetBstMidEndSepPunct{\mcitedefaultmidpunct}
{\mcitedefaultendpunct}{\mcitedefaultseppunct}\relax
\EndOfBibitem
\bibitem[Amato \latin{et~al.}(2015)Amato, Meade, Slutsky, Brochard, Costa,
  Schoenfeld, Stewart, Briel, Talmor, Mercat, Richard, Carvalho, and
  Brower]{RN84}
Amato,~M.~B.; Meade,~M.~O.; Slutsky,~A.~S.; Brochard,~L.; Costa,~E.~L.;
  Schoenfeld,~D.~A.; Stewart,~T.~E.; Briel,~M.; Talmor,~D.; Mercat,~A.;
  Richard,~J.~C.; Carvalho,~C.~R.; Brower,~R.~G. Driving pressure and survival
  in the acute respiratory distress syndrome. \emph{N Engl J Med}
  \textbf{2015}, \emph{372}, 747--55\relax
\mciteBstWouldAddEndPuncttrue
\mciteSetBstMidEndSepPunct{\mcitedefaultmidpunct}
{\mcitedefaultendpunct}{\mcitedefaultseppunct}\relax
\EndOfBibitem
\bibitem[Mellenthin \latin{et~al.}(2019)Mellenthin, Seong, Roy, Bartolák-Suki,
  Hamlington, Bates, and Smith]{RN15}
Mellenthin,~M.~M.; Seong,~S.~A.; Roy,~G.~S.; Bartolák-Suki,~E.;
  Hamlington,~K.~L.; Bates,~J.~H.; Smith,~B.~J. Using injury cost functions
  from a predictive single-compartment model to assess the severity of
  mechanical ventilator-induced lung injuries. \emph{Journal of Applied
  Physiology} \textbf{2019}, \emph{127}, 58--70\relax
\mciteBstWouldAddEndPuncttrue
\mciteSetBstMidEndSepPunct{\mcitedefaultmidpunct}
{\mcitedefaultendpunct}{\mcitedefaultseppunct}\relax
\EndOfBibitem
\bibitem[Mori(2016)]{RN34}
Mori,~K. From macro-scale to micro-scale computational anatomy: a perspective
  on the next 20 years. \emph{Med Image Anal} \textbf{2016}, \emph{33},
  159--164\relax
\mciteBstWouldAddEndPuncttrue
\mciteSetBstMidEndSepPunct{\mcitedefaultmidpunct}
{\mcitedefaultendpunct}{\mcitedefaultseppunct}\relax
\EndOfBibitem
\bibitem[Hamlington \latin{et~al.}(2016)Hamlington, Smith, Allen, and
  Bates]{RN11}
Hamlington,~K.~L.; Smith,~B.~J.; Allen,~G.~B.; Bates,~J.~H. Predicting
  ventilator-induced lung injury using a lung injury cost function.
  \emph{Journal of Applied Physiology} \textbf{2016}, \emph{121},
  106--114\relax
\mciteBstWouldAddEndPuncttrue
\mciteSetBstMidEndSepPunct{\mcitedefaultmidpunct}
{\mcitedefaultendpunct}{\mcitedefaultseppunct}\relax
\EndOfBibitem
\bibitem[Smith \latin{et~al.}(2015)Smith, Lundblad, Kollisch-Singule, Satalin,
  Nieman, Habashi, and Bates]{RN17}
Smith,~B.~J.; Lundblad,~L.~K.; Kollisch-Singule,~M.; Satalin,~J.; Nieman,~G.;
  Habashi,~N.; Bates,~J.~H. Predicting the response of the injured lung to the
  mechanical breath profile. \emph{Journal of applied physiology}
  \textbf{2015}, \emph{118}, 932--940\relax
\mciteBstWouldAddEndPuncttrue
\mciteSetBstMidEndSepPunct{\mcitedefaultmidpunct}
{\mcitedefaultendpunct}{\mcitedefaultseppunct}\relax
\EndOfBibitem
\bibitem[Chiew \latin{et~al.}(2011)Chiew, Chase, Shaw, Sundaresan, and
  Desaive]{RN73}
Chiew,~Y.~S.; Chase,~J.~G.; Shaw,~G.~M.; Sundaresan,~A.; Desaive,~T.
  Model-based PEEP optimisation in mechanical ventilation. \emph{Biomedical
  engineering online} \textbf{2011}, \emph{10}, 111\relax
\mciteBstWouldAddEndPuncttrue
\mciteSetBstMidEndSepPunct{\mcitedefaultmidpunct}
{\mcitedefaultendpunct}{\mcitedefaultseppunct}\relax
\EndOfBibitem
\bibitem[Ellwein~Fix \latin{et~al.}(2018)Ellwein~Fix, Khoury, Moores~Jr,
  Linkous, Brandes, and Rozycki]{RN13}
Ellwein~Fix,~L.; Khoury,~J.; Moores~Jr,~R.~R.; Linkous,~L.; Brandes,~M.;
  Rozycki,~H.~J. Theoretical open-loop model of respiratory mechanics in the
  extremely preterm infant. \emph{PloS one} \textbf{2018}, \emph{13},
  e0198425\relax
\mciteBstWouldAddEndPuncttrue
\mciteSetBstMidEndSepPunct{\mcitedefaultmidpunct}
{\mcitedefaultendpunct}{\mcitedefaultseppunct}\relax
\EndOfBibitem
\bibitem[Rees \latin{et~al.}(2006)Rees, Allerød, Murley, Zhao, Smith,
  Kjærgaard, Thorgaard, and Andreassen]{RN14}
Rees,~S.~E.; Allerød,~C.; Murley,~D.; Zhao,~Y.; Smith,~B.~W.; Kjærgaard,~S.;
  Thorgaard,~P.; Andreassen,~S. Using physiological models and decision theory
  for selecting appropriate ventilator settings. \emph{Journal of clinical
  monitoring and computing} \textbf{2006}, \emph{20}, 421\relax
\mciteBstWouldAddEndPuncttrue
\mciteSetBstMidEndSepPunct{\mcitedefaultmidpunct}
{\mcitedefaultendpunct}{\mcitedefaultseppunct}\relax
\EndOfBibitem
\bibitem[Serov \latin{et~al.}(2016)Serov, Salafia, Grebenkov, and
  Filoche]{RN16}
Serov,~A.~S.; Salafia,~C.; Grebenkov,~D.~S.; Filoche,~M. The role of morphology
  in mathematical models of placental gas exchange. \emph{Journal of Applied
  Physiology} \textbf{2016}, \emph{120}, 17--28\relax
\mciteBstWouldAddEndPuncttrue
\mciteSetBstMidEndSepPunct{\mcitedefaultmidpunct}
{\mcitedefaultendpunct}{\mcitedefaultseppunct}\relax
\EndOfBibitem
\bibitem[Nguyen \latin{et~al.}(2014)Nguyen, Bernstein, and Bates]{RN18}
Nguyen,~B.; Bernstein,~D.~B.; Bates,~J.~H. Controlling mechanical ventilation
  in acute respiratory distress syndrome with fuzzy logic. \emph{Journal of
  critical care} \textbf{2014}, \emph{29}, 551--556\relax
\mciteBstWouldAddEndPuncttrue
\mciteSetBstMidEndSepPunct{\mcitedefaultmidpunct}
{\mcitedefaultendpunct}{\mcitedefaultseppunct}\relax
\EndOfBibitem
\bibitem[Roth \latin{et~al.}(2017)Roth, Ismail, Yoshihara, and Wall]{RN19}
Roth,~C.~J.; Ismail,~M.; Yoshihara,~L.; Wall,~W.~A. A comprehensive
  computational human lung model incorporating inter‐acinar dependencies:
  Application to spontaneous breathing and mechanical ventilation.
  \emph{International journal for numerical methods in biomedical engineering}
  \textbf{2017}, \emph{33}, e02787\relax
\mciteBstWouldAddEndPuncttrue
\mciteSetBstMidEndSepPunct{\mcitedefaultmidpunct}
{\mcitedefaultendpunct}{\mcitedefaultseppunct}\relax
\EndOfBibitem
\bibitem[Reynolds \latin{et~al.}(2010)Reynolds, Ermentrout, and Clermont]{RN20}
Reynolds,~A.; Ermentrout,~G.~B.; Clermont,~G. A mathematical model of pulmonary
  gas exchange under inflammatory stress. \emph{Journal of theoretical biology}
  \textbf{2010}, \emph{264}, 161--173\relax
\mciteBstWouldAddEndPuncttrue
\mciteSetBstMidEndSepPunct{\mcitedefaultmidpunct}
{\mcitedefaultendpunct}{\mcitedefaultseppunct}\relax
\EndOfBibitem
\bibitem[Bates(2009)]{RN51}
Bates,~J.~H. \emph{Lung mechanics: an inverse modeling approach}; Cambridge
  University Press, 2009\relax
\mciteBstWouldAddEndPuncttrue
\mciteSetBstMidEndSepPunct{\mcitedefaultmidpunct}
{\mcitedefaultendpunct}{\mcitedefaultseppunct}\relax
\EndOfBibitem
\bibitem[Molkov \latin{et~al.}(2014)Molkov, Shevtsova, Park, Ben-Tal, Smith,
  Rubin, and Rybak]{RN75}
Molkov,~Y.~I.; Shevtsova,~N.~A.; Park,~C.; Ben-Tal,~A.; Smith,~J.~C.;
  Rubin,~J.~E.; Rybak,~I.~A. A closed-loop model of the respiratory system:
  focus on hypercapnia and active expiration. \emph{PloS one} \textbf{2014},
  \emph{9}, e109894\relax
\mciteBstWouldAddEndPuncttrue
\mciteSetBstMidEndSepPunct{\mcitedefaultmidpunct}
{\mcitedefaultendpunct}{\mcitedefaultseppunct}\relax
\EndOfBibitem
\bibitem[Molkov \latin{et~al.}(2017)Molkov, Rubin, Rybak, and Smith]{RN76}
Molkov,~Y.~I.; Rubin,~J.~E.; Rybak,~I.~A.; Smith,~J.~C. Computational models of
  the neural control of breathing. \emph{Wiley Interdisciplinary Reviews:
  Systems Biology and Medicine} \textbf{2017}, \emph{9}, e1371\relax
\mciteBstWouldAddEndPuncttrue
\mciteSetBstMidEndSepPunct{\mcitedefaultmidpunct}
{\mcitedefaultendpunct}{\mcitedefaultseppunct}\relax
\EndOfBibitem
\bibitem[Jolliffe and Stephenson(2012)Jolliffe, and Stephenson]{RN50}
Jolliffe,~I.~T.; Stephenson,~D.~B. \emph{Forecast verification: a
  practitioner's guide in atmospheric science}; John Wiley \& Sons, 2012\relax
\mciteBstWouldAddEndPuncttrue
\mciteSetBstMidEndSepPunct{\mcitedefaultmidpunct}
{\mcitedefaultendpunct}{\mcitedefaultseppunct}\relax
\EndOfBibitem
\bibitem[Sottile \latin{et~al.}(2018)Sottile, Albers, Higgins, Mckeehan, and
  Moss]{RN35}
Sottile,~P.~D.; Albers,~D.; Higgins,~C.; Mckeehan,~J.; Moss,~M.~M. The
  Association Between Ventilator Dyssynchrony, Delivered Tidal Volume, and
  Sedation Using a Novel Automated Ventilator Dyssynchrony Detection Algorithm.
  \emph{Critical Care Medicine} \textbf{2018}, \emph{46}, E151--E157\relax
\mciteBstWouldAddEndPuncttrue
\mciteSetBstMidEndSepPunct{\mcitedefaultmidpunct}
{\mcitedefaultendpunct}{\mcitedefaultseppunct}\relax
\EndOfBibitem
\bibitem[Tobin(2010)]{RN29}
Tobin,~M.~J. \emph{Principles and practice of mechanical ventilation}; McGraw
  Hill Professional, 2010\relax
\mciteBstWouldAddEndPuncttrue
\mciteSetBstMidEndSepPunct{\mcitedefaultmidpunct}
{\mcitedefaultendpunct}{\mcitedefaultseppunct}\relax
\EndOfBibitem
\bibitem[Guerin(2011)]{RN33}
Guerin,~C. The preventive role of higher PEEP in treating severely hypoxemic
  ARDS. \emph{Minerva Anestesiol} \textbf{2011}, \emph{77}, 835--45\relax
\mciteBstWouldAddEndPuncttrue
\mciteSetBstMidEndSepPunct{\mcitedefaultmidpunct}
{\mcitedefaultendpunct}{\mcitedefaultseppunct}\relax
\EndOfBibitem
\bibitem[Cavalcanti \latin{et~al.}(2017)Cavalcanti, Suzumura, Laranjeira,
  de~Moraes~Paisani, Damiani, Guimar\~{a}es, Romano, de~Moraes~Regenga,
  Taniguchi, and Teixeira]{RN57}
Cavalcanti,~A.~B.; Suzumura,~E.~A.; Laranjeira,~L.~N.; de~Moraes~Paisani,~D.;
  Damiani,~L.~P.; Guimar\~{a}es,~H.~P.; Romano,~E.~R.; de~Moraes~Regenga,~M.;
  Taniguchi,~L. N.~T.; Teixeira,~C. Effect of lung recruitment and titrated
  positive end-expiratory pressure (PEEP) vs low PEEP on mortality in patients
  with acute respiratory distress syndrome: a randomized clinical trial.
  \emph{Jama} \textbf{2017}, \emph{318}, 1335--1345\relax
\mciteBstWouldAddEndPuncttrue
\mciteSetBstMidEndSepPunct{\mcitedefaultmidpunct}
{\mcitedefaultendpunct}{\mcitedefaultseppunct}\relax
\EndOfBibitem
\bibitem[Wheeler and Bernard(2007)Wheeler, and Bernard]{RN39}
Wheeler,~A.~P.; Bernard,~G.~R. Acute lung injury and the acute respiratory
  distress syndrome: a clinical review. \emph{Lancet} \textbf{2007},
  \emph{369}, 1553--1564\relax
\mciteBstWouldAddEndPuncttrue
\mciteSetBstMidEndSepPunct{\mcitedefaultmidpunct}
{\mcitedefaultendpunct}{\mcitedefaultseppunct}\relax
\EndOfBibitem
\bibitem[Albers \latin{et~al.}(2019)Albers, Levine, Mamykina, and
  Hripcsak]{RN46}
Albers,~D.~J.; Levine,~M.~E.; Mamykina,~L.; Hripcsak,~G. The Parameter
  Houlihan: a solution to high-throughput identifiability indeterminacy for
  brutally ill-posed problems. \emph{Mathematical biosciences} \textbf{2019},
  \emph{316}, 108242\relax
\mciteBstWouldAddEndPuncttrue
\mciteSetBstMidEndSepPunct{\mcitedefaultmidpunct}
{\mcitedefaultendpunct}{\mcitedefaultseppunct}\relax
\EndOfBibitem
\bibitem[Westwick and Kearney(2003)Westwick, and Kearney]{RN67}
Westwick,~D.~T.; Kearney,~R.~E. \emph{Identification of nonlinear physiological
  systems}; John Wiley \& Sons, 2003; Vol.~7\relax
\mciteBstWouldAddEndPuncttrue
\mciteSetBstMidEndSepPunct{\mcitedefaultmidpunct}
{\mcitedefaultendpunct}{\mcitedefaultseppunct}\relax
\EndOfBibitem
\bibitem[Schoukens \latin{et~al.}(2016)Schoukens, Vaes, and Pintelon]{RN68}
Schoukens,~J.; Vaes,~M.; Pintelon,~R. Linear system identification in a
  nonlinear setting: Nonparametric analysis of the nonlinear distortions and
  their impact on the best linear approximation. \emph{IEEE Control Systems
  Magazine} \textbf{2016}, \emph{36}, 38--69\relax
\mciteBstWouldAddEndPuncttrue
\mciteSetBstMidEndSepPunct{\mcitedefaultmidpunct}
{\mcitedefaultendpunct}{\mcitedefaultseppunct}\relax
\EndOfBibitem
\bibitem[Hripcsak and Albers(2013)Hripcsak, and Albers]{RN42}
Hripcsak,~G.; Albers,~D.~J. Next-generation phenotyping of electronic health
  records. \emph{J Am Med Inform Assoc} \textbf{2013}, \emph{20}, 117--21\relax
\mciteBstWouldAddEndPuncttrue
\mciteSetBstMidEndSepPunct{\mcitedefaultmidpunct}
{\mcitedefaultendpunct}{\mcitedefaultseppunct}\relax
\EndOfBibitem
\bibitem[Albers \latin{et~al.}(2018)Albers, Levine, Stuart, Mamykina, Gluckman,
  and Hripcsak]{RN43}
Albers,~D.~J.; Levine,~M.~E.; Stuart,~A.; Mamykina,~L.; Gluckman,~B.;
  Hripcsak,~G. Mechanistic machine learning: how data assimilation leverages
  physiologic knowledge using Bayesian inference to forecast the future, infer
  the present, and phenotype. \emph{Journal of the American Medical Informatics
  Association} \textbf{2018}, \emph{25}, 1392--1401\relax
\mciteBstWouldAddEndPuncttrue
\mciteSetBstMidEndSepPunct{\mcitedefaultmidpunct}
{\mcitedefaultendpunct}{\mcitedefaultseppunct}\relax
\EndOfBibitem
\bibitem[Hripcsak and Albers(2018)Hripcsak, and Albers]{RN44}
Hripcsak,~G.; Albers,~D.~J. High-fidelity phenotyping: richness and freedom
  from bias. \emph{Journal of the American Medical Informatics Association}
  \textbf{2018}, \emph{25}, 289--294\relax
\mciteBstWouldAddEndPuncttrue
\mciteSetBstMidEndSepPunct{\mcitedefaultmidpunct}
{\mcitedefaultendpunct}{\mcitedefaultseppunct}\relax
\EndOfBibitem
\bibitem[Smith(2013)]{RN45}
Smith,~R.~C. \emph{Uncertainty quantification: theory, implementation, and
  applications}; Siam, 2013; Vol.~12\relax
\mciteBstWouldAddEndPuncttrue
\mciteSetBstMidEndSepPunct{\mcitedefaultmidpunct}
{\mcitedefaultendpunct}{\mcitedefaultseppunct}\relax
\EndOfBibitem
\bibitem[Albers \latin{et~al.}(2019)Albers, Blancquart, Levine, Seylabi, and
  Stuart]{RN47}
Albers,~D.~J.; Blancquart,~P.-A.; Levine,~M.~E.; Seylabi,~E.~E.; Stuart,~A.
  Ensemble Kalman methods with constraints. \emph{Inverse Problems}
  \textbf{2019}, \emph{35}, 095007\relax
\mciteBstWouldAddEndPuncttrue
\mciteSetBstMidEndSepPunct{\mcitedefaultmidpunct}
{\mcitedefaultendpunct}{\mcitedefaultseppunct}\relax
\EndOfBibitem
\bibitem[Albers \latin{et~al.}(2019)Albers, Levine, Sirlanci, and Stuart]{RN40}
Albers,~D.; Levine,~M.; Sirlanci,~M.; Stuart,~A. A Simple Modeling Framework
  For Prediction In The Human Glucose-Insulin System. \emph{arXiv preprint
  arXiv:1910.14193} \textbf{2019}, \relax
\mciteBstWouldAddEndPunctfalse
\mciteSetBstMidEndSepPunct{\mcitedefaultmidpunct}
{}{\mcitedefaultseppunct}\relax
\EndOfBibitem
\bibitem[Law \latin{et~al.}(2015)Law, Stuart, and Zygalakis]{RN48}
Law,~K.; Stuart,~A.; Zygalakis,~K. Data assimilation. \emph{Cham, Switzerland:
  Springer} \textbf{2015}, \relax
\mciteBstWouldAddEndPunctfalse
\mciteSetBstMidEndSepPunct{\mcitedefaultmidpunct}
{}{\mcitedefaultseppunct}\relax
\EndOfBibitem
\bibitem[Asch \latin{et~al.}(2016)Asch, Bocquet, and Nodet]{RN49}
Asch,~M.; Bocquet,~M.; Nodet,~M. \emph{Data assimilation: methods, algorithms,
  and applications}; SIAM, 2016\relax
\mciteBstWouldAddEndPuncttrue
\mciteSetBstMidEndSepPunct{\mcitedefaultmidpunct}
{\mcitedefaultendpunct}{\mcitedefaultseppunct}\relax
\EndOfBibitem
\bibitem[Gelman \latin{et~al.}(2013)Gelman, Carlin, Stern, Dunson, Vehtari, and
  Rubin]{RN70}
Gelman,~A.; Carlin,~J.~B.; Stern,~H.~S.; Dunson,~D.~B.; Vehtari,~A.;
  Rubin,~D.~B. \emph{Bayesian data analysis}; CRC press, 2013\relax
\mciteBstWouldAddEndPuncttrue
\mciteSetBstMidEndSepPunct{\mcitedefaultmidpunct}
{\mcitedefaultendpunct}{\mcitedefaultseppunct}\relax
\EndOfBibitem
\bibitem[Nelder and Mead(1965)Nelder, and Mead]{RN69}
Nelder,~J.~A.; Mead,~R. A simplex method for function minimization. \emph{The
  computer journal} \textbf{1965}, \emph{7}, 308--313\relax
\mciteBstWouldAddEndPuncttrue
\mciteSetBstMidEndSepPunct{\mcitedefaultmidpunct}
{\mcitedefaultendpunct}{\mcitedefaultseppunct}\relax
\EndOfBibitem
\bibitem[Cressoni \latin{et~al.}(2016)Cressoni, Gotti, Chiurazzi, Massari,
  Algieri, Amini, Cammaroto, Brioni, Montaruli, and Nikolla]{RN58}
Cressoni,~M.; Gotti,~M.; Chiurazzi,~C.; Massari,~D.; Algieri,~I.; Amini,~M.;
  Cammaroto,~A.; Brioni,~M.; Montaruli,~C.; Nikolla,~K. Mechanical power and
  development of ventilator-induced lung injury. \emph{Anesthesiology: The
  Journal of the American Society of Anesthesiologists} \textbf{2016},
  \emph{124}, 1100--1108\relax
\mciteBstWouldAddEndPuncttrue
\mciteSetBstMidEndSepPunct{\mcitedefaultmidpunct}
{\mcitedefaultendpunct}{\mcitedefaultseppunct}\relax
\EndOfBibitem
\bibitem[Amato \latin{et~al.}(2015)Amato, Meade, Slutsky, Brochard, Costa,
  Schoenfeld, Stewart, Briel, Talmor, Mercat, Richard, Carvalho, and
  Brower]{RN86}
Amato,~M.~B.; Meade,~M.~O.; Slutsky,~A.~S.; Brochard,~L.; Costa,~E.~L.;
  Schoenfeld,~D.~A.; Stewart,~T.~E.; Briel,~M.; Talmor,~D.; Mercat,~A.;
  Richard,~J.~C.; Carvalho,~C.~R.; Brower,~R.~G. Driving pressure and survival
  in the acute respiratory distress syndrome. \emph{N Engl J Med}
  \textbf{2015}, \emph{372}, 747--55\relax
\mciteBstWouldAddEndPuncttrue
\mciteSetBstMidEndSepPunct{\mcitedefaultmidpunct}
{\mcitedefaultendpunct}{\mcitedefaultseppunct}\relax
\EndOfBibitem
\bibitem[Chiumello \latin{et~al.}(2016)Chiumello, Carlesso, Brioni, and
  Cressoni]{RN87}
Chiumello,~D.; Carlesso,~E.; Brioni,~M.; Cressoni,~M. Airway driving pressure
  and lung stress in ARDS patients. \emph{Crit Care} \textbf{2016}, \emph{20},
  276\relax
\mciteBstWouldAddEndPuncttrue
\mciteSetBstMidEndSepPunct{\mcitedefaultmidpunct}
{\mcitedefaultendpunct}{\mcitedefaultseppunct}\relax
\EndOfBibitem
\bibitem[Aoyama \latin{et~al.}(2018)Aoyama, Pettenuzzo, Aoyama, Pinto,
  Englesakis, and Fan]{RN88}
Aoyama,~H.; Pettenuzzo,~T.; Aoyama,~K.; Pinto,~R.; Englesakis,~M.; Fan,~E.
  Association of driving pressure with mortality among ventilated patients with
  acute respiratory distress syndrome: a systematic review and meta-analysis.
  \emph{Critical care medicine} \textbf{2018}, \emph{46}, 300--306\relax
\mciteBstWouldAddEndPuncttrue
\mciteSetBstMidEndSepPunct{\mcitedefaultmidpunct}
{\mcitedefaultendpunct}{\mcitedefaultseppunct}\relax
\EndOfBibitem
\bibitem[Gattinoni \latin{et~al.}(1987)Gattinoni, Pesenti, Avalli, Rossi, and
  Bombino]{RN89}
Gattinoni,~L.; Pesenti,~A.; Avalli,~L.; Rossi,~F.; Bombino,~M. Pressure-volume
  curve of total respiratory system in acute respiratory failure: computed
  tomographic scan study. \emph{American Review of Respiratory Disease}
  \textbf{1987}, \emph{136}, 730--736\relax
\mciteBstWouldAddEndPuncttrue
\mciteSetBstMidEndSepPunct{\mcitedefaultmidpunct}
{\mcitedefaultendpunct}{\mcitedefaultseppunct}\relax
\EndOfBibitem
\bibitem[Smith \latin{et~al.}(2020)Smith, Roy, Cleveland, Mattson, Okamura,
  Charlebois, Hamlington, Novotny, Knudsen, and Ochs]{RN90}
Smith,~B.~J.; Roy,~G.~S.; Cleveland,~A.; Mattson,~C.; Okamura,~K.;
  Charlebois,~C.~M.; Hamlington,~K.~L.; Novotny,~M.~V.; Knudsen,~L.; Ochs,~M.
  Three Alveolar Phenotypes Govern Lung Function in Murine Ventilator-Induced
  Lung Injury. \emph{Frontiers in Physiology} \textbf{2020}, \emph{11},
  660--660\relax
\mciteBstWouldAddEndPuncttrue
\mciteSetBstMidEndSepPunct{\mcitedefaultmidpunct}
{\mcitedefaultendpunct}{\mcitedefaultseppunct}\relax
\EndOfBibitem
\end{mcitethebibliography}


\providecommand{\latin}[1]{#1}
\makeatletter
\providecommand{\doi}
  {\begingroup\let\do\@makeother\dospecials
  \catcode`\{=1 \catcode`\}=2 \doi@aux}
\providecommand{\doi@aux}[1]{\endgroup\texttt{#1}}
\makeatother
\providecommand*\mcitethebibliography{\thebibliography}
\csname @ifundefined\endcsname{endmcitethebibliography}
  {\let\endmcitethebibliography\endthebibliography}{}
\begin{mcitethebibliography}{0}
\providecommand*\natexlab[1]{#1}
\providecommand*\mciteSetBstSublistMode[1]{}
\providecommand*\mciteSetBstMaxWidthForm[2]{}
\providecommand*\mciteBstWouldAddEndPuncttrue
  {\def\EndOfBibitem{\unskip.}}
\providecommand*\mciteBstWouldAddEndPunctfalse
  {\let\EndOfBibitem\relax}
\providecommand*\mciteSetBstMidEndSepPunct[3]{}
\providecommand*\mciteSetBstSublistLabelBeginEnd[3]{}
\providecommand*\EndOfBibitem{}
\mciteSetBstSublistMode{f}
\mciteSetBstMaxWidthForm{subitem}{(\alph{mcitesubitemcount})}
\mciteSetBstSublistLabelBeginEnd
  {\mcitemaxwidthsubitemform\space}
  {\relax}
  {\relax}

\end{mcitethebibliography}

\end{document}


\clearpage

\small{

\begin{figure}[h!]
\centerline{\includegraphics[width = 1\textwidth]{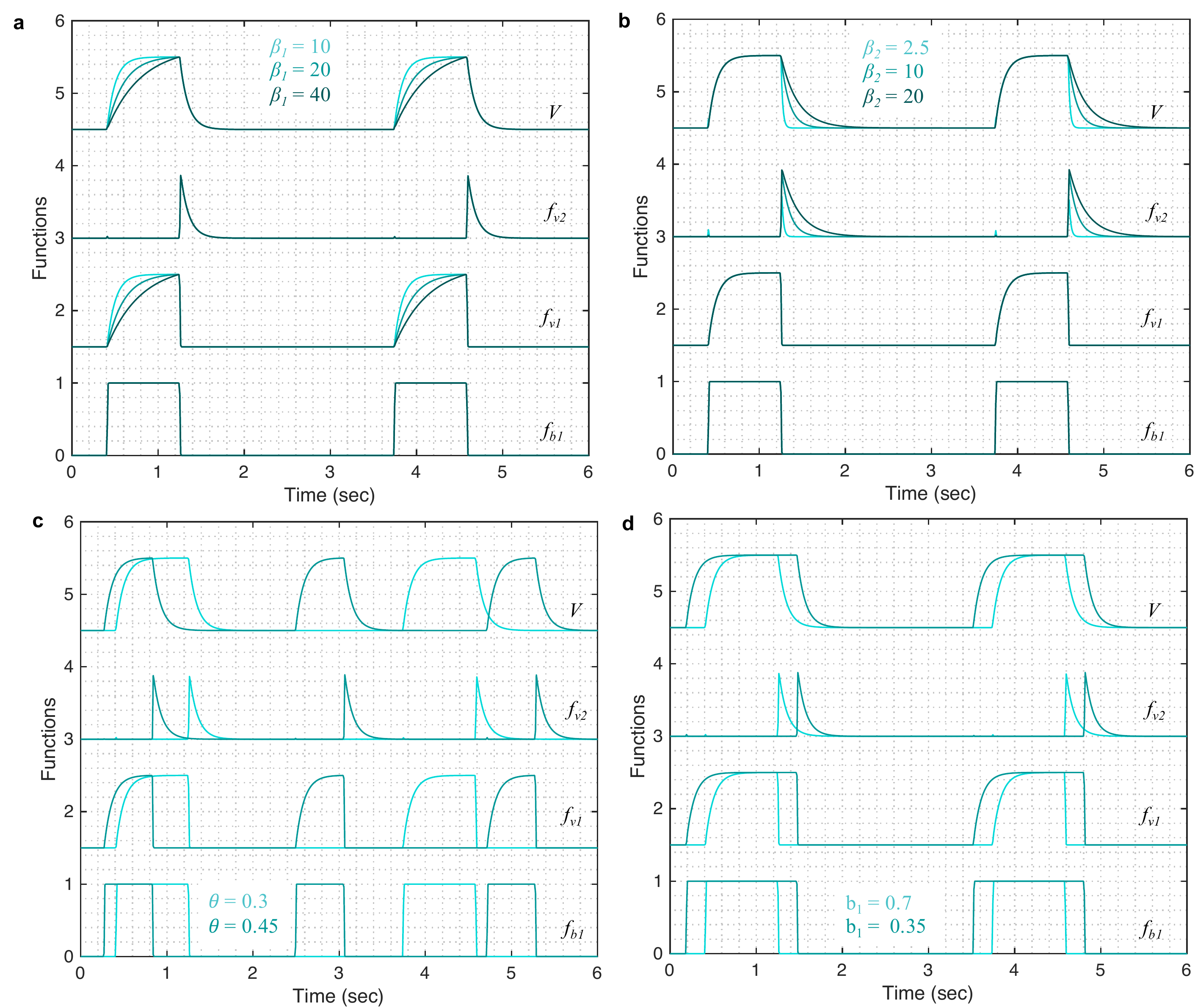}}
\caption{Figure S1: Showing the effect of parameter variations onto the submodels that form the volume model for the cases shown in Fig.~3. Changing (a) $\beta_1$ and (b) $\beta_2$ changes the gradient of the rising and falling signals, respectively. Increased value of these parameters increases the transient time for the signal to reach the same volume level. (c) Change in the respiratory frequency ($\theta$) changes the periodicity of the breath while (d) parameter $b_1$ changes the I:E ratio (inspiratory to expiratory time ratio). Here, $V$ is the output of the model, which was calculated using Eqns.~(1)-(5) while considering $\theta$ = 0.3, $a_1$ = 200, $b_1$ = 0.7, $\phi_1$ = 0, $\beta_1$ = 10, $\beta_2$ = 10, $A_{v}$ = 1. Y-axis was normalized to represent all the submodels in a sequential manner. }
\label{figS1}
\end{figure}

\begin{figure}[h!]
\centerline{\includegraphics[width = 1\textwidth]{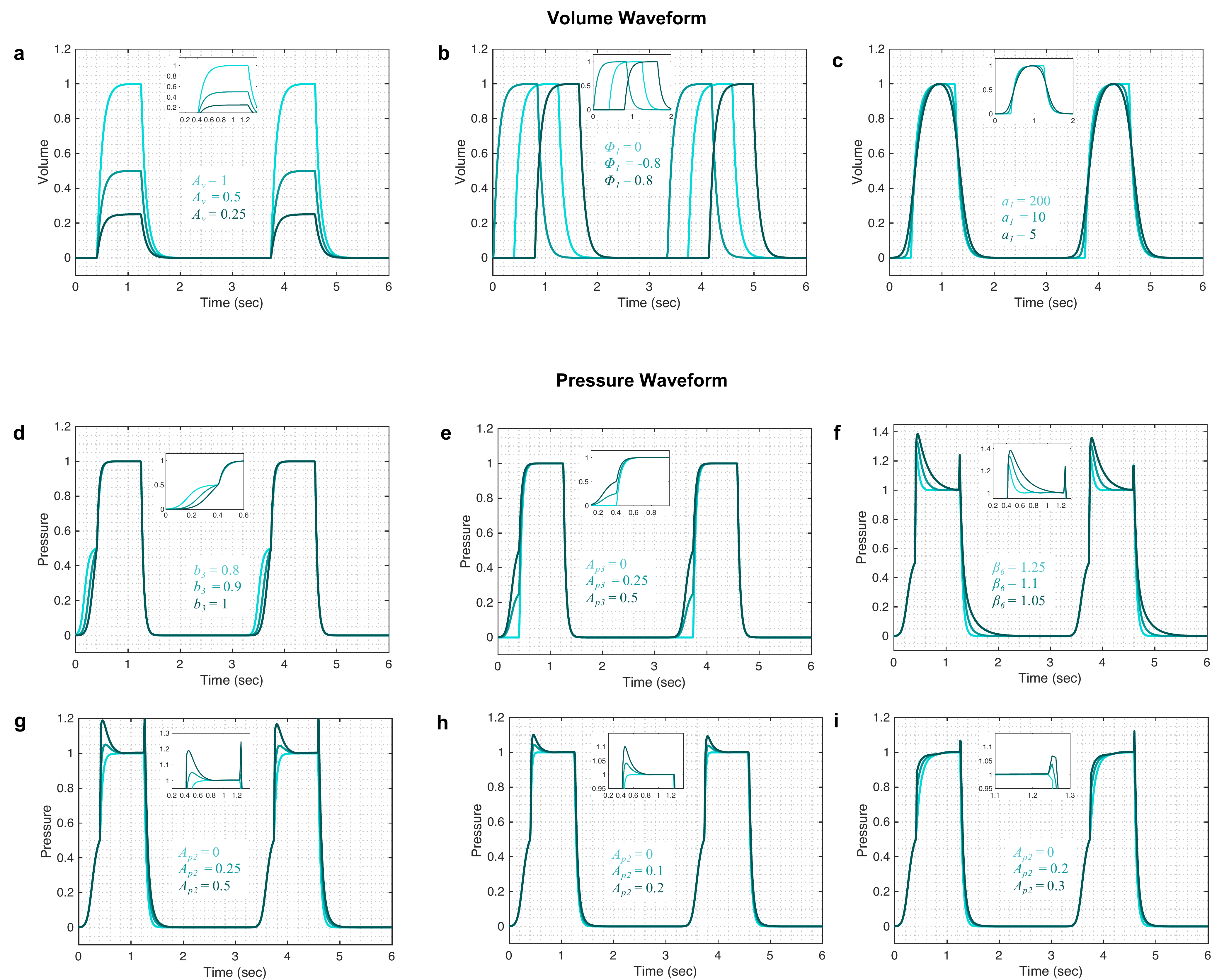}}
\caption{Figure S2: (a-c) Further variations in the volume waveform can be achieved by changing the (a) $A_{v}$, (b) $\phi_1$ and (c) $a_1$ parameters. (d-i) Further variations in the pressure waveform can be achieved by changing the (d) $b_3$, and (e) $A_{p_{3}}$ parameters; (f) $\beta_6$ parameter when $\beta_3$ = 2.5 and $A_{p_{2}}$ = 0.5; (g-i) $A_{p_{2}}$ parameter when (g) $\beta_3$ = 5, (h) $\beta_3$ = 2.5 (i) $\beta_3$ = 10; To simulate the response of the volume and pressure models, Eqns~(1)-(5) and Eqns~(6)-(18) were used, respectively, at the parameter values $\theta$ = 0.3, $a_1$ = 200, $b_1$ = 0.7, $\phi_1$ = 0, $\beta_1$ = 10, $\beta_2$ = 10, $A_{v}$ = 1, $a_2$ = 200, $b_2$ = 0.7, $\phi_2$ = 0, $a_3$ = 10, $b_3$ = 0.9, $\phi_3$ = -0.6, $\beta_3$ = $\beta_4$ = 5, $\beta_5$ = 1.001, $\beta_6$ = 1.1111, $A_{p_{1}}$ = 1, $A_{p_{2}}$ = 0, $A_{p_{3}}$ = 0.5, $A_{p_{4}}$ = 0. A zoomed-in view of each plot is shown inside the respective plot to highlight the changes in the waveform. }
\label{figS2}
\end{figure}

\begin{figure}[h!]
\centerline{\includegraphics[width = 0.8\textwidth]{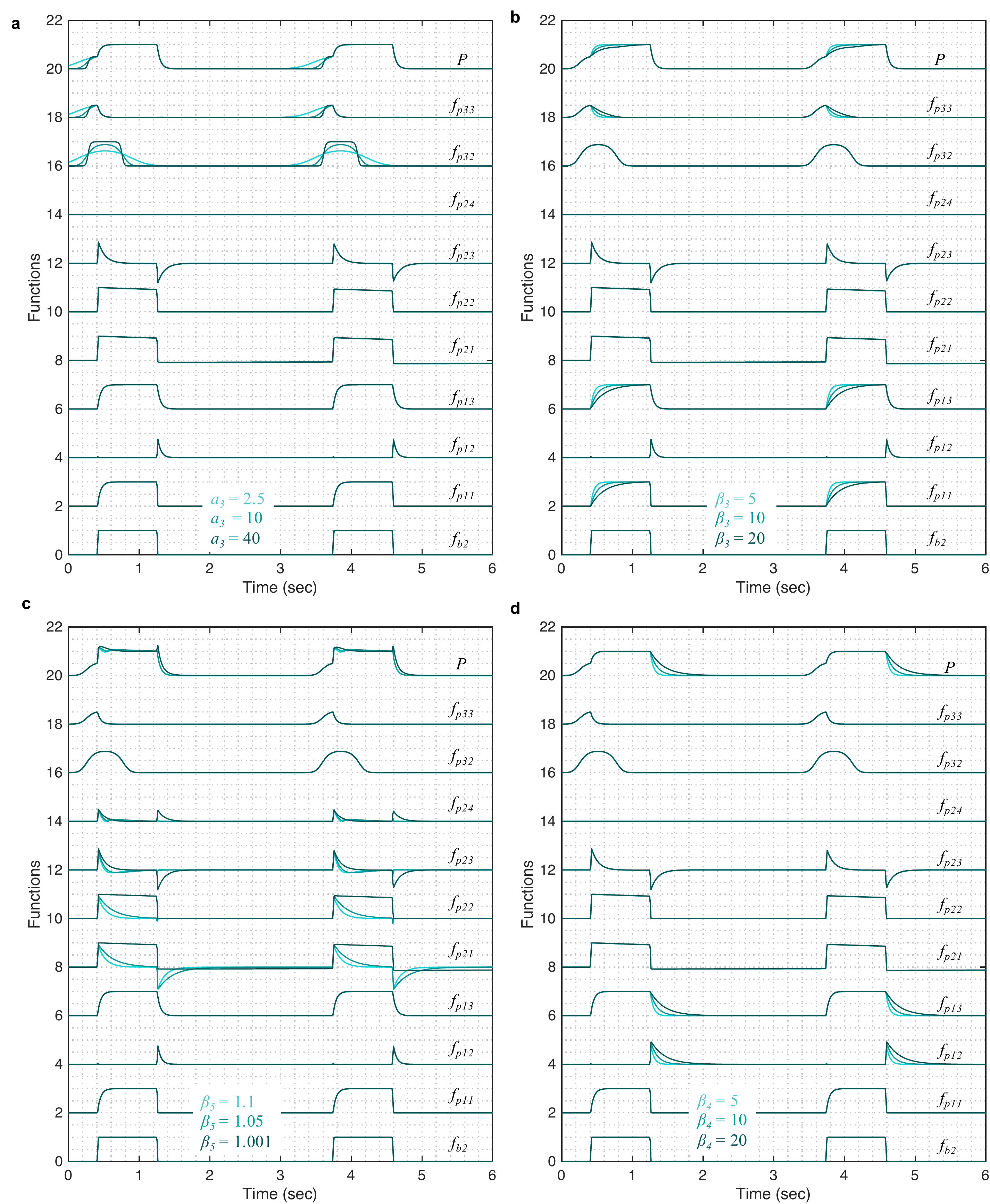}}
\caption{Figure S3 Showing the effect of parameter variations onto the submodels that make up the pressure model for the cases shown in Fig.~5. (a) Feature A1, which corresponds to the shape of the rising signal gradient at the beginning of inspiration, is produced via $f_{p_{33}}$ submodel where changes in parameter $a_3$ alters the gradient of the signal before inflection point. (b) Feature A2 is controlled via $f_{p_{11}}$ function, that is a part of $f_{p_{13}}$ submodel and changing $\beta_3$ changes the shape of feature A2. (c) Features B1 and B2, which correspond to peaks at the beginning and end of plateau respectively, are incorporated via $f_{p_{24}}$ submodel and their shapes can be altered by changing $\beta_5$ parameter ($A_{p_{4}}$ = 0.5) (d) Feature C, which corresponds to the shape of falling signal gradient, is controlled via $f_{p_{12}}$ function where changing $\beta_4$ changes the shape of this feature. Equations (6)-(18) were used to simulate the response of the pressure model while considering $\theta$ = 0.3, $a_2$ = 200, $b_2$ = 0.7, $\phi_2$ = 0, $a_3$ = 10, $b_3$ = 0.9, $\phi_3$ = -0.6, $\beta_3$ = $\beta_4$ = 5, $\beta_5$ = 1.001, $\beta_6$ = 1.1111, $A_{p_{1}}$ = 1, $A_{p_{2}}$ = 0, $A_{p_{3}}$ = 0.5, $A_{p_{4}}$ = 0. Y-axis was normalized to represent all the submodels in a sequential manner. }
\label{figS3}
\end{figure}

 \begin{figure}[htbp!]
\centerline{\subfigure{\includegraphics[width = 4.5cm]{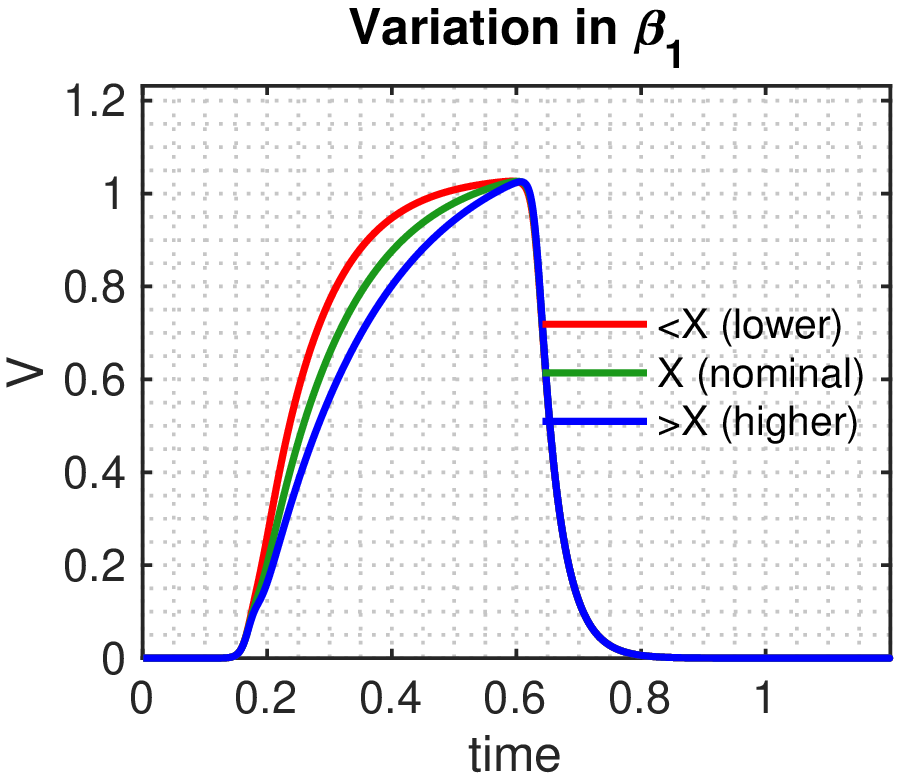}}
\subfigure{\includegraphics[width = 4.5cm]{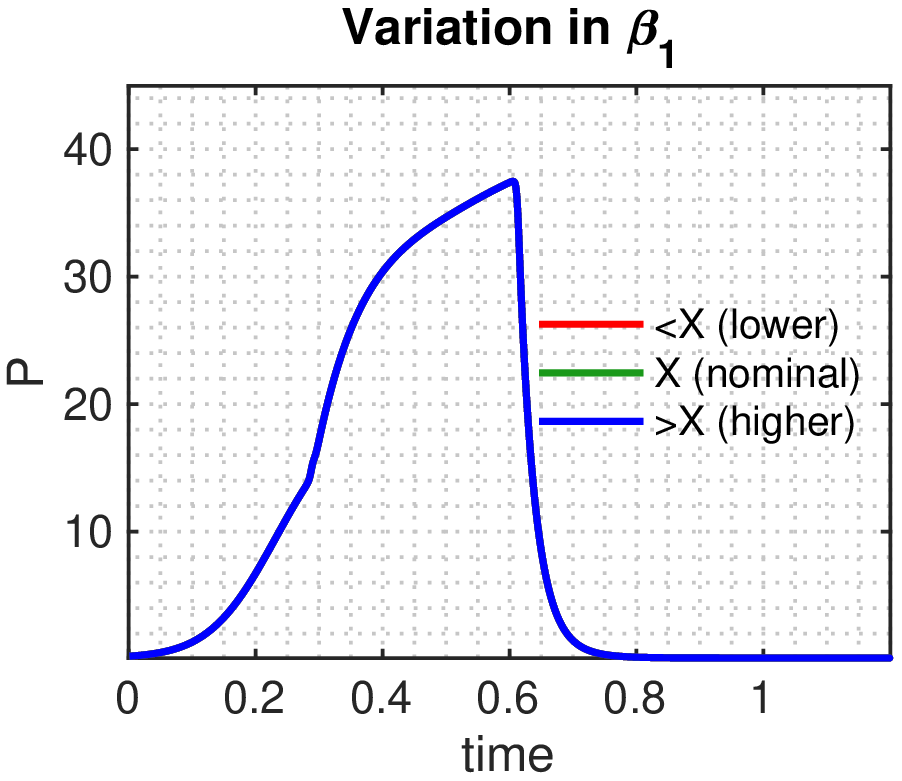}}
\subfigure{\includegraphics[width = 4.5cm]{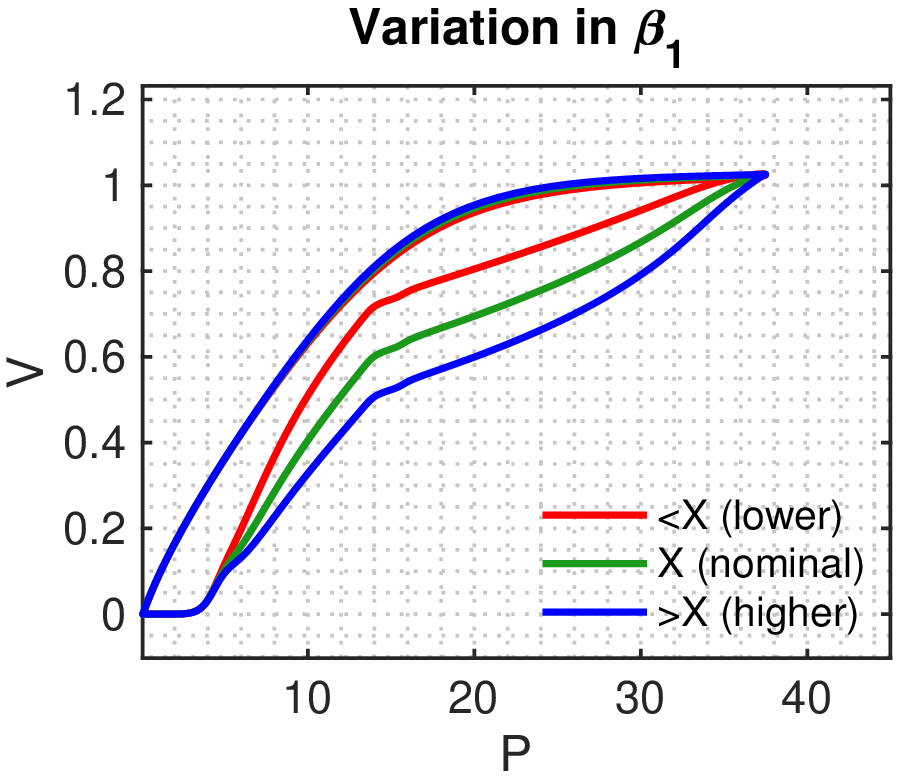}}}
\centerline{\subfigure{\includegraphics[width = 4.5cm]{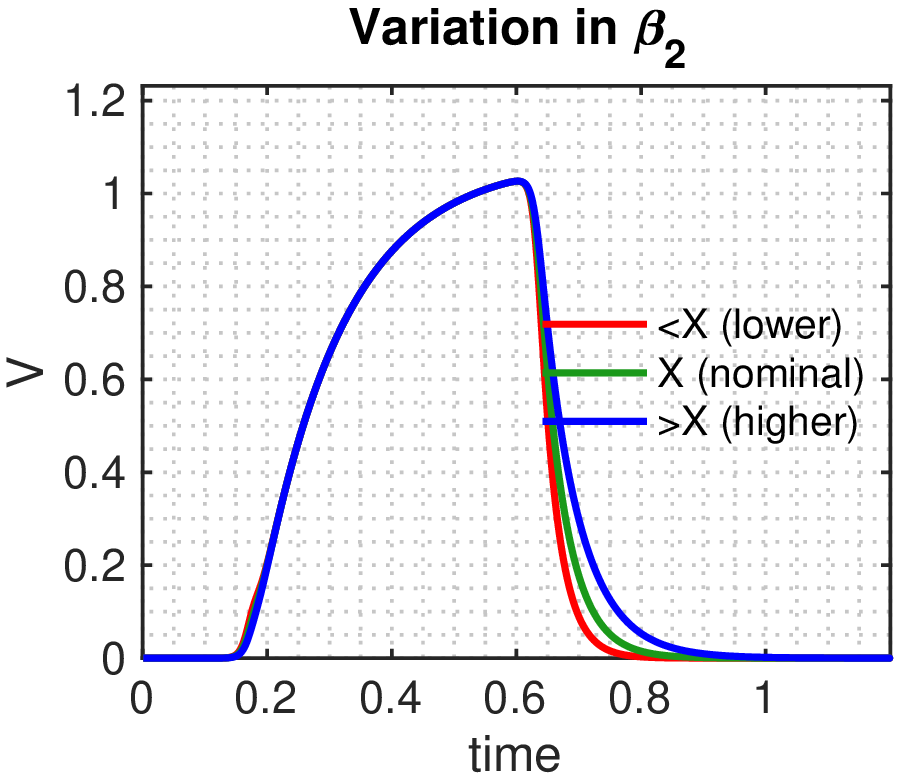}}
\subfigure{\includegraphics[width = 4.5cm]{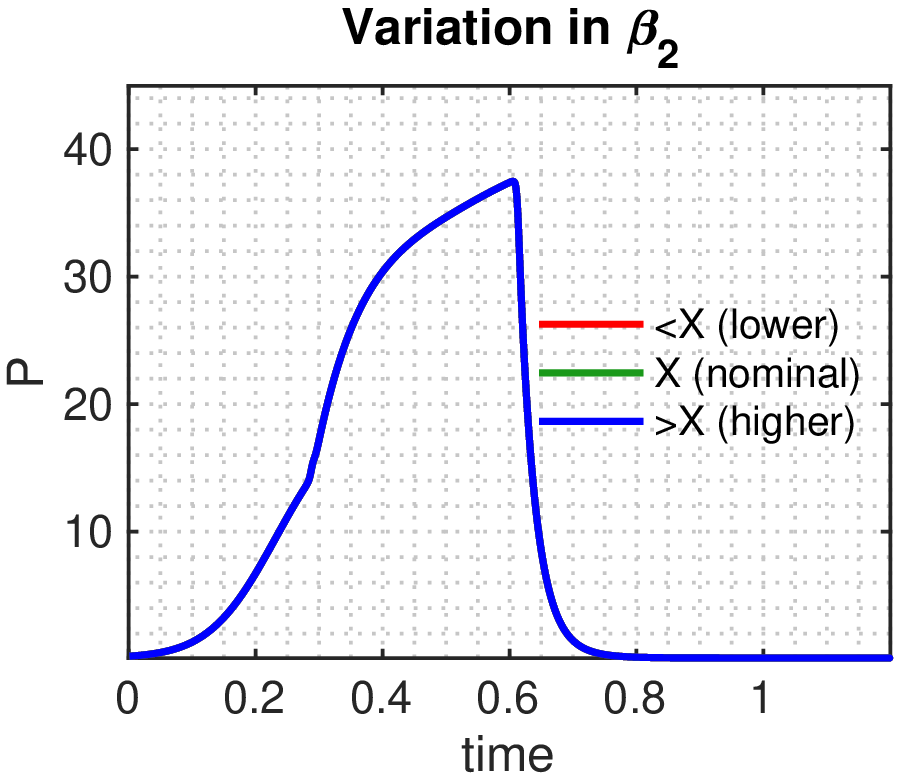}}
\subfigure{\includegraphics[width = 4.5cm]{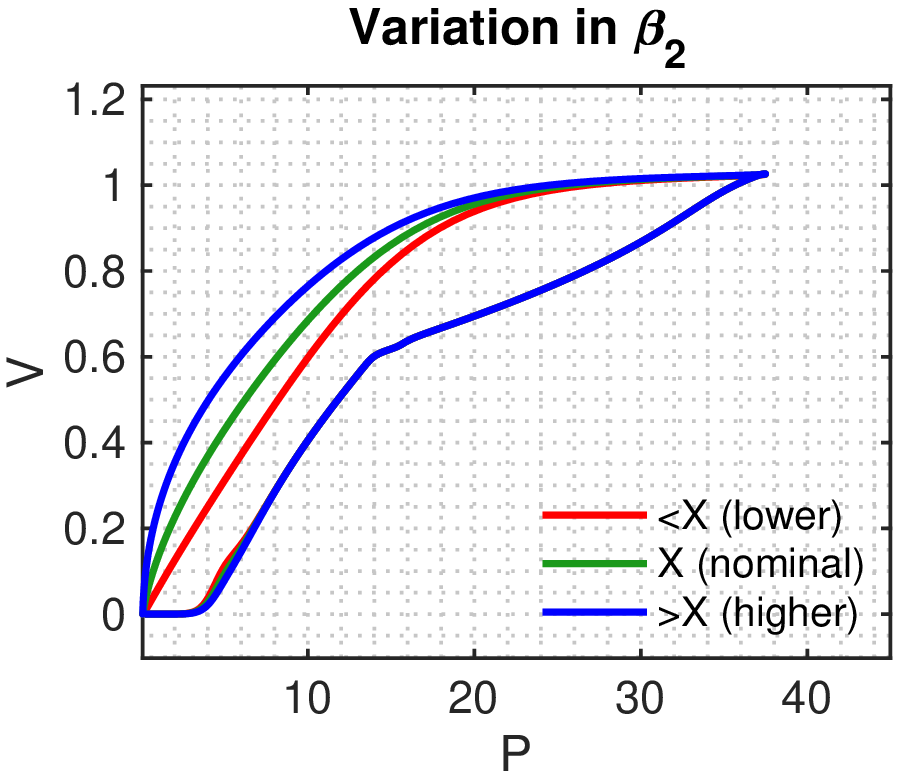}}}
\centerline{\subfigure{\includegraphics[width = 4.5cm]{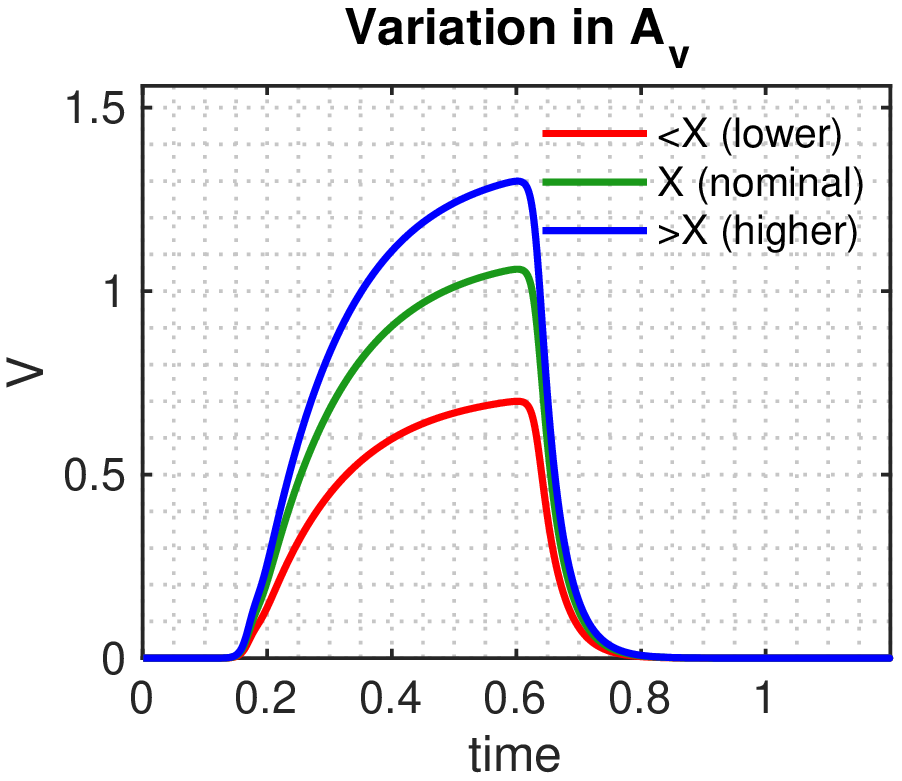}}
\subfigure{\includegraphics[width = 4.5cm]{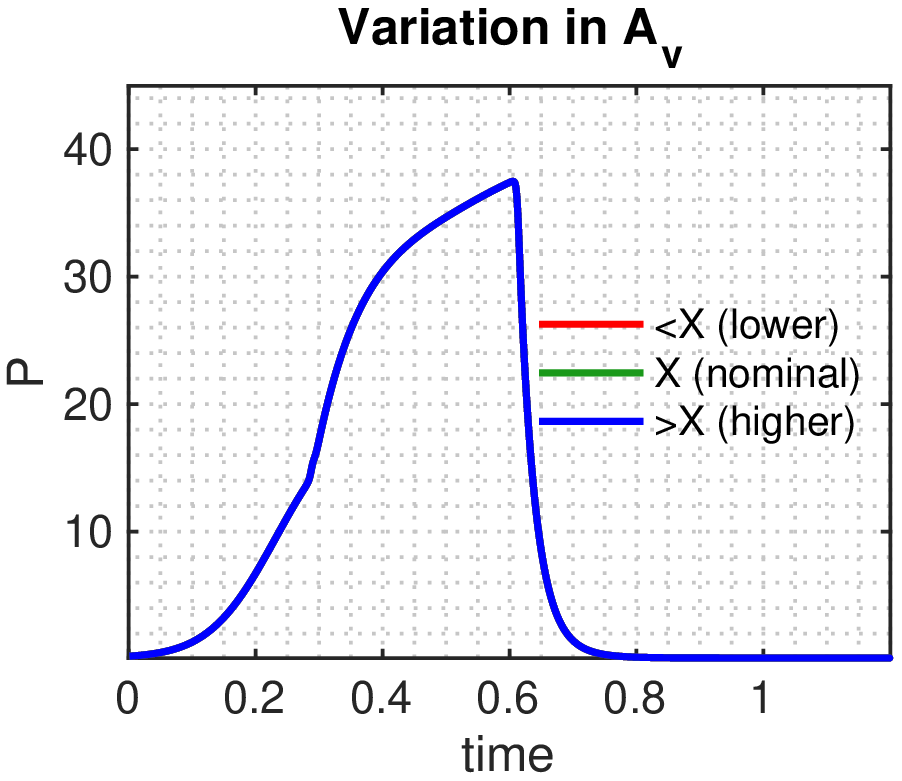}}
\subfigure{\includegraphics[width = 4.5cm]{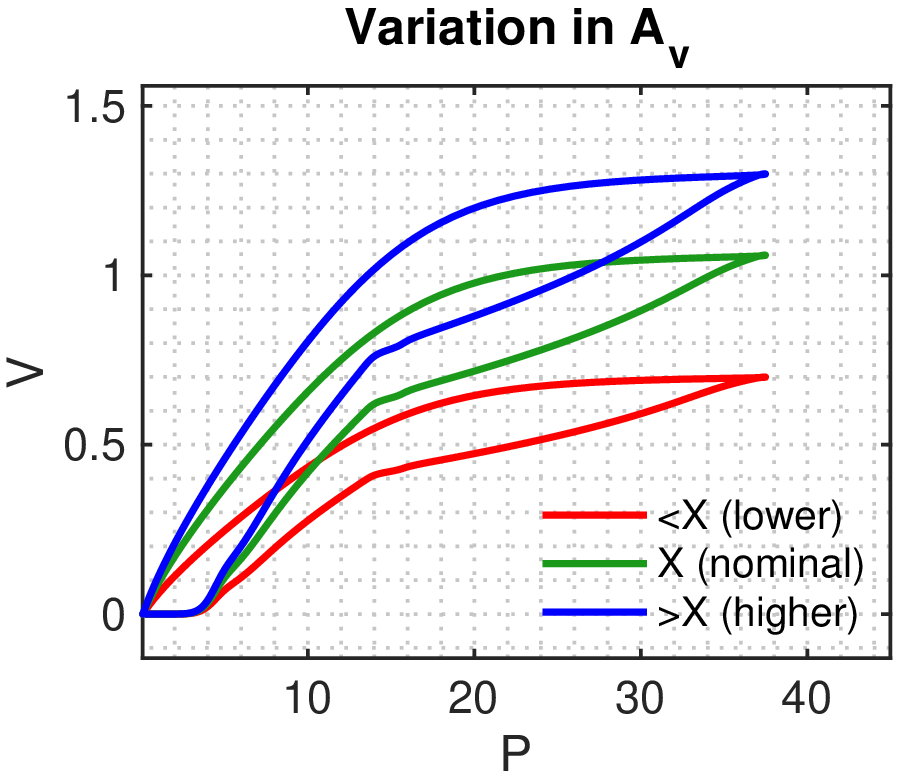}}}
 \label{FigS4}
 \caption{Figure S4: Three parameters in the volume model could have direct physiological meaning. These are $\beta_1$, $\beta_2$ and $A_v$. Simulated response of the volume and pressure models are shown when the specific parameter was varied while considering the nominal parameter values shown in Table~1 for the mouse model PCV, healthy case (Fig.~6). Here, variable $X$ corresponds to the nominal value of the parameter while $<X$ and $>X$ represent smaller and larger values than the nominal value, respectively. }
\end{figure}

 \begin{table}[h!]
 \begin{center}
 \begin{tabular}{ c }
 \includegraphics[width = 1\textwidth]{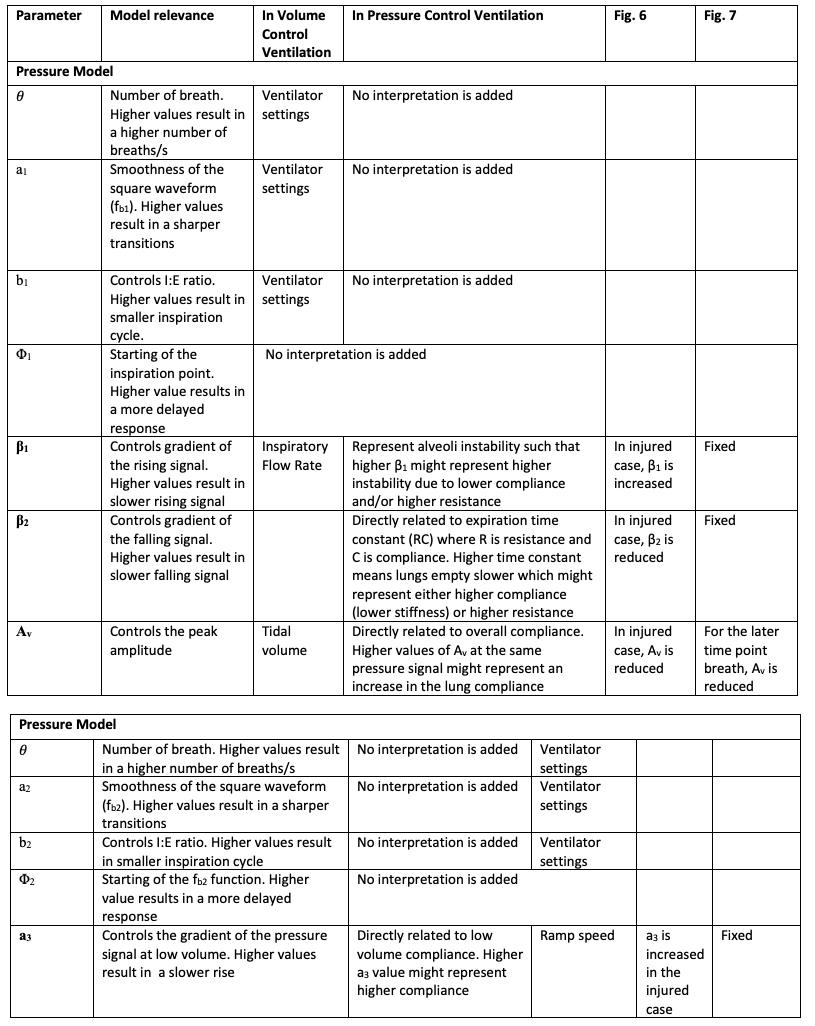}\\
 \end{tabular}
 \caption{Table S1: Interpretation of the volume and pressure model parameters. The parameters that are correlated with a known measures of lung physiology are in bold. }
 \label{tableS1}
 \end{center}
 \end{table}
 
 \begin{table}[h!]
 \begin{center}
 \begin{tabular}{ c }
 \includegraphics[width = 1\textwidth]{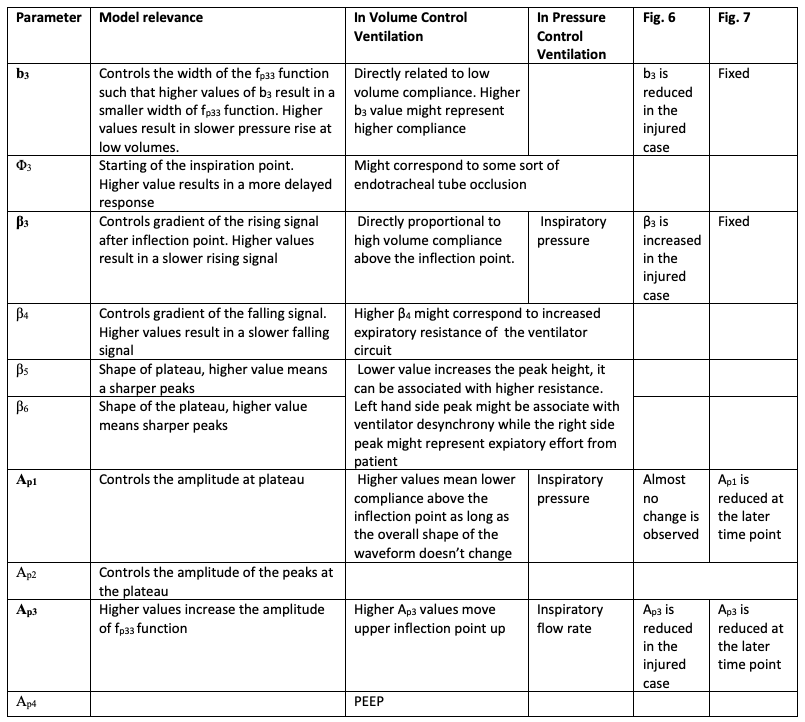}\\
 \end{tabular}
 \caption{}
 \label{tableS1}
 \end{center}
 \end{table}

\begin{figure}[t!]
\centerline{\includegraphics[width = 0.8\textwidth]{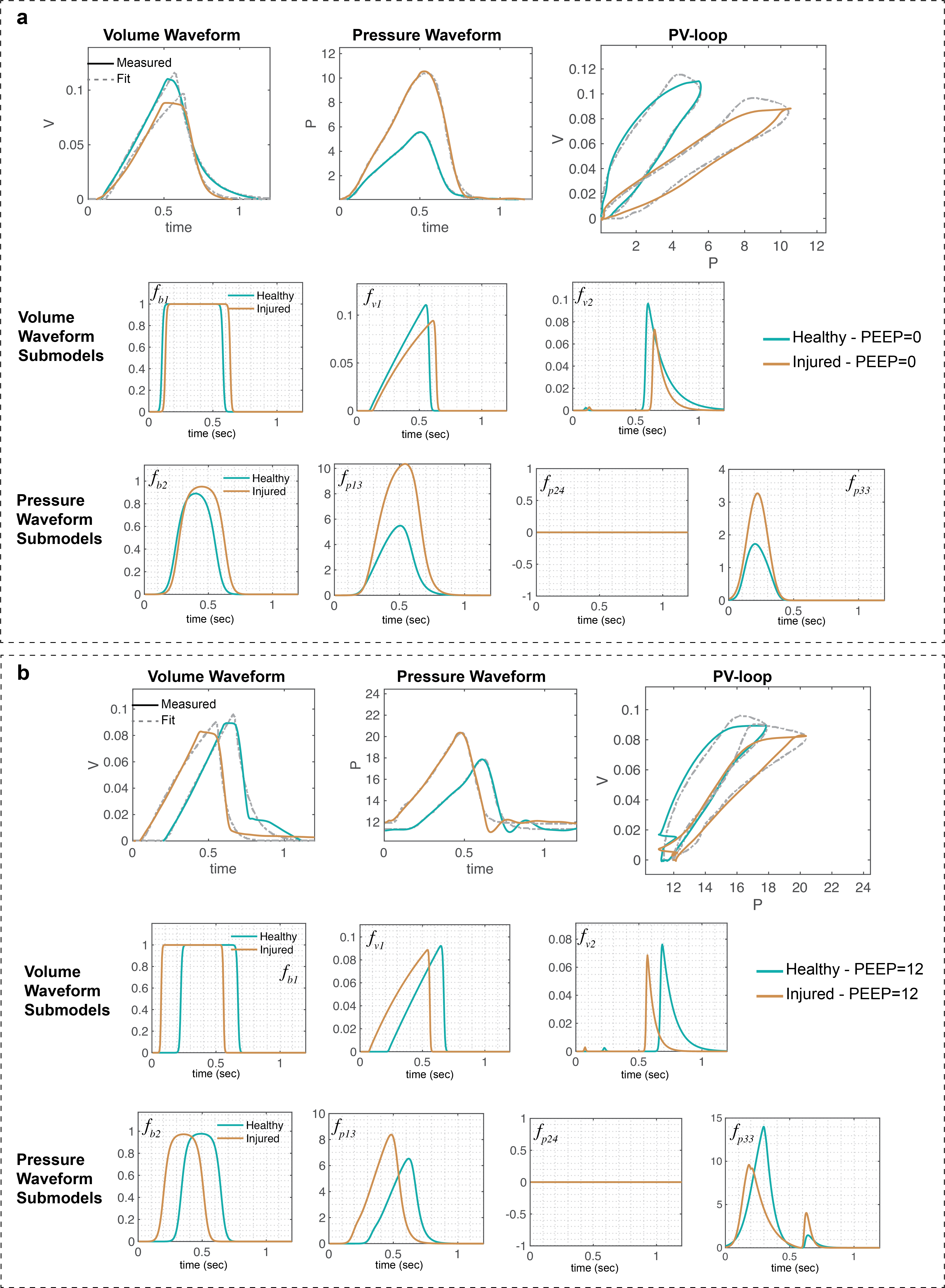}}
\caption{Figure S6: Experimental data from a representative mouse is compared with the model data in healthy and injured condition at (a) PEEP = 0 and (b) PEEP = 12. In each panel, in the first row the measured response is shown in solid lines while the model inferred response is shown in dashed lines. Changes in the volume and pressure submodels are shown in the second and third rows, respectively (in solid lines). The volume and pressure models shown in Eqns.~(1)-(5) and (6)-(18) were used to generate the best-fit model response using estimated mean parameter values shown in Table~S2, respectively. The respective uncertainties in the parameter values is shown in Table S2.}
\label{figS6}
\end{figure}

\begin{figure}[htbp!]%
\centerline{\subfigure{\includegraphics[width = 4.5cm]{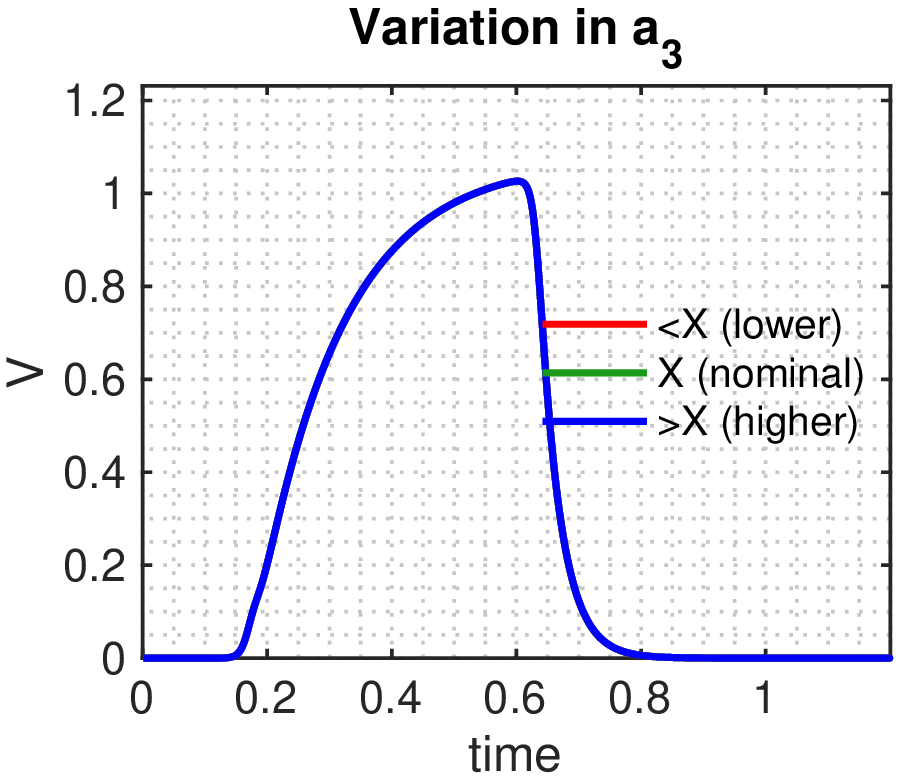}}
\subfigure{\includegraphics[width = 4.5cm]{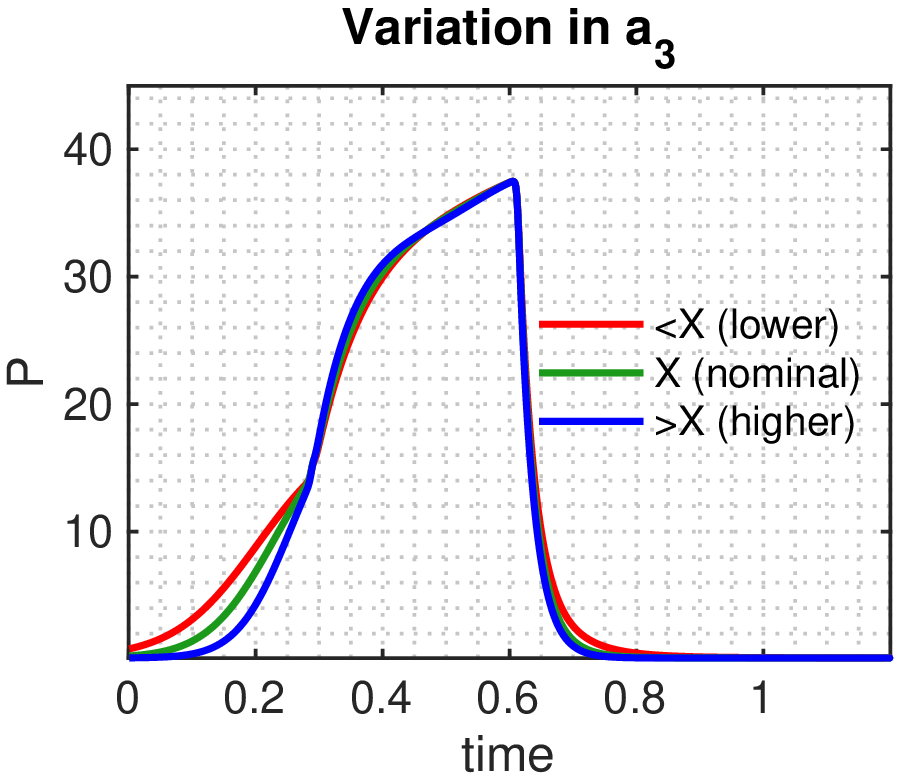}}
\subfigure{\includegraphics[width = 4.5cm]{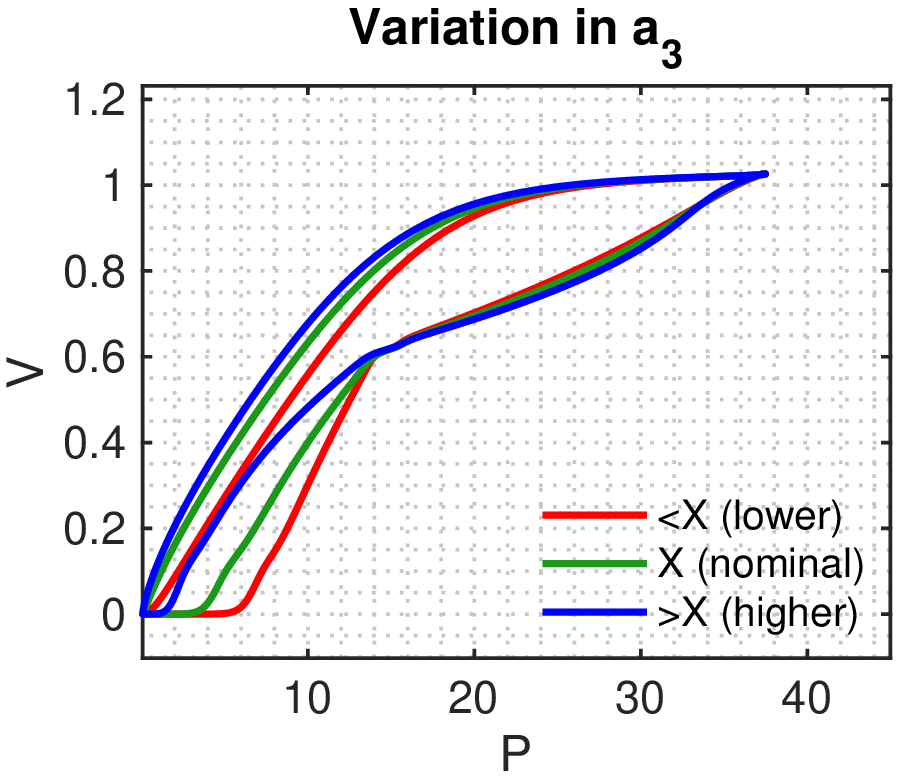}}}
\centerline{\subfigure{\includegraphics[width = 4.5cm]{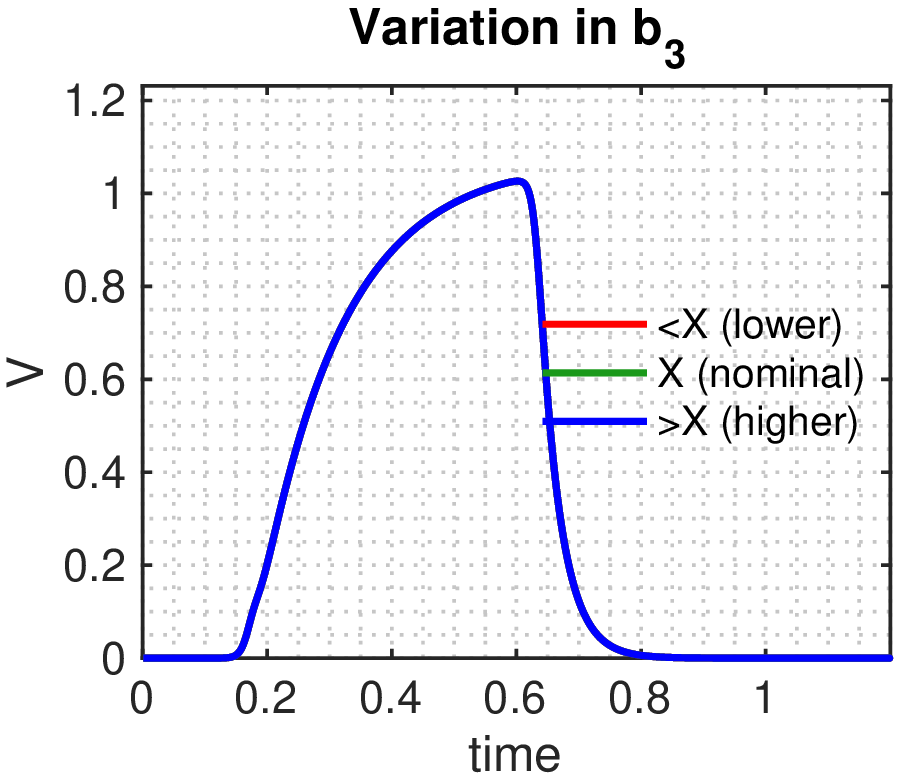}}
\subfigure{\includegraphics[width = 4.5cm]{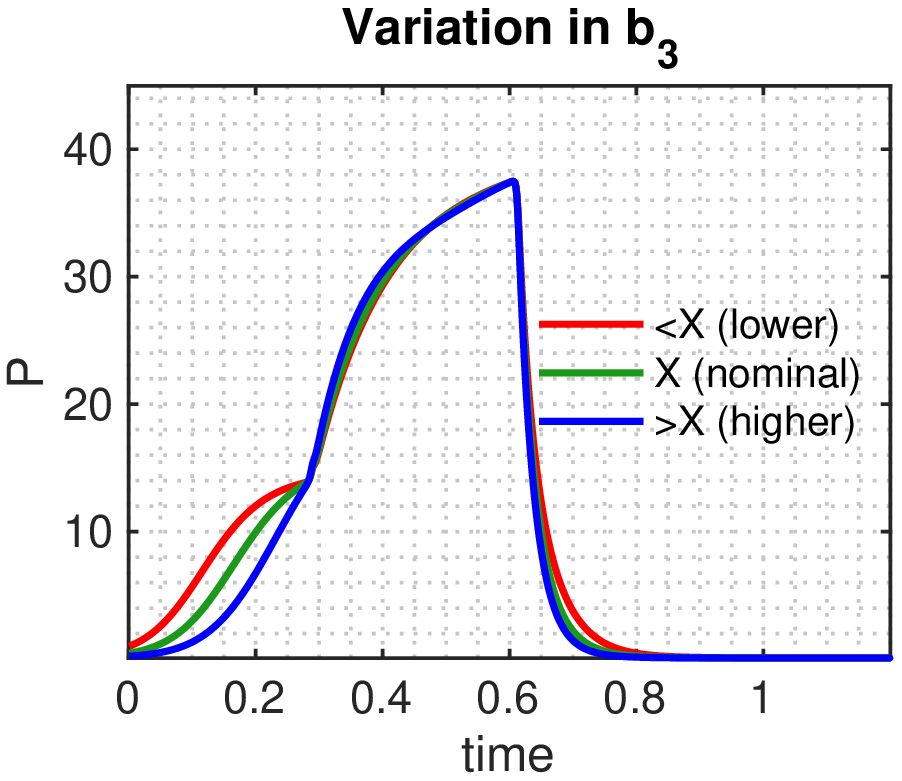}}
\subfigure{\includegraphics[width = 4.5cm]{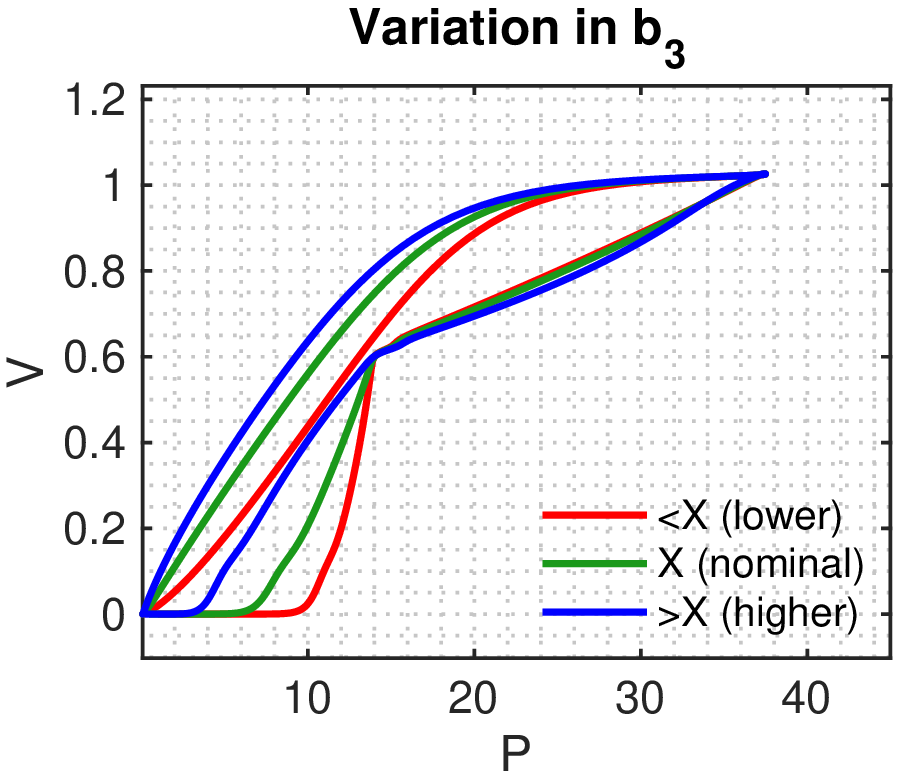}}}
\centerline{\subfigure{\includegraphics[width = 4.5cm]{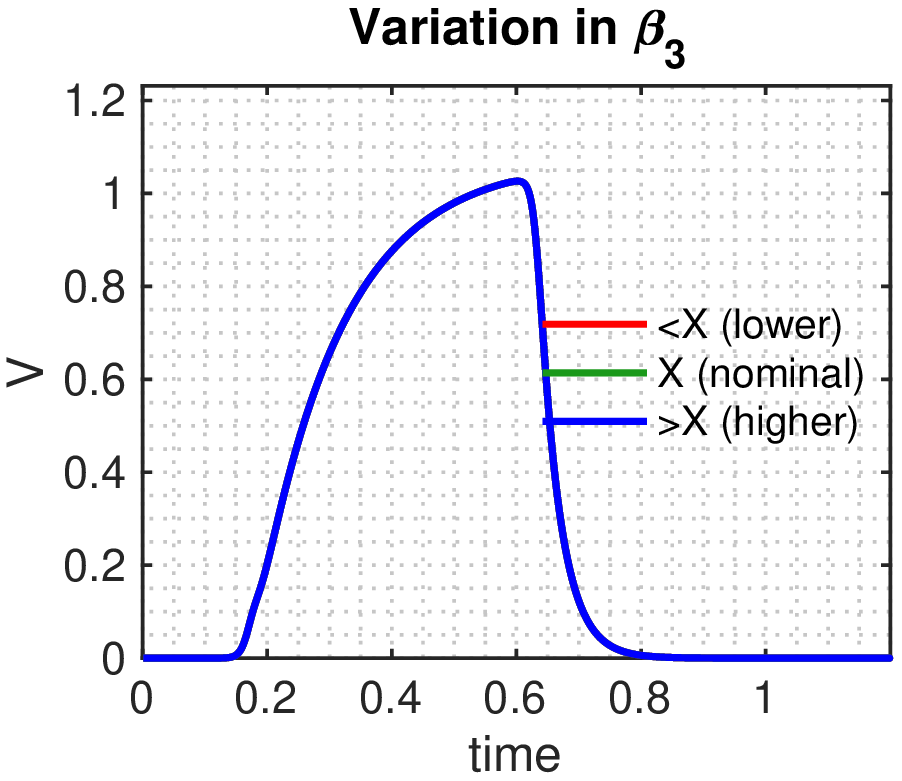}}
\subfigure{\includegraphics[width = 4.5cm]{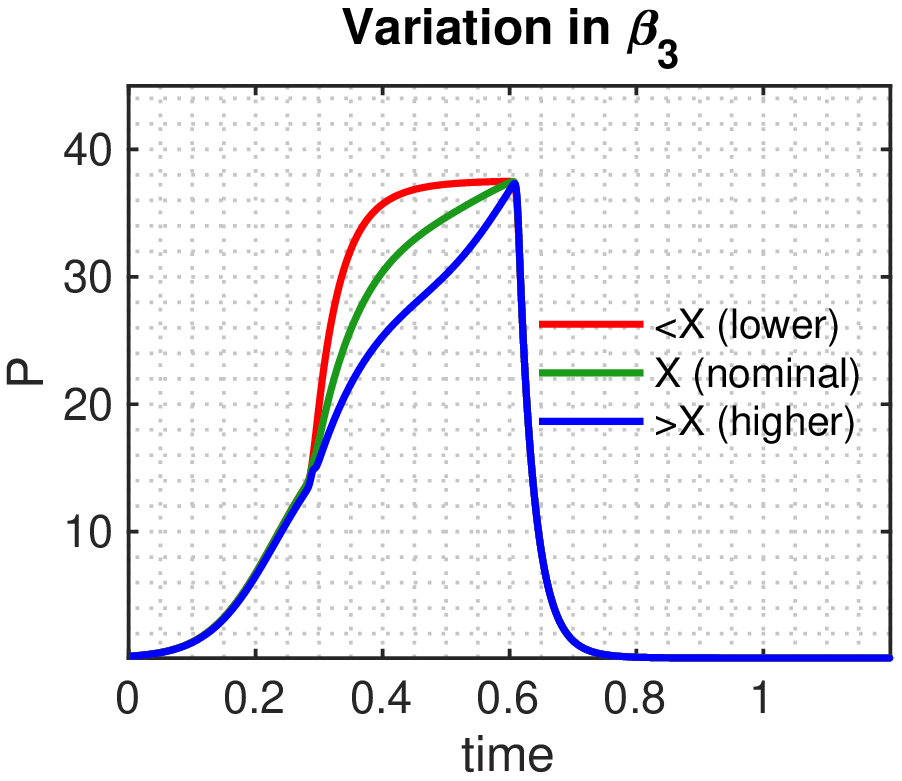}}
\subfigure{\includegraphics[width = 4.5cm]{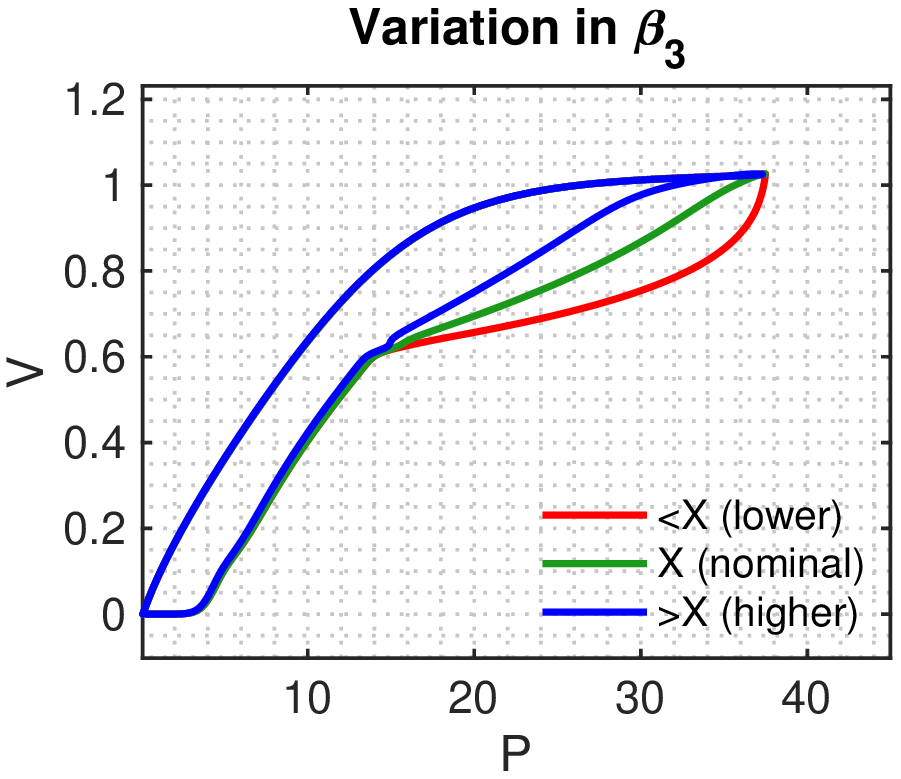}}}
\centerline{\subfigure{\includegraphics[width = 4.5cm]{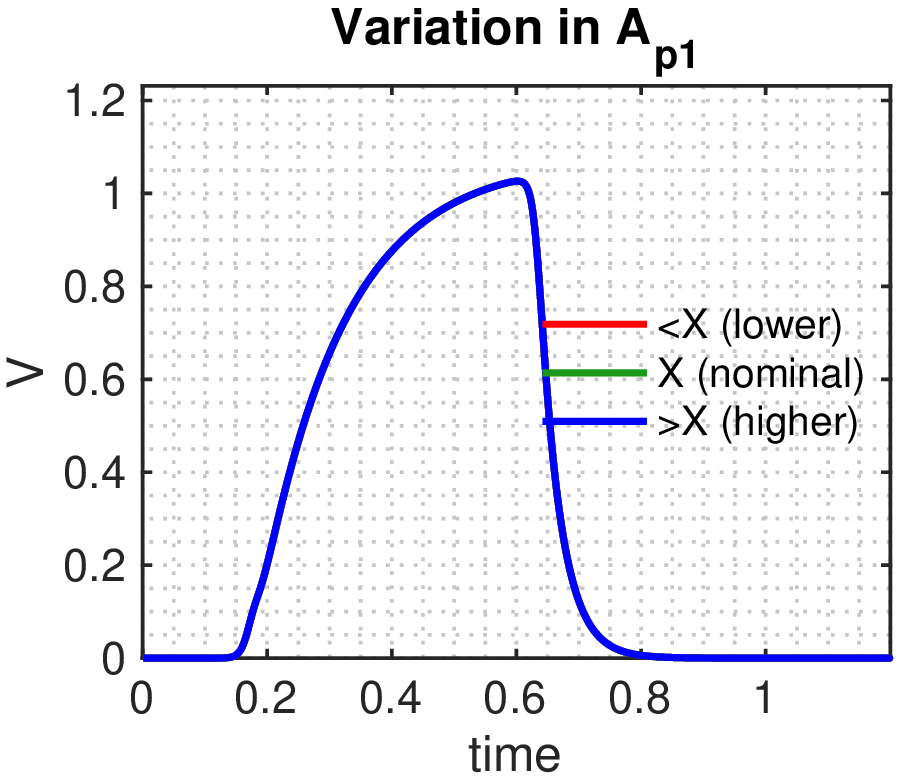}}
\subfigure{\includegraphics[width = 4.5cm]{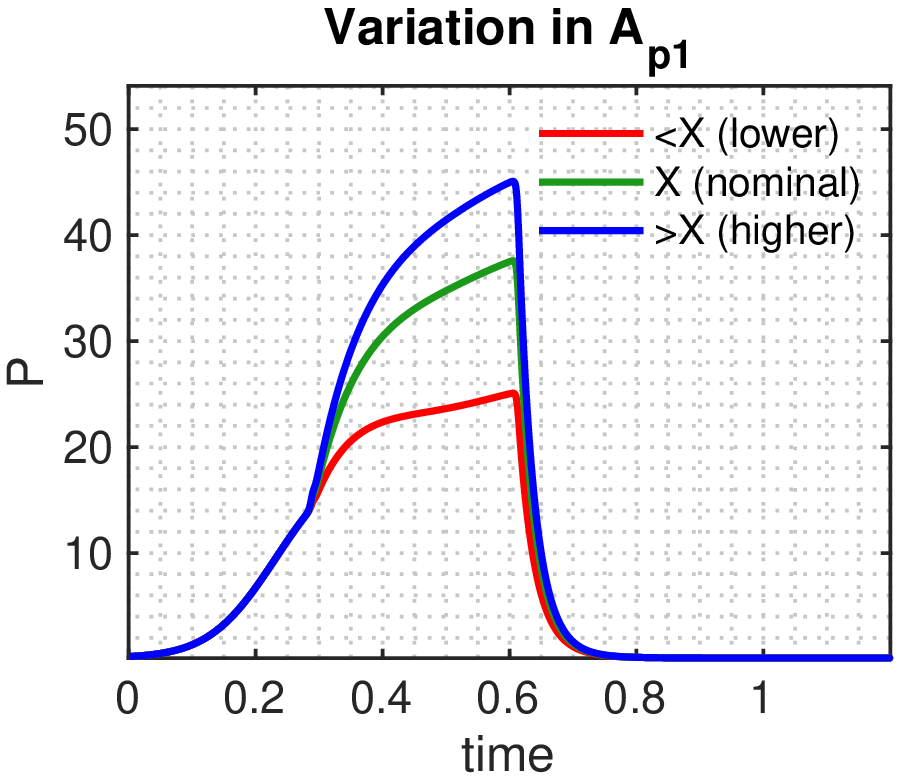}}
\subfigure{\includegraphics[width = 4.5cm]{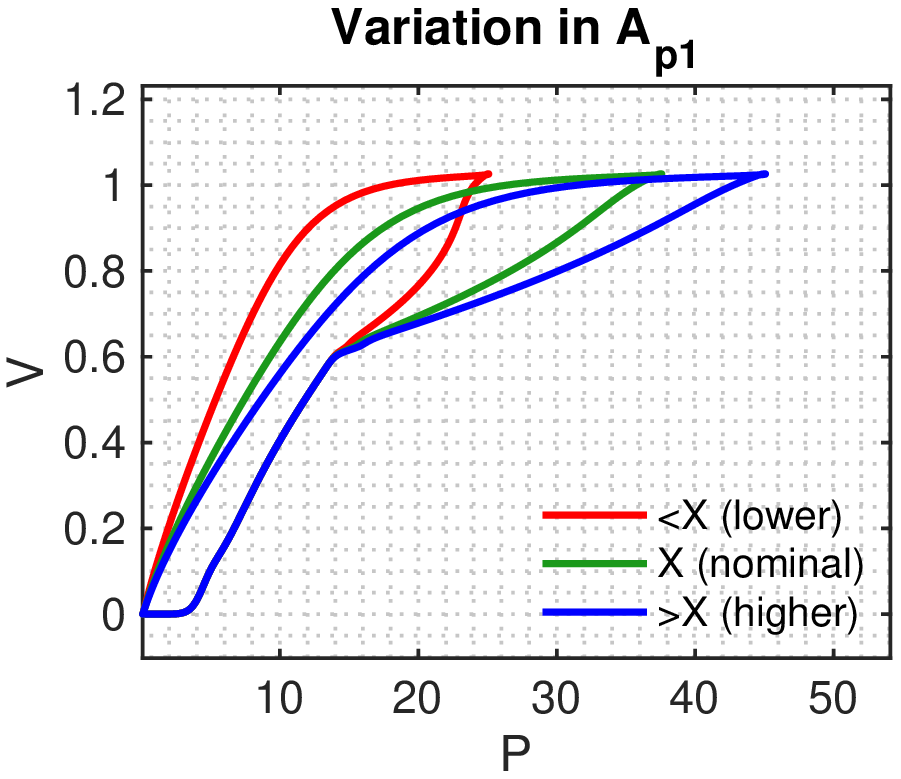}}}
\centerline{\subfigure{\includegraphics[width = 4.5cm]{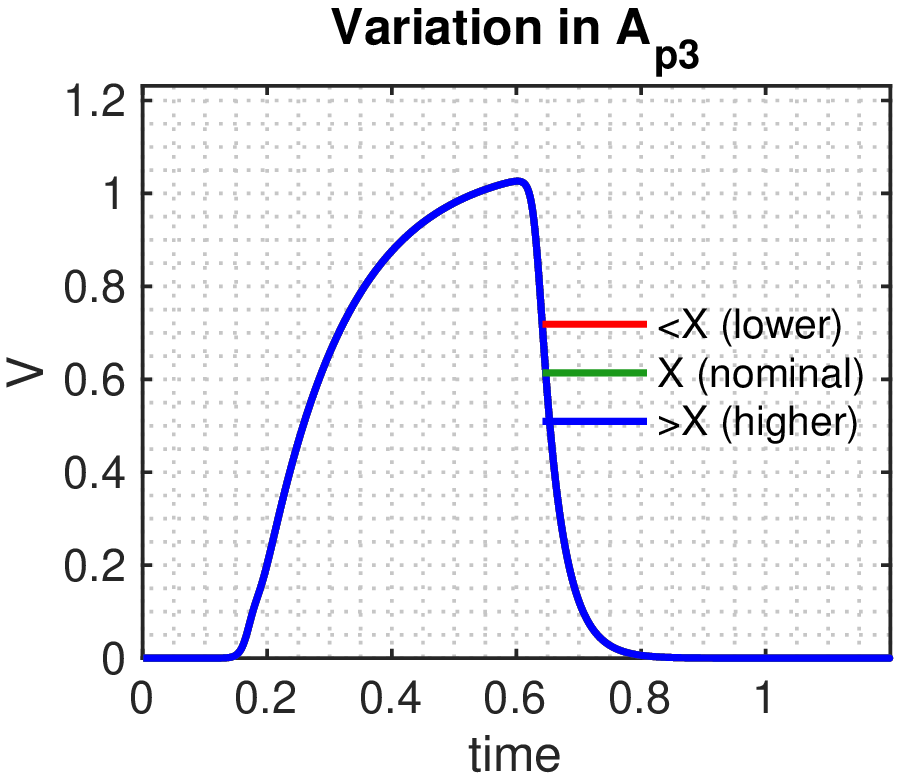}}
\subfigure{\includegraphics[width = 4.5cm]{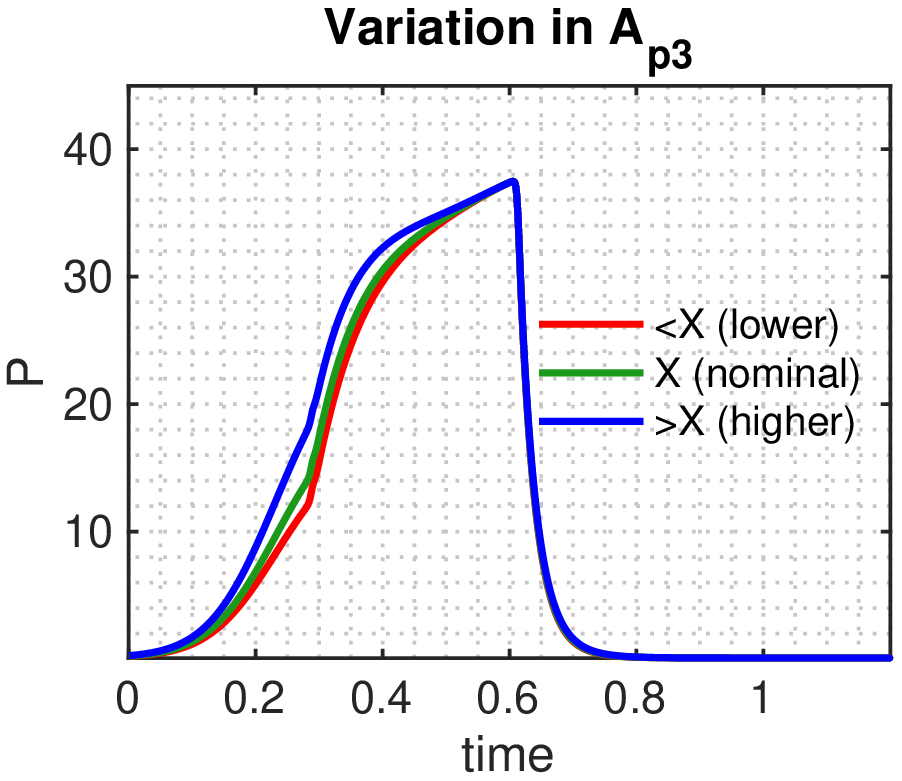}}
\subfigure{\includegraphics[width = 4.5cm]{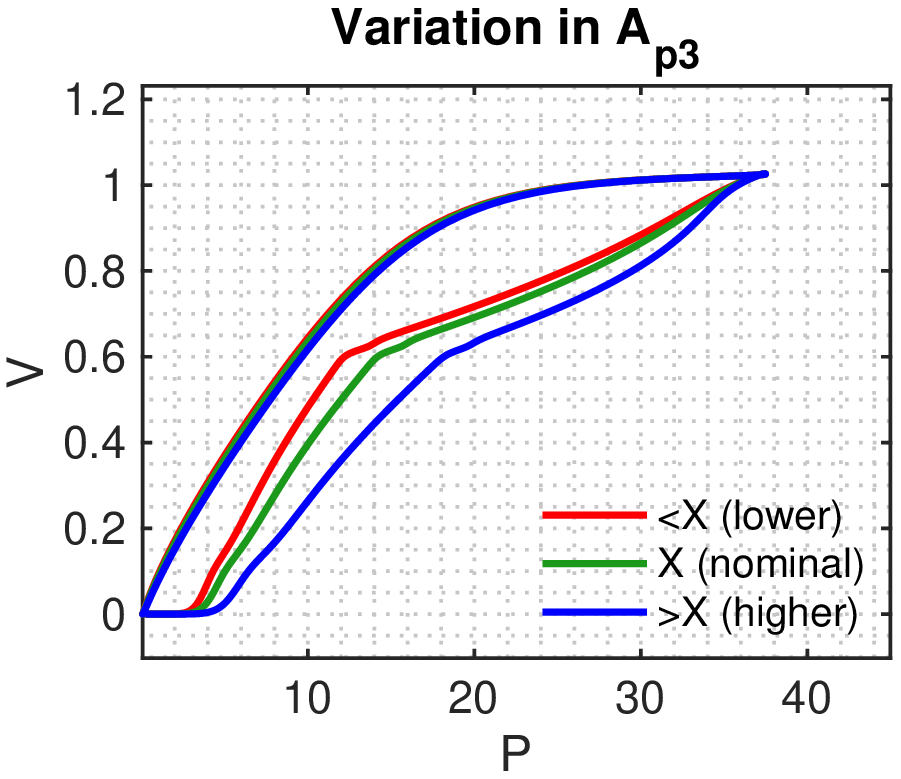}}}
 \label{FigS5}
 \caption{Figure S5: Five parameters in the pressure model could have direct physiological meaning. These are $a_3$, $b_3$, $\beta_3$, $A_{p_{1}}$ and $A_{p_{3}}$. Simulated response of the volume and pressure models are shown when the specific parameter was varied while considering the nominal parameter values shown in Table~1 for the mouse model PCV, healthy case (Fig.~6). Here, variable $X$ corresponds to the nominal value of the parameter while $<X$ and $>X$ represent smaller and larger values than the nominal value, respectively. }
\end{figure}

\begin{figure}[t!]
\centerline{\includegraphics[width = 1\textwidth]{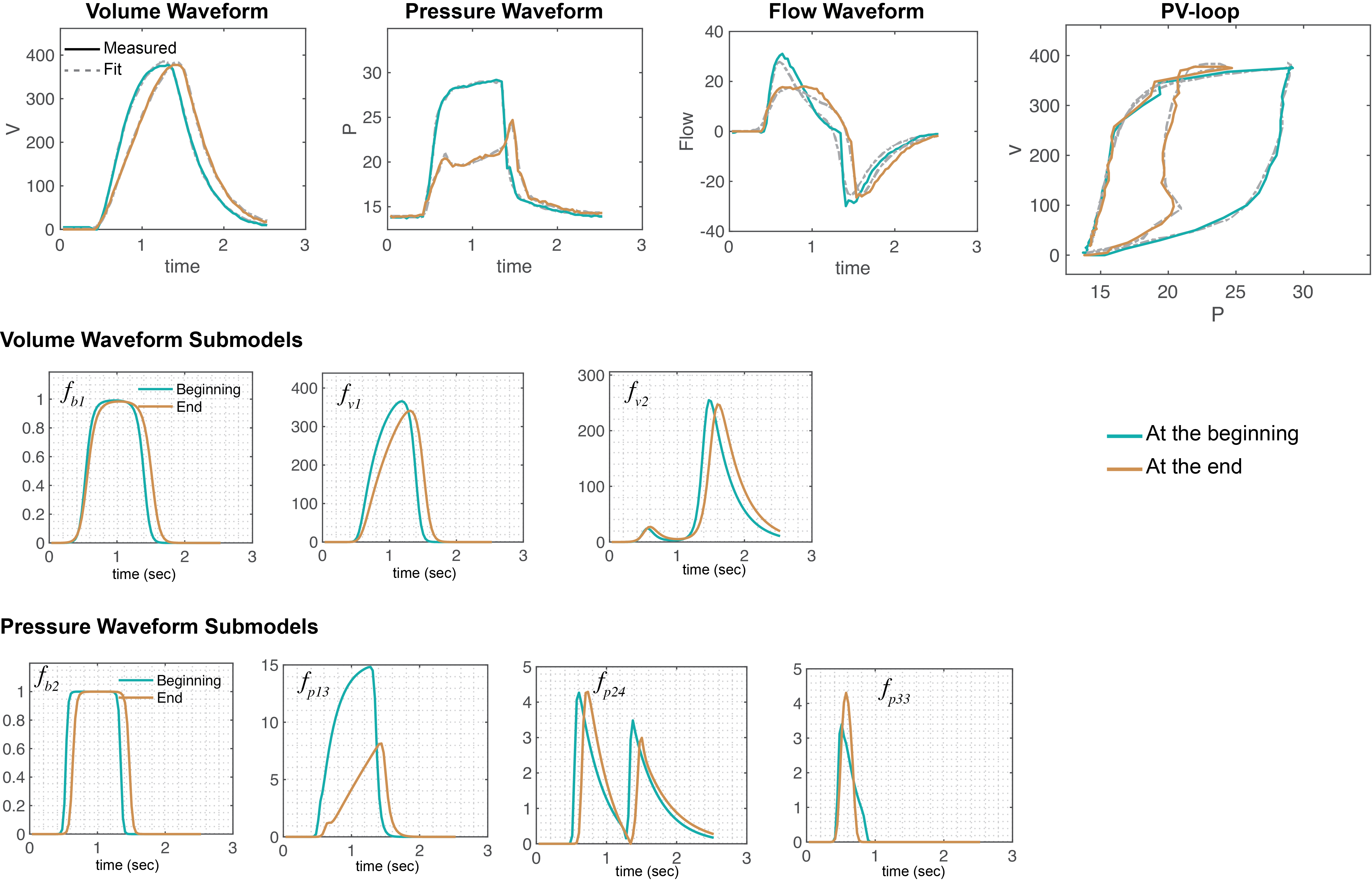}}
\caption{Figure S7: Clinical data of a human patient is compared with the model data at two different time points. In the first row, the measured response is shown in solid lines while the model inferred response is shown in dashed lines. Changes in the volume and pressure submodels are shown in the second and third rows, respectively (in solid lines). The volume and pressure models shown in Eqns.~(1)-(5) and (6)-(18) were used to generate the best-fit model response using estimated mean parameter values shown in Table~S2, respectively. The respective uncertainties in the parameter values is shown in Table S2.
}
\label{figS7}
\end{figure}
 \clearpage

 \begin{table}[t!]
 \begin{center}
 \begin{tabular}{ c }
 \includegraphics[width = 1\textwidth]{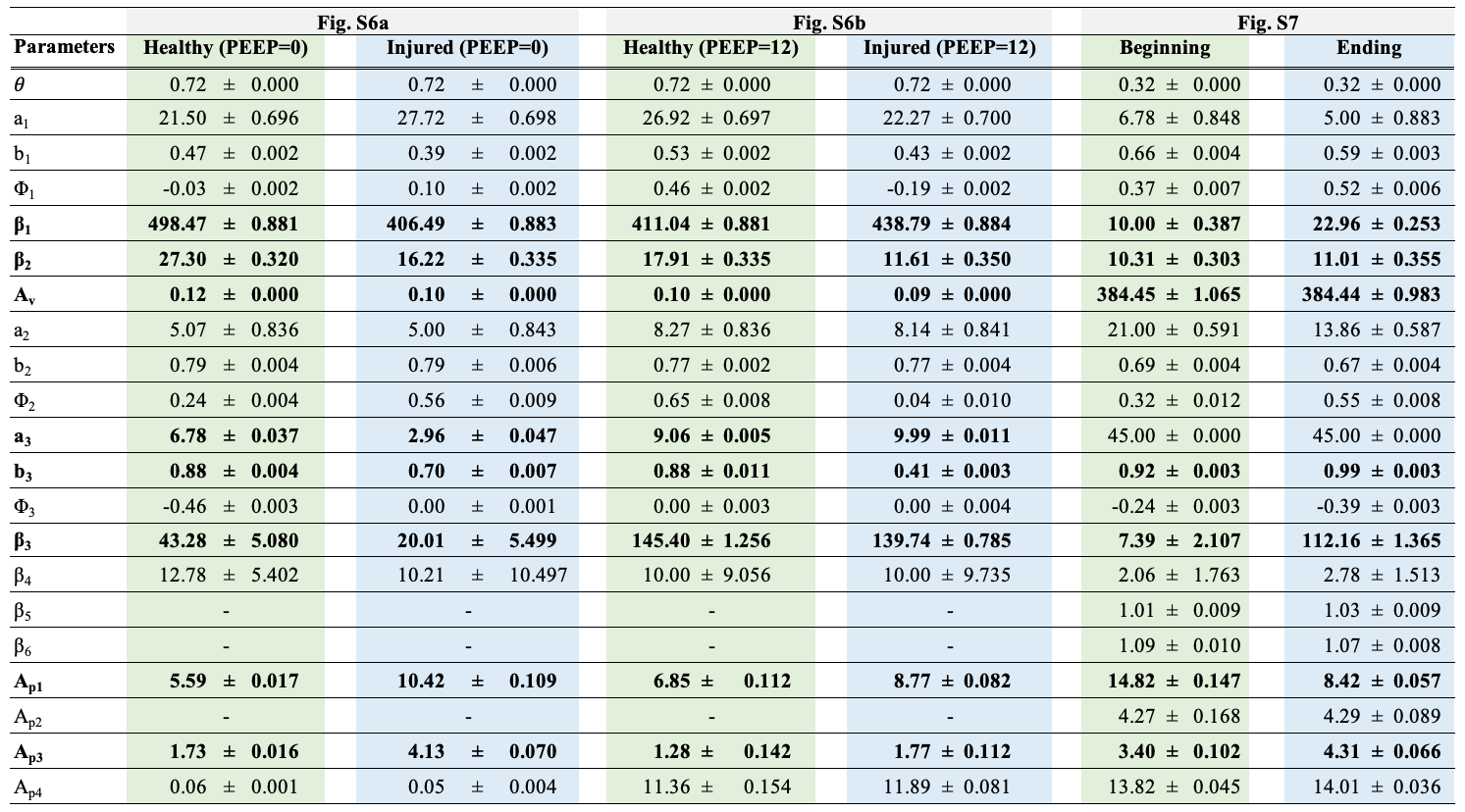}\\
 \end{tabular}
 \caption{Table S2: Estimated model parameters obtained from the optimization scheme for the results shown in Fig.~S6, S7. The error values were determined using the standard error of the mean. The parameters that are correlated with a known measures of lung physiology are in bold. N =1000.}
 \label{tableS3}
 \end{center}
 \end{table}

 \begin{table}[t!]
 \begin{center}
 \begin{tabular}{ c }
 \includegraphics[width = 1\textwidth]{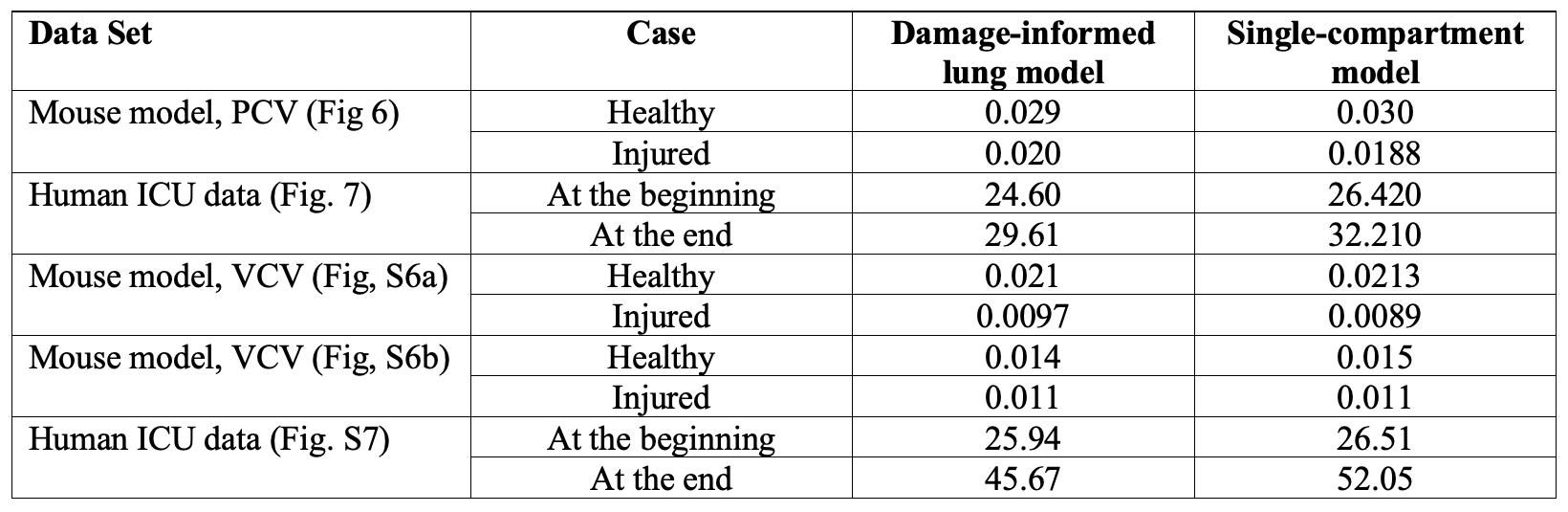}\\
 \end{tabular}
 \caption{Table S3: Comparing lung compliance values extracted by fitting the single-compartment model to the mouse and human data with the damaged-informed lung model. In later case, $A_{v}/A_{p_{1}}$ ratio was used to calculate the compliance using the parameters values shown in Table 1 and Table S1.}
 \label{tableS4}
 \end{center}
 \end{table}

}